\documentclass[twocolumn,times,Astrosymb]{aastex63}
\usepackage{xcolor}
\newcommand{\ovi}{\ion{O}{6}}
\newcommand{\cii}{\ion{C}{2}}
\newcommand{\siii}{\ion{Si}{2}}
\newcommand{\siiii}{\ion{Si}{3}}
\newcommand{\nv}{\ion{N}{5}}
\newcommand{\hi}{\ion{H}{1}}
\newcommand{\lya}{Ly$\alpha$}
\newcommand{\lyb}{Ly$\beta$}
\newcommand{\cosigrm}{COS-IGrM}
\newcommand{\Lstar}{L$_{\star}$}
\newcommand{\igrm}{IGrM}

\newcommand{\msun}{$\rm{M_{\odot}}$}

\newcommand{\kms}{$\rm{kms^{-1}}$}

\usepackage{multirow}
\usepackage{amsmath}
\usepackage{afterpage}

\accepted{\today}
\submitjournal{ApJ}

\shorttitle{COS-IGrM: Survey and Results}
\shortauthors{McCabe et al.}

\begin{document}

\title{Detection of a Multiphase Intragroup Medium: Results from the COS-IGrM Survey}

\correspondingauthor{Tyler McCabe}
\email{tyler.mccabe@asu.edu}

\author[0000-0002-5506-3880]{Tyler McCabe}
\author[0000-0002-2724-8298]{Sanchayeeta Borthakur}
\affiliation{School of Earth \& Space Exploration, Arizona State University,
	Tempe, AZ 85287-1404, USA}
\author[0000-0001-6670-6370]{Timothy Heckman}
\affiliation{Johns Hopkins University, Baltimore, MD 21218, USA}
\author[0000-0002-7982-412X]{Jason Tumlinson}
\affiliation{Space Telescope Science Institute, Baltimore, MD 21218, USA}
\author[0000-0002-3120-7173]{Rongmon Bordoloi}
\affiliation{Department of Physics, North Carolina State University, Raleigh, North Carolina, 27695, USA}
\author[0000-0003-2842-9434]{Romeel Dave}
\affiliation{Scottish Universities Physics Alliance, Institute for Astronomy, University of Edinburgh, Royal Observatory, Edinburgh EH9 3HJ, UK}

\begin{abstract}
We present the results of the COS Intragroup Medium (COS-IGrM) Survey that used the Cosmic Origins Spectrograph on the \textit{Hubble Space Telescope} to observe a sample of 18 UV bright quasars, each probing the intragroup medium (IGrM) of a galaxy group. We detect \lya, \cii, \nv, \siii, \siiii, and \ovi\ in multiple sightlines. The highest ionization species detected in our data is \ovi, which was detected in 8 out of 18 quasar sightlines. The wide range of ionization states observed provide evidence that the IGrM is patchy and multiphase. We find that the \ovi\ detections generally align with radiatively cooling gas between 10$^{5.8}$ and 10$^{6}$~K. 
The lack of \ovi\ detections in 10 of the 18 groups illustrates that \ovi\ may not be the ideal tracer of the volume filling component of the \igrm. Instead, it either exists at trace levels in a hot \igrm\ or is generated in the boundary between the hotter \igrm\ and cooler gas.
\end{abstract}



\section{Introduction} \label{sec:intro}
The majority of galaxies in the Universe exist in groups, where the dark matter halos cover mass ranges of $10^{12} \la \rm{M_{halo}} \la 10^{14.5}$~\msun \citep{Tully1987}. The diffuse, hot gas gravitationally bound to the group is commonly referred to as the intragroup medium (IGrM) and may constitute a significant entry into the missing baryon problem \citep{Persic1992, Fukugita2006, Spergel2007}. The effect on galaxy evolution of the IGrM and the halos of groups remains uncertain. The gas in galaxy group halos can be characterized through X-rays, the Sunyaev-Zel'dovich (SZ) effect, and through UV absorption lines from background quasars (QSOs).

Early \igrm\ detections were based on \emph{ROSAT} observations of high mass, elliptical rich groups \citep{Mulchaey1996Xray, Helsdon2000, Mulchaey2000}. These groups were believed to be more massive than spiral rich groups and hence more luminous in X-rays. From these observations, initial scaling relations \citep{Helsdon2000} were derived and the mass of the hot gas was determined to be comparable to the stellar mass of the galaxies \citep{Mulchaey1996Xray}. More recently, \cite{Bregman07} studied \ion{O}{7} absorption to distinguish it's origin as from the Milky Way's galactic halo or from the Local Group's IGrM. In searching for the IGrM, they found that the Milky Way halo models were preferred, but a contribution to the \ion{O}{7} absorption from the IGrM could not be conclusively ruled out. 

The thermal SZ effect, where cosmic microwave background (CMB) photons are scattered as a result of energetic, free electrons, provides an alternative means of observing the diffuse gas bound to dark matter halos. While the SZ effect is typically used to analyze galaxy clusters, recent studies using large stacks have led to detections around galaxy groups and individual galaxies \citep[and references therein]{Greco2015, Vikram2017, Bregman2018, Pratt2020, Tanimura2020}. As noted in \cite{LeBrun2015, Tumlinson2017} and \cite{Tanimura2020}, the gas content of galaxy halos down to $10^{11}$\msun\ comes into tension with existing X-ray observations as the self-similar scaling relations appear to fail. 

However, \cite{LeBrun2015} proposes that the discrepancy may result from the low resolution of the Plank SZ map and therefore might not be as robust at low radii ($r \la r_{500}$) when compared to X-ray observations. This effect was reproduced using X-ray simulations were convolved with the Plank beam \citep{LeBrun2015}. Cosmological ``zoom in'' simulations by \cite{VandeVoort2016} find that hot gas near the virial temperature causes more consistent X-ray luminosity scaling relations for halos with $\rm{M_{halo}} \ga 10^{13}$~\msun, while less massive halos show X-ray luminosities that are more strongly affected by star formation feedback. 

Ultraviolet (UV) absorption lines observed in the spectra of background QSOs remain one of the most robust methods to probe gas at intermediate temperatures, where the gas is not hot enough for X-ray emission. QSO absorption lines (QALs) have shown that a significant amount of baryons lie in the diffuse gas that makes up the intergalactic medium (IGM) \citep{Rauch1998, Shull2012}. This provides means of probing the composition of the IGrM since the large majority of galaxy groups are lower in mass and do not have the temperature and density necessary for X-ray emission. At the virial temperature of typical galaxy groups, \cite{Mulchaey1996} predicted the existence of broad, shallow \ovi\ absorption with Lyman series transitions without lower ions such as \ion{C}{4} and \nv\ based upon collisional ionization equilibrium (CIE) models. In this scenario, \ion{C}{4} and \nv\ are present at levels not currently detectable with current instruments such as the Cosmic Origins Spectrograph (COS; \cite{Green2012}) aboard the \emph{Hubble Space Telescope}.

With this background, studies by \cite{Tripp2000, TrippSavage2000} and \cite{Stocke2006} used background quasars to search for \ovi, but the data were inconclusive in correlating \ovi\ absorption with the IGrM. \cite{Stocke2014} conducted redshift surveys around 14 previously detected broad \lya\ and \ovi\ detections, which were indicative of gas above 10$^5$~K. They found galaxy groups around these QSO sightlines and concluded with 2$\sigma$ confidence that these absorbers were due to the group environment and not the nearest galaxy to the sightline. The possibility that the \ovi\ detections were tracing cooler clouds rather than the hot component of the \igrm\ was still a hypothesis and it lacked any direct observational confirmation. Other \ovi\ studies by \cite{Pointon2017} and \cite{Stocke2017} compared the detections in group environments to the circumgalactic medium (CGM) of isolated galaxies. These studies found that group environments contained \ovi\ absorption that could be modeled with broader components than isolated systems and concluded that \ovi\ was characteristic of the boundary between cooler CGM gas and the hotter IGrM. 

Studies by \cite{Tripp2008,Savage10,Savage12,Savage14} and \cite{Rosenwasser2018} detected \ovi\ absorption features that are consistent with multiphase gas at both cooler and hotter temperatures that could be produced by photoionization or collisional ionization, respectively.

However, from these studies, it is difficult to distinguish the origin of \ovi\ as being due to the boundary of multiphase gas in the \igrm\ or resulting from the CGM of member galaxies. The COS-Halos Survey \citep{Tumlinson2011,Tumlinson2013,Peeples2014,Werk2014,Werk16} analyzed spectra of 44 QSO-galaxy pairs and found \ovi\ in addition to a significant amount of metals in the halos of isolated galaxies. The COS-Halos Survey found a strong correlation of \ovi\ detections in the inner CGM of star forming galaxies leading to the idea that it may originate from large streams of cooling gas or from the hotter component of the CGM if a temperature gradient is assumed as opposed to a uniform halo at the virial temperature \citep{Werk2014, McQuinn2018}.

\cite{Heckman2002} and \cite{Bordoloi2017} (and references therein) show that \ovi\ observed in the intergalactic medium, CGM of galaxies, and the Milky Way halo can be explained by radiative cooling models. These models agree with observations and simulations showing complex, multiphase structures at the interfaces between hot and cold gas \citep{Oppenheimer2009,Churchill2012,Pachat2016, Narayanan2018,Ahoranta2020}.

Recently, \cite{Stocke2019} carried out a survey of 12 galaxy groups paired with background QSOs to look for \ovi\ associated with galaxy groups. They find that \ovi\ was not uniformly detected within the sample, leading to the idea that CGM-like clouds can escape individual galaxies and can be observed within the group. They do not find evidence that these clouds can easily escape the group, which means that galaxy groups might be ``closed-boxes'' for galaxy evolution. Lastly, they conclude that the gas traced through \ovi\ is not volume filling and that a hotter component is necessary for a complete baryon census in the group environment.

Here we present the COS-IGrM survey, designed to probe the IGrM of lower mass groups than those probed by \cite{Stocke2019}, where \ovi\ could be a better tracer of the \igrm\ due to lower virial temperatures. The COS-IGrM sample consists of 18 galaxy groups paired with background UV bright quasars (QSOs) and was selected without bias towards predefined sightlines with \ovi\ detections. This is the largest sample of low redshift ($z_{gp} \le 0.2$) galaxy groups ever probed for \ovi\ associated with the \igrm.

This paper is organized as follows: in \S\ref{sec:sample} we describe the \cosigrm\ sample, \S\ref{sec:observations} details the HST/COS observations along with the data reduction and analysis, \S\ref{sec:results} presents the results of the survey, \S\ref{sec:discussion} discusses the overall significance of our results and \S\ref{sec:conclusion} presents the conclusions of our survey.

\movetabledown=6cm
\begin{rotatetable*}
\begin{deluxetable*}{llcccccccccccccccl}
\tabletypesize{\scriptsize}
\tablecolumns{18}
\centering
\tablecaption{COS-IGrM Sample}
\tablehead{
\colhead{Sightline}		&		\colhead{Group}	&	\colhead{RA$_{gp}$} & \colhead{DEC$_{gp}$} &	\colhead{$z_{gp}$}	&	\colhead{RA$_{qso}$} &  \colhead{DEC$_{qso}$} &	\colhead{z$_{qso}$} &    \colhead{N$_{gp}$}		&		\colhead{$\sigma_{Tago}$}		&		\colhead{R$_{vir, Tago}$}		&		\colhead{$\rho_{QSO}$}   &   \colhead{log[M$_{\rm{halo}}$]}   &   \colhead{R$_{vir}$}    &   \colhead{$\sigma_{gp}$}  & \colhead{RA$_{cg}$}  & \colhead{DEC$_{cg}$} &   \colhead{Subset}    \\
\colhead{(1)}         			 & \colhead{(2)}       		 & \colhead{(3)}      			 &	\colhead{(4)}         			 & \colhead{(5)}      			  & \colhead{(6)}       		&	\colhead{(7)}   &   \colhead{(8)}   &   \colhead{(9)}   &   \colhead{(10)}   &   \colhead{(11)} &   \colhead{(12)}  &   \colhead{(13)}  &   \colhead{(14)}  &   \colhead{(15)}  &   \colhead{(16)}   &   \colhead{(17)}   &   \colhead{(18)}
}
\startdata
1	&	J0841+1406		& 130.493  & 14.100  &	   	0.1250   & 130.496  & 14.112  &	1.2514	&		3		&		81.3		&		638		&		95        &   13.32    &  567   &   229    &   130.502    &   14.097    &  CGM + IGrM  \\
2	&	J1017+4702 		& 154.246  & 47.049  &		0.1637   & 154.379  & 47.040  &	0.3350	&		3		&		91.9		&		710		&		926       &   13.33    &  573   &   232    &   154.302    &   47.025    &  IGrM\\
3  	&	J1020+1003		& 155.222  & 10.137  &		0.1229   & 155.235  & 10.059  &	0.6074	&		3		&		223.0		&		466		&		632       &   13.19    &  516   &   209    &   155.200    &   10.092    &  IGrM\\
4  	&	J1025+4808		& 156.286  & 48.110  &		0.1333   & 156.304  & 48.148  &	0.3317	&		4		&		135.7		&		315		&		340       &   13.31    &  564   &   228    &   156.293    &   48.164    &  CGM + IGrM \\
5	&	J1102+0521		& 165.676  & 5.296   &		0.1314   & 165.653  & 5.355   &	0.4987	&		4		&		210.1		&		528		&		536       &   13.23    &  532   &   215    &   165.695    &   5.355     &  IGrM \\
6	&	J1126+1204		& 171.691  & 12.122  &		0.1640   & 171.637  & 12.077  &	0.9759	&		5		&		76.0		&		686		&		707       &   13.61    &  709   &   287    &   171.641    &   12.094    &  CGM + IGrM\\
7 	&	J1127+2654		& 171.875  & 26.900  &		0.1521   & 171.902  & 26.914  &	0.3790	&		3		&		89.6		&		492		&		267       &   13.40    &  606   &   245    &   171.904    &   26.904    &  CGM + IGrM \\
8 	&	J1216+0712		& 184.140  & 7.148   &		0.1360   & 184.169  & 7.207   &	0.5864  &		3		&		242.7		&		404		&		572       &   13.35    &  579   &   234    &   184.161    &   7.174     &  IGrM \\
9	&	J1301+2819		& 195.206  & 28.410  &		0.1439   & 195.254  & 28.329  &	1.3597	&		3		&		141.2		&		613		&		836       &   13.33    &  574   &   232    &   195.243    &   28.361    &  IGrM \\
10	&	J1339+5355		& 204.814  & 53.990  &		0.1590   & 204.802  & 53.924  &	0.2933	&		4		&		85.1		&		523		&		653       &   13.43    &  620   &   251    &   204.852    &   53.969    &  IGrM \\
11	&	J1343+2538		& 206.031  & 25.700  &		0.0749   & 205.986  & 25.647  &	0.0866	&		3		&		43.0		&		291		&		346       &   12.85    &  396   &   160    &   205.975    &   25.675    &  IGrM \\
12	&	J1344+5546		& 206.195  & 55.802  &		0.1546   & 206.198  & 55.782  &	0.9369	&		3		&		177.2		&		570		&		194       &   13.44    &  625   &   253    &   206.210    &   55.795    &  CGM + IGrM \\
13	&	J1348+4303		& 207.343  & 43.017  &		0.0947   & 207.228  & 43.053  &	0.2748	&		7		&		207.7		&		574		&		580       &   13.60    &  705   &   285    &   207.276    &   43.047    &  IGrM \\
14	&	J1408+5657		& 212.222  & 56.911  &		0.1302   & 212.226  & 56.962  &	0.3363	&		3		&		43.2		&		403		&		427       &   13.46    &  632   &   256    &   212.257    &   56.980    &  IGrM \\
15	&	J1424+4214		& 216.247  & 42.261  &		0.0995   & 216.231  & 42.235  &	0.3162	&		3		&		46.3		&		459		&		189       &   13.17    &  506   &   205    &   216.219    &   42.251    &  CGM + IGrM \\
16	&	J1426+1955		& 216.465  & 19.914  &		0.1091   & 216.555  & 19.924  &	0.2133	&		3		&		111.1		&		594		&		616       &   12.89    &  407   &   165    &   216.504    &   19.867    &  IGrM \\
17	&	J1428+3225		& 217.304  & 32.404  &		0.1308   & 217.246  & 32.419  &	0.6270	&		4		&		123.6		&		777		&		433       &   13.32    &  567   &   229    &   217.320    &   32.431    &  IGrM \\
18	&	J1617+0854		& 244.430  & 8.913   &		0.0993   & 244.349  & 8.904   & 0.2064  &       6		&		69.3		&		392		&		533       &   13.36    &  585   &   237    &   244.395    &   8.884     &  IGrM
\enddata
\tablecomments{Columns (10) and (15) have units of $\rm{kms^{-1}}$; \ Columns (11), (12), and (14) have units of kpc; \ Column (13) has units of \msun. RA$_{cg}$ and DEC$_{cg}$ are the coordinates of the closest member galaxy to the QSO sightline. }
\label{tab:sample}
\end{deluxetable*}
\end{rotatetable*}
\normalsize

\begin{figure}
\centering
\plotone{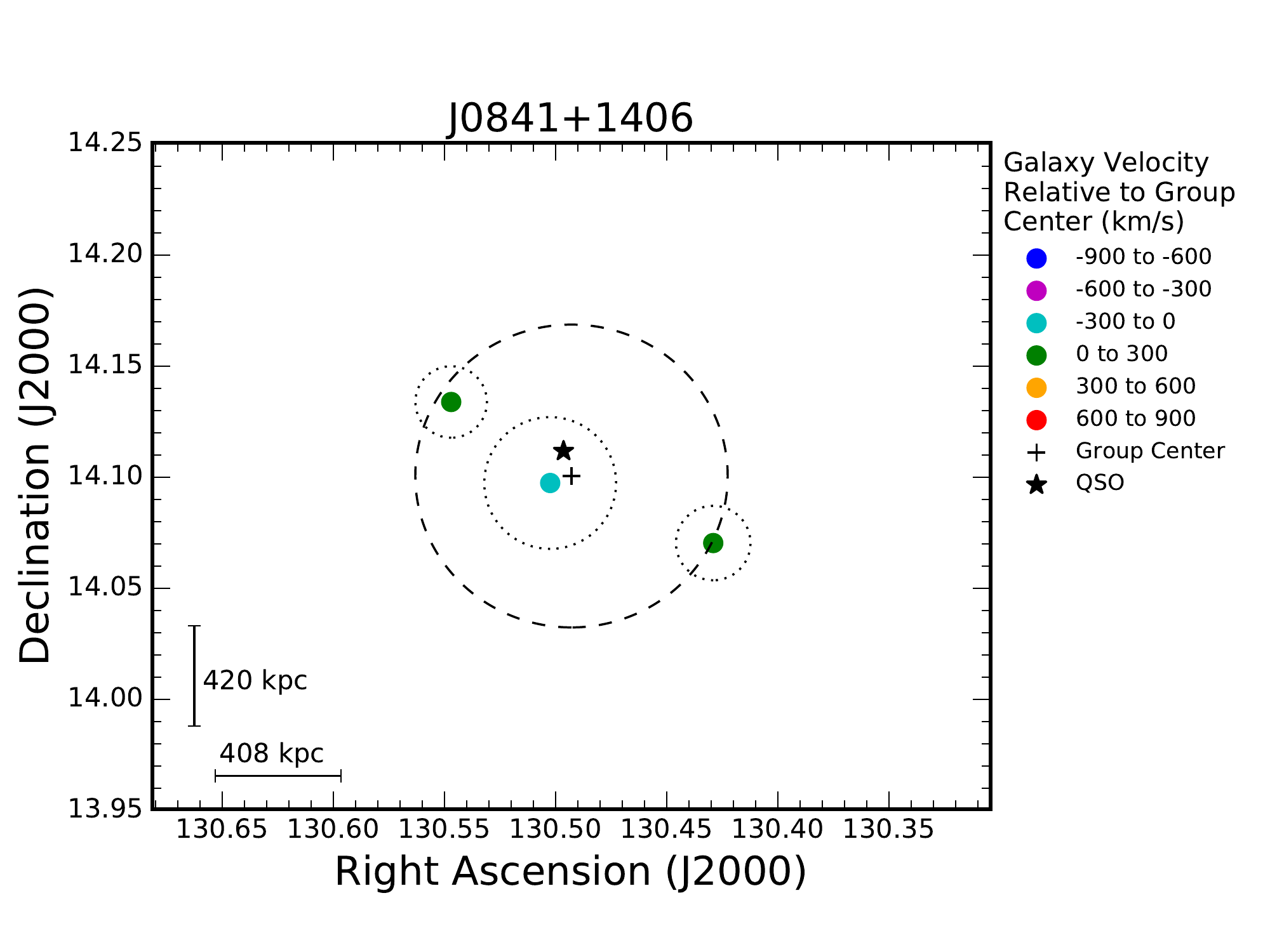}
\caption{Environment of galaxy group J0841+1406. The color of the points represent the velocity of the member galaxies relative to the center of the group. The thick, dashed line represents the virial radius of the group and the thin, dotted lines represent the virial radii of the member galaxies.}
\label{fig:env}
\end{figure}

\section{Sample} \label{sec:sample}

The COS-IGrM sample\footnote{The data is available at MAST: \dataset[10.17909/t9-wqg9-9043]{\doi{10.17909/t9-wqg9-9043}}} is composed of 18 galaxy groups\footnote{In the HST proposal, there were 19 sightlines; however, one sightline had an insufficient signal to noise to use for this analysis.} each with a background QSO with a GALEX far ultraviolet (FUV) magnitude brighter than 19. The sample was created by cross referencing the \cite{Tago2010} galaxy group catalog with the catalog of unique GALEX Data Release 5 QSOs \citep{Bianchi2011}. Four additional criteria were implemented to create a robust sample from the \cite{Tago2010} group catalog:
\begin{enumerate}
\item The groups must have at least three spectroscopically confirmed members;
\vspace{-0.25cm}
\item The group redshifts must be between $0.075 \le z_{gp} \le 0.2$ for \ovi\ to be within the COS bandpass; 
\vspace{-0.25cm}
\item The QSO redshift, $z_{QSO} > z_{gp} + 0.1$ to eliminate confusion between absorption features from the group and from the QSO;
\vspace{-0.25cm}
\item The QSO impact parameter must be less than 1.5 times the group's virial radius ($1.5\rm{R_{vir}}$).
\end{enumerate}

In order to ensure that the groups in the COS-IGrM sample are physical groups, the \cite{Yang2007} galaxy group catalog was used to look for confirmed groups in the same location and with similar halo mass as provided by the \cite{Tago2010} catalog. This additional check aided in confidently identifying galaxy groups with as little as three spectroscopically confirmed members. The environment of one group in our sample is shown in Figure \ref{fig:env} along with the location of the background QSO. The remaining 17 group environments are listed in Appendix \ref{ap:env}. Since galaxy groups were identified from Sloan Digital Sky Survey (SDSS) spectroscopic group catalogs, our sample is biased towards groups with luminous galaxies, $L \ga L_*$.

We also include sightlines from \cite{Stocke2019} in our analysis to extend the sample to larger halo masses. In order to make sample parameters consistent with those from \cite{Stocke2019}, we re-defined the group parameters such as the group halo mass, virial radius, and velocity dispersion through the following relations from their paper:

\begin{equation}
M_{gp} = 310\times \left( \frac{L_{gp}}{L_*}\right) \times 10^{10} ~M_{\odot}
\end{equation}
Where $L_{gp}$ is total the r-band luminosity of the group members calculated via the r-band magnitudes from the \cite{Tago2010} catalog. 
\begin{equation}
R_{vir} = 957\times \left(\frac{M_{gp}}{10^{14}}\right)^{1/3} \rm{kpc}
\end{equation}
and
\begin{equation}
\sigma _{gp} = 387\times \left( \frac{M_{gp}}{10^{14}M_{\odot} } \right)^{1/3}~\rm{kms^{-1}} 
\end{equation}

In Equation 2, the virial radius is defined as the limit where the overdensity of the medium is equal to 200$\rho_{crit}$ as described in \cite{Shull2012} and \cite{Stocke2019}. The full properties of our galaxy group sample and each corresponding background QSO are listed in Table \ref{tab:sample} along with the adopted values for the halo mass, virial radius, and velocity dispersion.

While each sightline probes the IGrM, some also pass through within the CGM of member galaxies. Therefore, we divide our sample into two sub-samples - one with sightlines passing through the CGM of member galaxies and the other where the sightline is at impact parameters larger than the viral radius of the member galaxies (assuming an isolated halo). This assumption may over-estimate the size of the CGM of group members; however, it remains the most reliable radius estimate without requiring extensive cosmological simulations. The first group contains six sightlines and are referred to as the CGM~+~IGrM. The remaining 12 sightlines fall in the latter category. In the absence of a deeper spectroscopic survey, our limiting magnitude allows us to claim that the ``pure" IGrM sightlines do not go through the CGM of any $\sim$L$_*$ galaxies. While it is possible that occasionally a much smaller galaxy could be close to the QSO sightline, care was taken to identify any possible galaxy candidates at the same redshift near the QSO sightline. Statistically, we do not find a significant number of possible member galaxies. This was followed up by recent multi object spectroscopy of two of the fields with the MMT and the Gemini Observatory, which confirmed this result , i.e., very few new galaxies were detected to be part of the groups and none very close to the QSO. These results will be discussed further in a future paper (McCabe et al. in prep).
In Table \ref{tab:sample}, we refer to the sub-grouping for each sightline/group as either \igrm\ or CGM~+~IGrM depending on the location of the QSO sightline.

\section{HST/COS Observations} \label{sec:observations}
The 18 QSOs in our sample were observed with the G130M grating of COS aboard the \textit{Hubble Space Telescope}. Data for 16 of the QSO sightlines were obtained under program 13314, while the remaining were obtained through archival data. The observations were designed to achieved a signal to noise (S/N) greater than 10 per resolution element for each sightline. This resolution is necessary in order to observe broad and shallow absorption lines that are expected to be associated with hot media. The spectra covered a observed frame wavelength from 1070$-$1465 \AA\ corresponding to a rest-frame wavelength range of 946$-$1295 \AA\ for the median redshift of the sample of $z$=0.1311.

The QSO spectra were created by coadding the individual exposures. The spectra were binned by three pixels, corresponding to half the resolution element, to enhance the S/N before any analysis was performed. As the first step, we identified the absorption lines associated with the groups. To do so, absorption lines in the entire spectra were identified. This is critical to ensure that the absorption lines associated with the target system are not blended with intervening or Milky Way absorbers. Special care has been given to the identification of metal-line species, in particular to \ovi. Fortunately, the redshift range of our sample resulted in the observed wavelength of \ovi\ shortward of 1215~$\rm \AA$, thus eliminating the possibility of any contamination from weak \lya\ absorbers in the IGM.

Continua were fit through an automated pipeline created and described by \cite{Tumlinson2011} and \cite{Werk2012}. A few sightlines exhibited complicated continua where the automated system failed and, as a result, those data were reduced individually by the authors following the prescriptions in \cite{Sembach1992} and \cite{Sembach2004}. The continuum fitting was done in a way consistent with the automated system. 

Our data covered absorption lines from species such as \lya\  $\lambda$1215, \lyb\ $\lambda$1025, \cii\ $\lambda$1036, \nv\ $\lambda$, \siii\ $\lambda$1190 $\lambda$1193 $\lambda$1260, \siiii\ $\lambda$1206 and \ovi\ $\lambda\lambda$1031 $\lambda\lambda$1037, which trace gas from $10^{2-6}$~K. Absorption features were searched within $\pm$800 $\rm{kms^{-1}}$ of the group's systematic redshift. Features beyond this range were not considered to be physically related to the group.

Features with an equivalent width greater than 3$\sigma$ were considered detections; otherwise, a 3$\sigma$ upper limit was estimated. The uncertainty corresponding to each equivalent width measurement was determined through the RMS noise of the data within the measurement window. Each feature was fit with a Voigt profile in order to determine the column density, doppler `$b$' parameter, and velocity centroid. For sightlines with multiple absorption systems, we determined the total column density by linearly adding up the components. Unless stated otherwise, the column density represented in the figures refers to the total column density along the line of sight.

\begin{figure}[]
\centering
\plotone{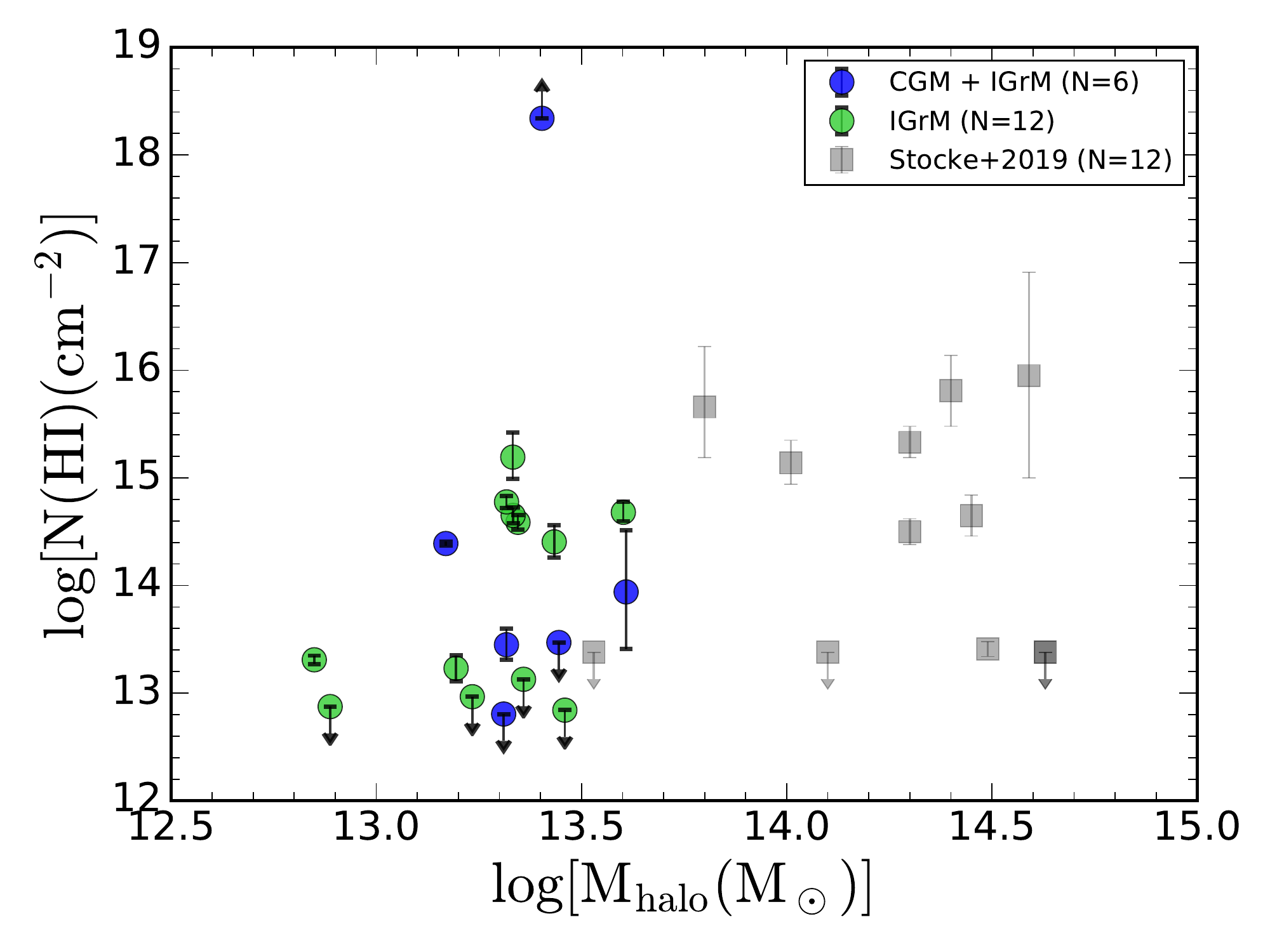}
\caption{\lya\ column density as a function of group halo mass. The \cosigrm\ sightlines are shown as filled circles and the data from \cite{Stocke2019} are are grey squares. The \cosigrm\ sample is further split into sightlines that are expected to probe the CGM and the IGrM of galaxy groups (blue data points) and those that probe just the \igrm\ of the group (green data points).}
\label{fig:lyamass}
\end{figure}

\begin{figure*}[]
\centering
\plottwo{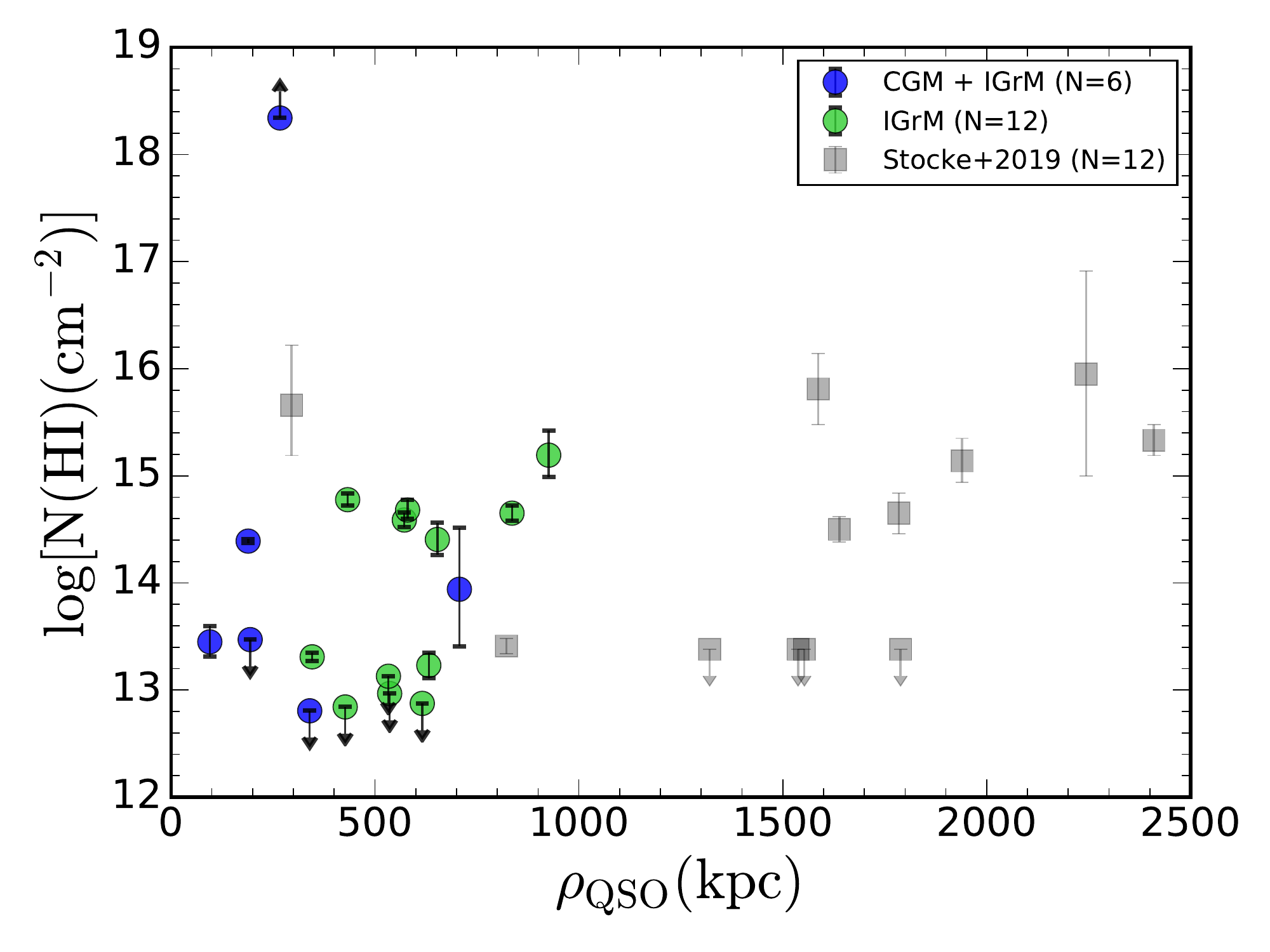}{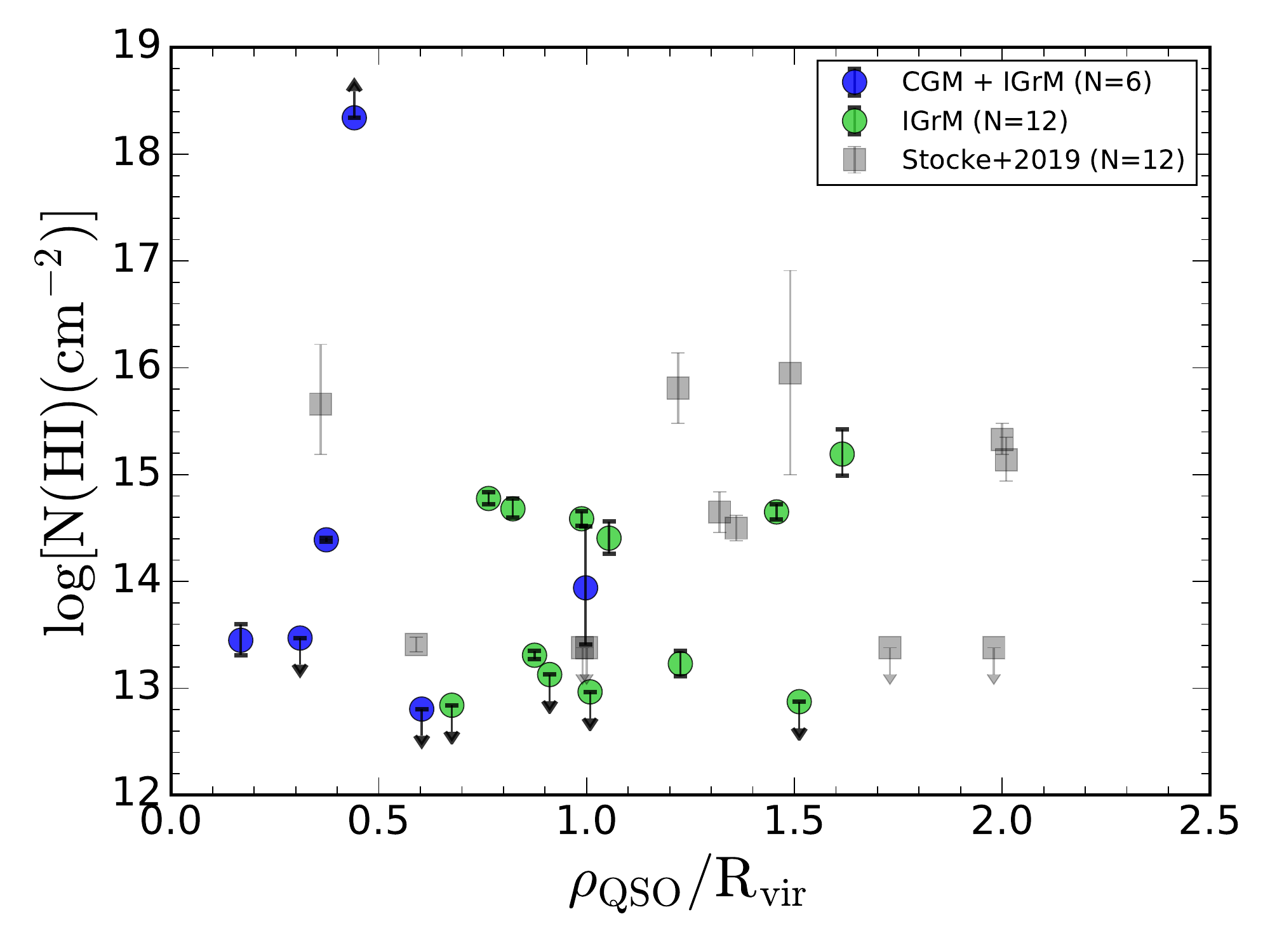}
\caption{\lya\ column densities as a function of projected QSO impact parameter (left) and the QSO impact parameter normalized by the group's virial radius (right). The colors are the same as for Figure \ref{fig:lyamass}. }
\label{fig:lyarho}
\end{figure*}

\section{Results} \label{sec:results}
We detected \lya, \lyb, \cii, \nv, \siii, \siiii\ and \ovi\ throughout the 18 sightlines. The detection rate of \lya\ is the highest at 67$\pm$5\% (12/18) followed by \ovi\ at 44$\pm$5\% (8/18). We also detected low ionization species such as \siii\ and \cii\ at detection rates of 6$\pm$5\% (1/18) and 28$\pm$5\% (5/18) respectively; the intermediate species \siiii\ at 28$\pm$5\% (5/18); and and high ionization \nv\ at 11$\pm$5\% (2/18).

Table \ref{tab:lines} presents measurements for each of these species for the \cosigrm\ sample\footnote{Table 3 is published in machine-readable format in the online journal.}. In the following subsections, we discuss and analyze the properties and distribution of each species. In our analysis, we also include data from the \cite{Stocke2019} IGrM survey, which covered higher mass groups. This allows us to search for trends over a larger range of halo masses and group sizes.  One significant difference to note between the two surveys is that unlike the \citeauthor{Stocke2019} sample, we do not eliminate QSO sightlines that fall within 0.25~$\rm{R_{gp}}$ in our sample selection.

\subsection{\ion{H}{1} \lya\ Absorption} \label{sec:lya}

We detected \lya\ absorption features in 12 of our 18 galaxy groups, with four groups having accompanying \lyb. Figure \ref{fig:lyamass} presents our \ion{H}{1} column density measurements as a function of the halo mass of the group. We found that lower mass halos exhibit a slightly narrower range of \lya\ column densities compared to higher mass halos. We see no evidence of varying column densities of \lya\ absorption between CGM~+~IGrM and IGrM sightlines. We find that there appears to be two main populations of data points: one with moderate column densities, log[N(\ion{H}{1})]$\sim$ 14.5-15, and another set clustered around log[N(\ion{H}{1})]$\sim$ 13. These groupings may indicate that the QSO sightlines are passing through patchy, non-uniform \ion{H}{1} clouds as opposed to a continuous distribution with a decreasing density gradient. 

One sightline, J1127+2654 (Figure 19), was seen to have saturated \lya\ and \lyb\ absorption. The column density of this absorption feature should be treated as a lower limit due to the absorption line occupying the flat regime of the curve of growth. This sightline probes the IGrM as well as the CGM of the closest galaxy to the sightline, which is at $\sim$70~\kms\ from the group's systematic velocity and an impact parameter of $\sim$119~kpc. This saturated \ion{H}{1} feature is composed of three components centered at -59, 32, and 116~\kms, respectively from the systemic velocity of the group, with the middle component being the strongest. The low impact parameter of $\sim$119~kpc from the closest galaxy to the sightline suggests that we are likely probing the CGM of this galaxy. The location of the QSO with respect to the group member in the full environment plot in Appendix \ref{ap:env}.

Figure \ref{fig:lyarho} shows the distribution of \lya\ column density as a function of impact parameter ($\rho_{\rm QSO}$) in the left panel and as a function of normalized impact parameter in the right panel ($\rho_ {\rm QSO}/\rm R_{vir}$). We overplot the \cite{Stocke2019} sample as grey squares in Figure \ref{fig:lyarho}. We find no statistically significant correlation between the column density of \lya\ absorbers and the QSO impact parameter using the Kendall's Tau correlation test provided in the \textsc{Astronomy SURvival Analysis} (ASURV) package \citep{ASURV1,ASURV2,ASURV3}. Using the ASURV Kendall's Tau test, we observed no correlation between the column density of \lya\ absorption and IGrM or CGM~+~IGrM sightlines. Most of the stronger \lya\ absorbers (log[N(\ion{H}{1})]$\sim$ 14.5-15) are seen in sightlines that pass through only the \igrm. 

The origins of cooler, partially neutral gas are not well understood. Possible scenarios include remnants of tidally stripped structures \citep{Davis1997,Bekki2009, Borthakur10, Nestor11, Gauthier13,Fossati2019, chen19, Peroux19}, in-situ condensation \citep{Voit19}, outflowing material from star forming galaxies \citep{Veilleux2005, Tripp2011, Nielsen18, frye19} and/or cold gas accretion from the intergalactic medium \citep{Keres2005, vogt15, Bielby17, borthakur19}. These processes are all capable of producing strong \lya\ absorption. On the other hand, lower column density absorbers and non-detections are seen in sightlines irrespective of whether they probe the CGM or just the IGrM. The presence of weak \lya\ absorbers ($\le 10^{14}~\rm{cm}^2$), including several non-detections, indicates that the sightlines pass through an ionized medium. 
In those sightlines, we do not find \ion{O}{6} or \ion{N}{5}, indicating that the medium must be at temperatures greater than 10$^6$~K, assuming collisional ionization equilibrium.
This phenomena might be related to the inability for galaxies in groups to continue the gas accretion necessary to fuel star formation. This has been observed in galaxy clusters \citep{Yoon2013, Gim21}, and when scaled to the group environment, could indicate the beginning of the preprocessing and quenching processes \citep{Zabludoff1998,McGee2009,Wetzel2013,Schawinski2014,Crossett2017,kacprzak21}.\\

\subsection{Low and intermediate-ionization tracing tracing cool/warm gas} \label{sec:low-ionization}

Apart from \lya, we also observe other transitions like \siii\, \cii\ and \siiii\ tracing gas up to the ionization potentials of 33.5~eV. \cii\ and \siiii\ are the most commonly detected low and intermediate-ionization species that are seen in five of the eighteen sightlines.  This is consistent with other studies of the CGM and IGM \citep{collins09,  shull09, lehner12, lehner15, richter16, borthakur16}. All of these absorbers are associated with strong, most likely saturated \lya\ absorbers.

\begin{figure}
\centering
\plotone{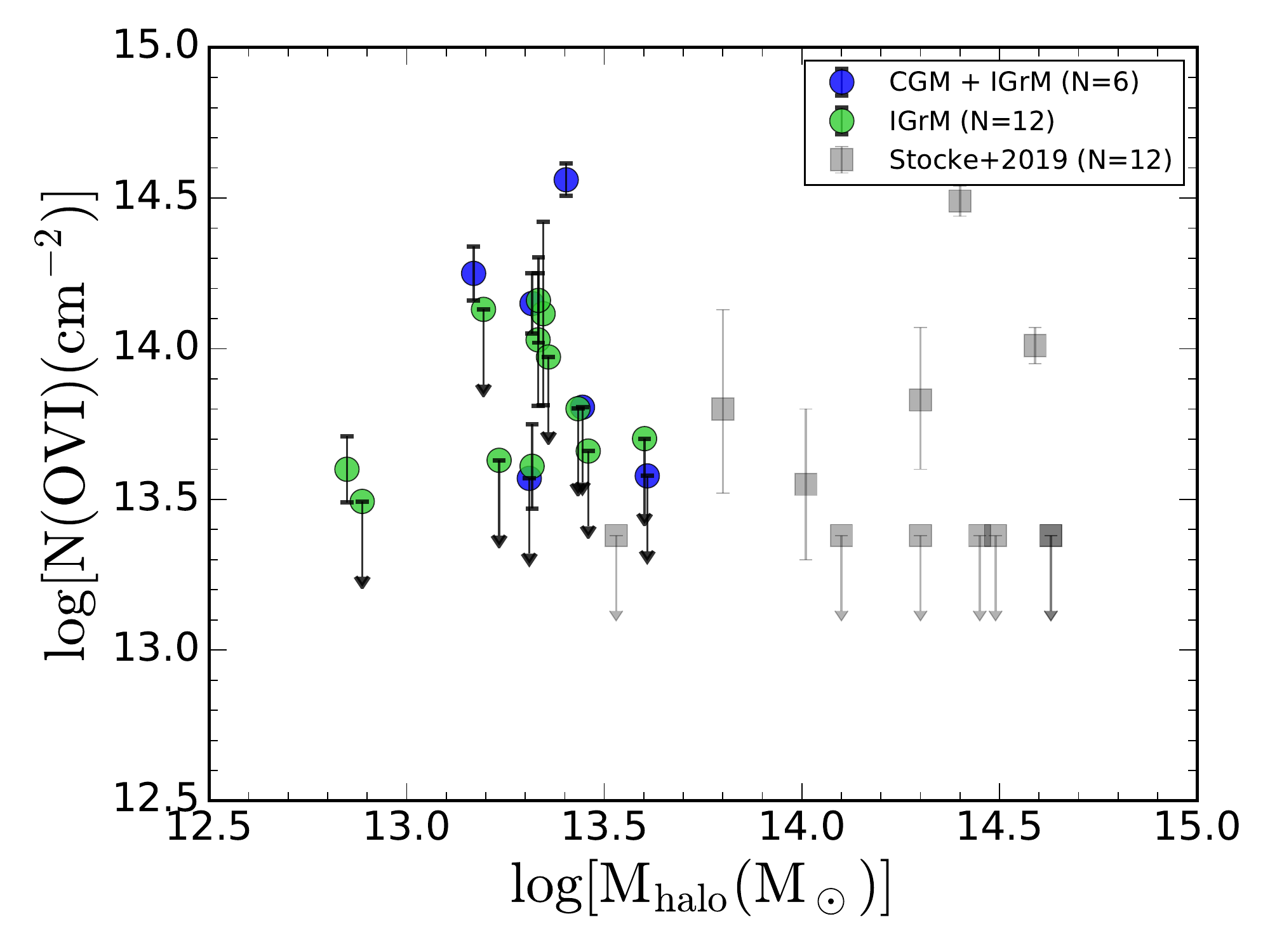}
\caption{\ovi\ column density detections and 3$\sigma$ upper limits as a function of group halo mass for the \cosigrm\ sample. The blue points indicate QSO sightlines that pass within the virial radius of individual group members, while the green points show sightlines that we expect to only probe \igrm. The gray points are from \cite{Stocke2019}, which systematically selects higher mass groups.}
\label{fig:ovimass}
\end{figure}

\subsection{\ovi\ and \ion{N}{5} absorption tracing highly ionized gas} \label{sec:ovi}
We detect \ovi\ absorbers in 8 and \ion{N}{5} absorbers in 2 out of our 18 sightlines. Each \nv\ absorption feature was also present with \ovi\ absorption. Of the eight detections, five of the sightlines were ``pure'' \igrm\ sightlines, while three sightlines were CGM~+~IGrM. We detected both the transitions of the \ovi\ doublet for two sightlines. Five sightlines showed the stronger of the two transition at \ovi\ $\lambda$1031~$\rm\AA$, while one sightline showed absorption at \ovi\ $\lambda$1037~$\rm\AA$ with an intervening absorption line at the expected position of \ovi\ $\lambda$1031~$\rm\AA$. As noted earlier, the redshift range places the \ovi\ doublet at observed wavelengths lower than 1215$ \rm \AA$ and hence we do not expect any misidentification of lower redshift \lya\ absorbers as \ovi. Figure \ref{fig:ovimass} shows the column density of \ovi\ absorbers from the \cosigrm\ survey as well the survey by \cite{Stocke2019}. Over the entire halo mass range, the \ovi\ and \nv\ detection rates are 44$\pm$5\% and 11$\pm$5\%, respectively.

\begin{figure*}
\centering
\plottwo{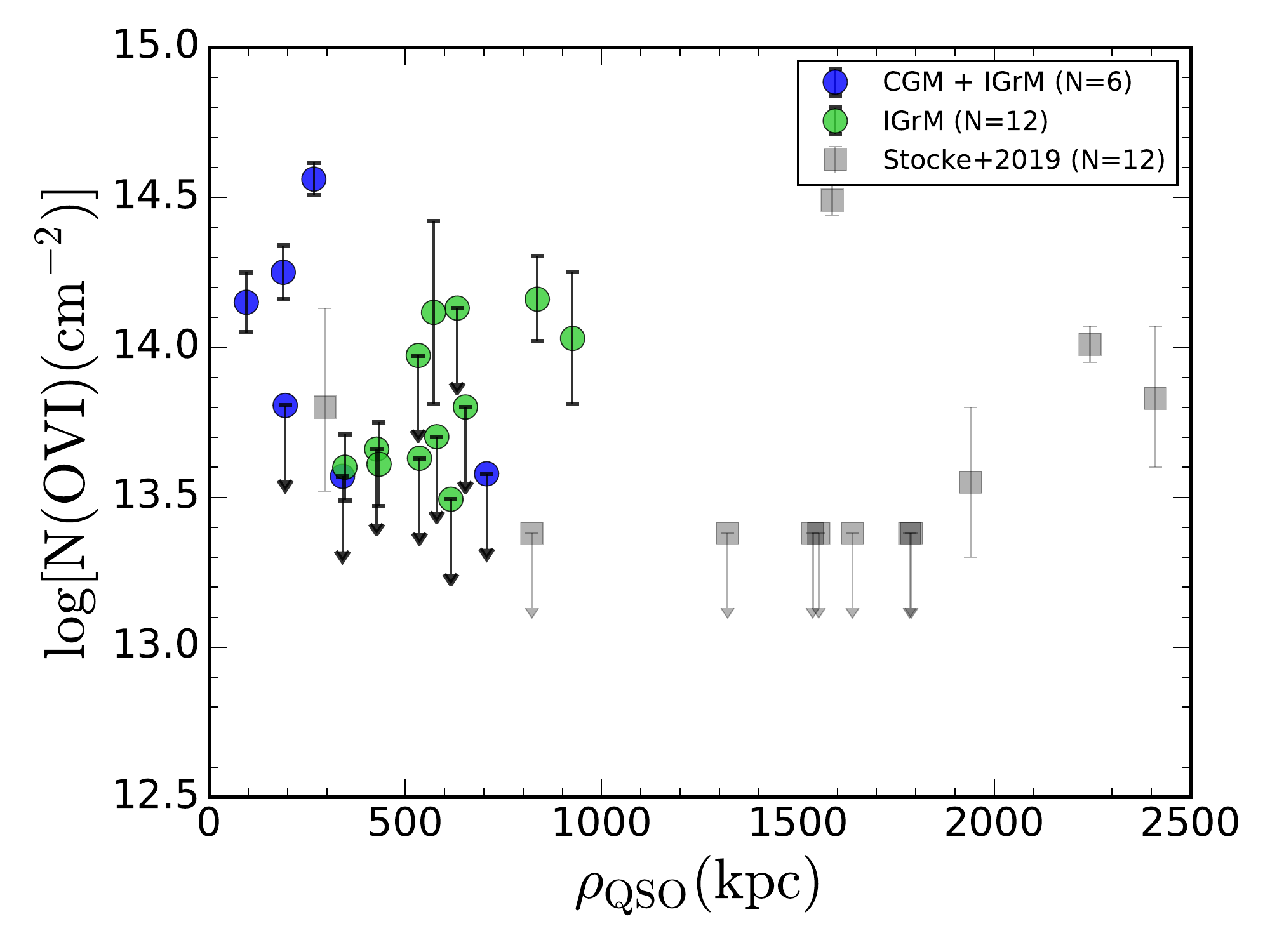}{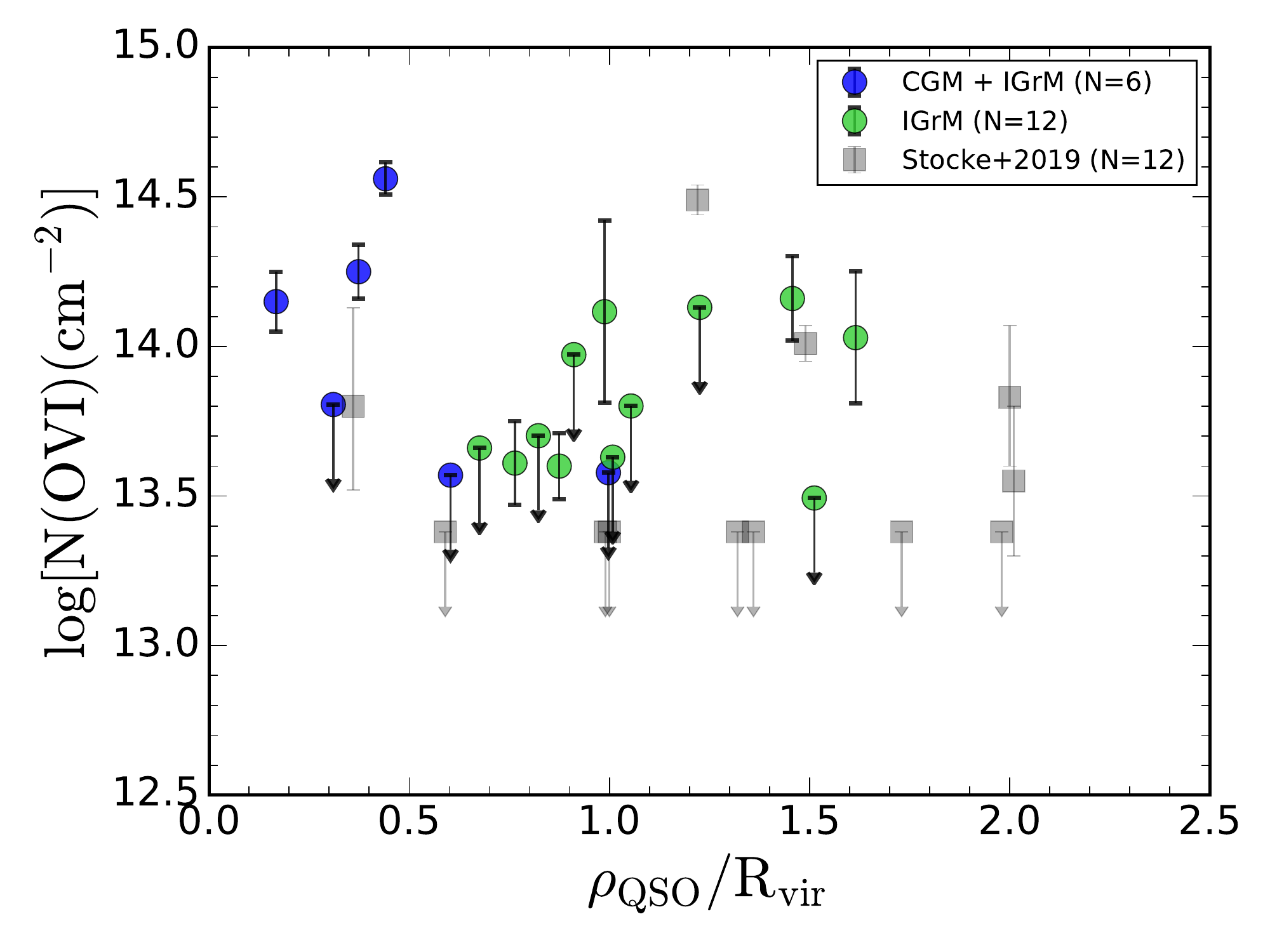}
\caption{\ovi\ column density detections and 3$\sigma$ upper limits as a function of projected QSO impact parameter (left) and the QSO impact parameter normalized by the group's virial radius (right). The colors are the same as in Figure \ref{fig:ovimass}.}
\label{fig:ovirho}
\end{figure*}

Figure \ref{fig:ovirho} shows the \ovi\ detections as a function of QSO impact parameter from the center of the group. We observe a flat distribution of detections from $0.1-1.5~\rm{R_{vir}}$, which suggests that the sightlines may be probing gas that is not at the virial temperature. Since X-ray studies \citep{helsdon_ponman00, Mulchaey2000,Robson2020} show temperature gradients in galaxy groups, the observed flat distribution of \ovi\ detections provides evidence that \ovi\ is not tracing the bulk component of the \igrm. 

Another indication that the OVI absorbers in our sample is tracing a mix of hot and cool gas is the fact that while all the systems that show \ovi\ also show \lya, but the kinematics can be quite different. For example, six of the eight sightlines with \ovi\ detections show \lya\ absorption with the same (or slightly offset) velocity centroid, while the remaining two sightlines have \lya\ at a much larger ($\ga 200$~\kms). A single ionization process could not produce both \lya\ and \ovi\ at the levels detected in some cases. Therefore, for these two different species to be observed within a close velocity offset, multiple clouds must be present, which indicates hot and cool gas in close proximity. 

\subsection{Absorber Kinematics} \label{sec:kinematics}

\begin{figure*}
\centering
\plottwo{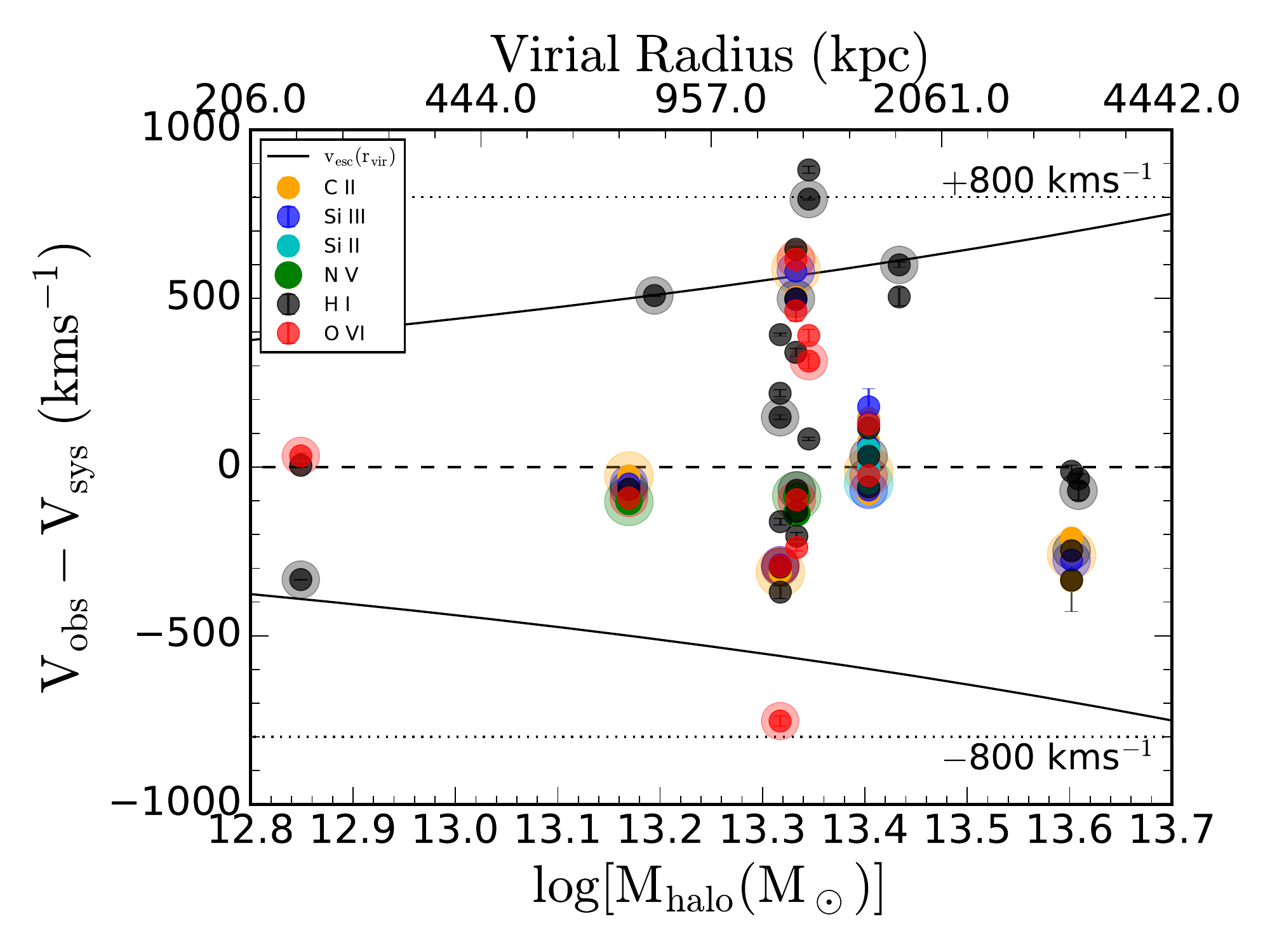}{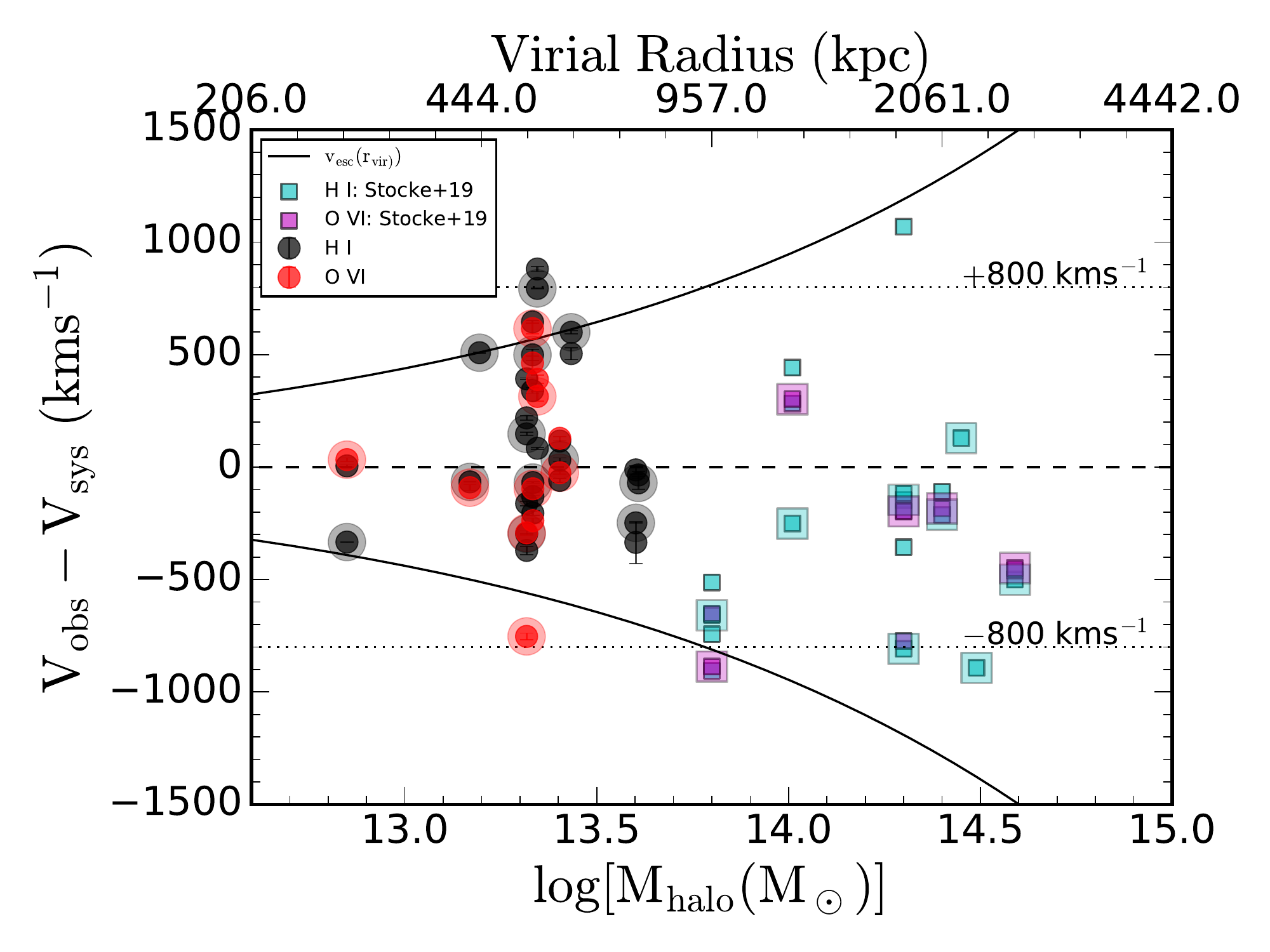}
\caption{\emph{Left: }Observed line of sight velocity relative to the group center for all detections compared to the halo mass of each group. The solid line represents the escape velocity of the group's gravitational potential. The absorption features with the largest column density are marked by a lighter halo over the data point to indicate the dominant component of the transition. 
\emph{Right: }Same as the left panel, but only for \lya\ and \ovi\ including the data from \cite{Stocke2019}. }
\label{fig:escapevel}
\end{figure*}

In this section, we use the kinematics of the absorbers to explore further the nature and distribution of gas as traced by absorption. 
First, we use the velocity spread of the absorbers to ascertain if the absorbing gas is bound to the group. Figure \ref{fig:escapevel} (left) shows the absorption lines detected in the COS spectra at velocity relative to the systemic velocity of the group\footnote{This velocity offset is only from the line of sight velocity. As a result, if the other two velocity components were known, then the fraction of unbound absorbers could increase.}, which is depicted by the dashed line at $v-v_{sys}=0$. Each species is color coded with the dominant absorption feature indicated by a larger halo around the data point. In total, there are 70 absorbers depicted in the plot. 29 and 12 of those are \lya\ and \ovi, respectively, while the remaining represent the other species discussed above.
The solid lines show the escape velocity as a function of halo mass and virial radius. The right panel of Figure \ref{fig:escapevel} includes the data from \cite{Stocke2019}, which extends the dynamic range of halo masses. 

The vast majority of the absorption features are bound to the gravitational potential of the groups. There are 9 out of 70 absorbers from five sightlines (J0841+1406, J1017+4702, J1020+1003, J1216+0712, and J1339+5355) that have sufficient velocities, relative to the group, to escape the gravitational potential. These are composed of 5 \lya, 2 \ovi, 1 \cii, and 1 \siiii\ absorbers. The same trend is observed with the data from \cite{Stocke2019} as only two absorption features (1 \lya\ and 1 \ovi) are observed at high enough velocity offsets to escape the group.

Among the 12 \ovi\ absorbers detected in 8 sightlines, 83\% (10/12) are gravitationally bound to their group halo and only two absorbers show velocities greater than the escape velocity. One of the unbound \ovi\ absorbers is seen in the sightline towards J0841+1406 passes through the CGM of a 4.4~L$_*$ galaxy at 145~kpc.  The velocity offset between the \ovi\ absorber and this is $\sim 650$~\kms.
Interestingly, the \ovi\ absorber does not have a corresponding \lya\ absorber. The \lya\ absorber is seen in this group is more than 600~\kms\ offset from the \ovi\ absorption feature and is offset by $\sim 150$~\kms\ to the closest member galaxy. Hence, the \ovi\ absorber could be tracing infalling or outflowing gas. The large doppler width of the \ovi\ absorber of $b=81.6$~\kms\ suggests that this absorber is tracing WHIM-like material \citep{Cen1999,Dave2001}. While this doppler width is consistent with WHIM-like material, we cannot conclusively rule out the possibility that this absorber is not related to the overall gas phase. As stated in \cite[and references therein]{Oppenheimer2009}, \ovi\ with these doppler widths cannot be only a result of thermal broadening, but also requires a kinematic origin. This leaves some uncertainty as to the exact gas phase due to the lack of other metal-line transitions.

The second \ovi\ absorber with a large velocity offset relative to the group is in the sightline towards J1017+4702 (Figure 2.2). The velocity offset of this \ovi\ absorber is sufficient to escape the gravitational potential of the group. This sightline also exhibits a saturated \lya\ profile with a column density, logN(HI)$>$15.2. Since \lyb\ is blended with an intervening absorber, we cannot utilize it to help constrain the column density. Interestingly, in this case, the sightline does not pass within the virial radius of any spectroscopically confirmed L$_*$ galaxy and is at 926~kpc ($\equiv \rm 1.6~R_{vir}$) from the group center. However, there is one galaxy with matching photometric redshift at 90~kpc from the sightline, which might be the host of this saturated absorption system. It also has a neighbor with similar photometric redshift at an impact parameter from the QSO sightline of 212~kpc. Future spectroscopic redshift measurements are needed to confirmation the association between this neighboring galaxy and the absorption features present in this sightline. If the photometric redshifts are confined, then these galaxies would most likely be part of the group.

\begin{figure*}
\centering
\plottwo{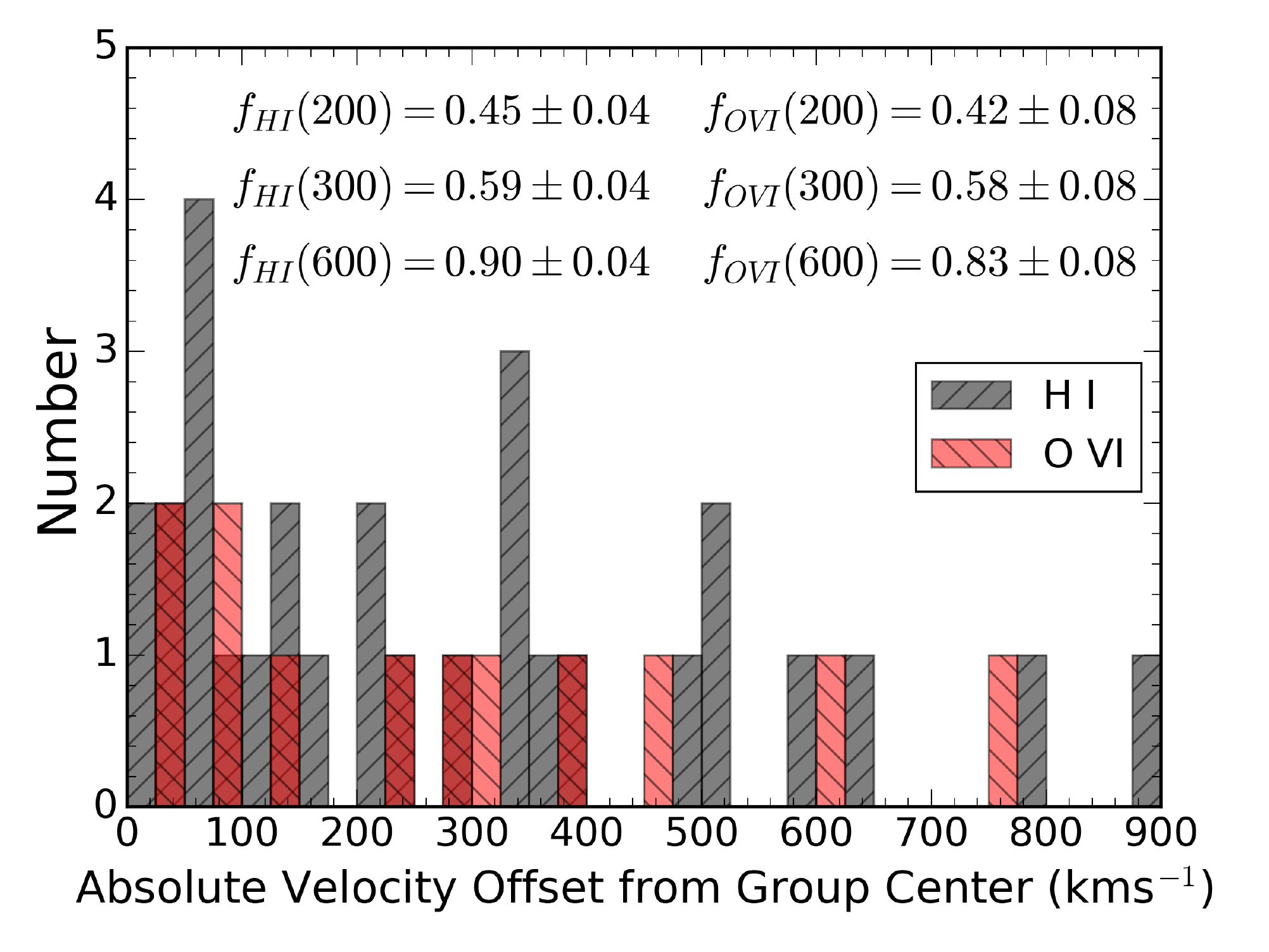}{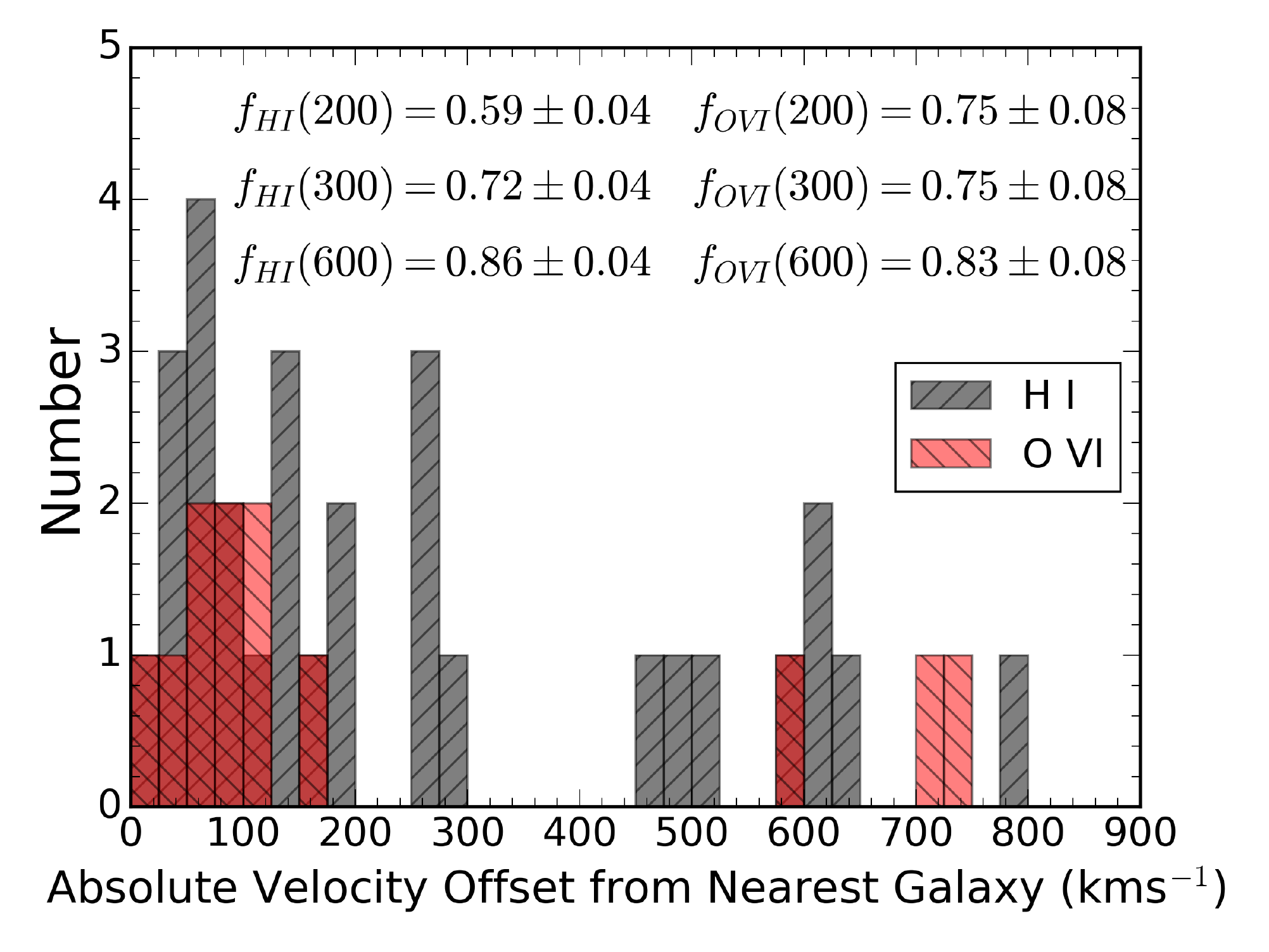}
\caption{\emph{Left: }Absolute velocity offset of \ion{H}{1} (black) and \ovi\ (red) absorbers from the center the group. The fraction of absorbers within 200, 300, and 600~\kms\ for both ions are displayed for all 29 \lya\ and 12 \ovi\ absorbers. 
\emph{Right: }Absolute velocity offset of \ion{H}{1} (black) and \ovi\ (red) absorbers from the nearest galaxy to the sightline. The fraction of absorbers within 200, 300, and 600~\kms\ for both ions are displayed. For both ions, more than 50\% are within 200~\kms\ of the closest galaxy to the sightline, which suggests that some of these absorbers may originate from galactic outflows.}
\label{fig:abshist}
\end{figure*}

Figure \ref{fig:abshist} (left) shows a histogram of all of the \ion{H}{1} and \ovi\ absorbers as a function of absolute velocity offset from the group center. This histogram quantitatively shows that the majority of both \ion{H}{1} and \ovi\ absorbers are gravitationally bound to the group, while $\sim$10\% are observed to have velocities high enough to escape the group potential. These could either indicate infalling clouds or outflows. Therefore, we conclude that the large majority of the absorbers are tracing the cooler, gravitationally bound gas, which is centered well within the group's escape velocity. A similar conclusion was made by the \citet{Stocke2019} study, which concluded that galaxy groups primarily act as  ``closed boxes'' for galactic evolution at low redshifts. However, the IGrM should still experience ``outside-in'' enrichment from the IGM \citep{Tegmark1993,Scanniapieco2002,Oppenheimer2012}. While the source of the initial IGM enrichment at early epochs is model dependent, each model of ``outside-in'' enrichment predicts that structures can regain metals that were expelled at earlier times.

We also investigate the velocity of the absorbers with respect to the nearest, spectroscopically confirmed member galaxy in projection. This is illustrated in the right panel of Figure \ref{fig:abshist} where \ion{H}{1} and \ovi\ absorption features are shown by a histogram as a function of absolute velocity from the nearest galaxy, which could range between 0 and 1600~\kms. For both \ion{H}{1} and \ovi\ absorbers, we see that more than 50\% are within 200~\kms\ of the closest galaxy to the QSO sightline. The fraction increases to 70\% within 300~\kms, which is less than the escape velocity of an L$_*$ galaxy. 
This indicates that these absorption features may originate from gas in the CGM of member galaxies in the group. On the other hand, the absorbers at higher velocity offsets are most likely tracing patchy components of the IGrM, or inflows/outflows from individual group members.

\subsection{Nature of IGrM} \label{sec:multiphase}
\begin{figure}
\centering
\plotone{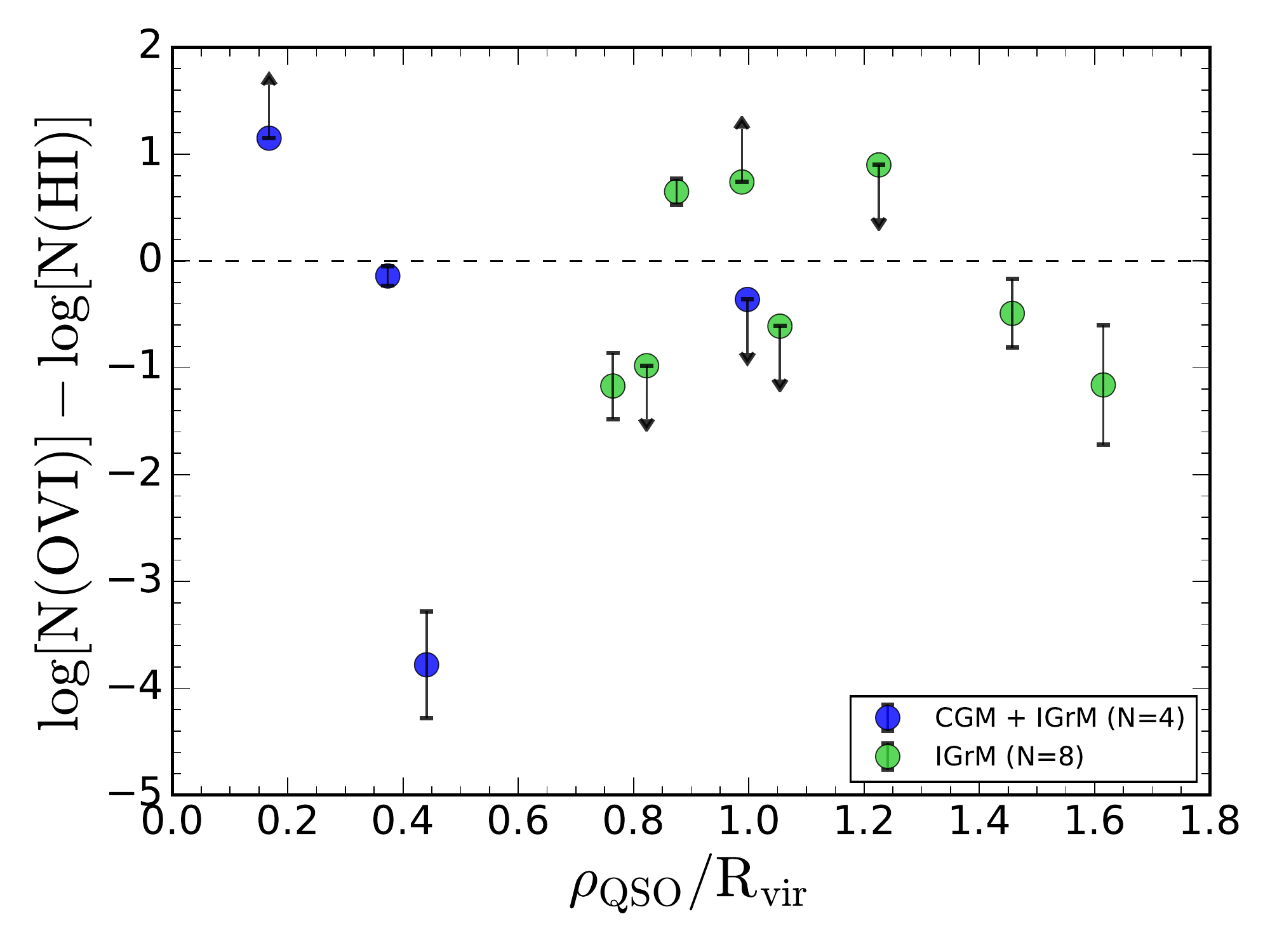}
\caption{Ratio of \ovi\ to \ion{H}{1} column density as a function of normalized impact parameter for each of the 18 sightlines in the COS-IGrM sample.   }
\label{fig:ovi_lya_ratio}
\end{figure}

In order to look at the overall ionization state of the IGrM, the ratio of \ovi\ to \ion{H}{1} was examined for each of the 18 sightlines in the COS-IGrM sample (Figure \ref{fig:ovi_lya_ratio}). For these ratios, only the components of \ion{H}{1} found at the same velocity as \ovi\ were used. If there were no \ovi\ or \ion{H}{1} detections, then an upper limit was used. Sightlines with no \lya\ or \ovi\ absorption were not included in this analysis. The sightlines with \ovi\ column densities greater than the \ion{H}{1} columns shows that there is highly ionized gas throughout the IGrM, while the lack of a correlation between the ratio of column densities and impact parameter shows that there is no significant dependence of ionization state on normalized impact parameter. 

Out of the 12 sightlines that show \lya\ absorption, 9 of those sightlines show evidence of multiple metal-line species detected in absorption that allows us to model the ionization state of the gas. Of these 9 sightlines, 7 clearly depict multiphase gas, where various metal-line species are present in varying levels suggesting that the components have very different ionization states. 
The presence of these multiple components in most of the sightlines indicates that the absorption is associate with pockets of gas that maybe cooler than the rest of the media (and possibly more dense if they are in pressure equilibrium). Therefore, we believe that our data is primarily tracing a complex multiphase media, which cannot be described by a single ionization process. In Table \ref{tab:ionization}, we present the probable ionization process for each group, based upon the observed spectra. 

For ionization modeling, the primary interest was to determine if any of the absorption lines from the COS-IGrM sample are consistent with photoionization, collisional ionization, or inconsistent with either process. For CIE modeling, the ratio of \lya\ to \ovi\ absorption at 50\% solar was examined over a range of temperatures. If the observed column density ratio was consistent with CIE predictions, then it was noted that the absorption features were consistent with CIE. Since CIE predicts broad, shallow \ovi\ without the presence of lower ionization state transitions, only those sightlines that had \lya\ and \ovi\ were examined for consistency with CIE models. 

The photoionization modeling was inherently less certain due to unresolved, blended components and a lack of multiple metal line species in the majority of sightlines. We used CLOUDY \citep{Ferland2013} with a Haart-Madau background and a total hydrogen density grid (log[n(h)] from [-5,-2]~particles~$cm^{-3}$) in 0.5 dex increments. The total neutral column density was fixed to the observed \lya\ column density. If a point in the grid existed where the column density ratio each metal species was consistent within the same density grid point, then we stated that the absorption components were consistent with photoionization. 


\startlongtable
\begin{deluxetable*}{ll}
\tabletypesize{\scriptsize}
\tablecolumns{2}
\tablecaption{Results of ionization modeling for sightlines with identified absorption features.}

\tablehead{
\colhead{Group}  & \colhead{Ionization Model} 	\\[-0.3cm]
}
\startdata
J0841+1406 &  Consistent with CIE at $\sim$10$^{5.4}$~K or at $\sim$10$^{5.7}$~K\\
J1017+4702 &  \ion{C}{2}/\ion{Si}{3} ratio consistent with photoionization between $10^{-3} < n(h) < 10^{-2.5} cm^{-3}$\\
           &  Higher velocity components inconsistent with photoionization\\
J1020+1003 &  Insufficient data for modeling  \\
J1126+1204 &  Insufficient data for modeling  \\
J1127+2654 &  Inconsistent with photoionization (1)\\
J1216+0712 &  Consistent with CIE at $\sim$10$^{5.2-5.3}$~K \\
J1301+2819 &  Lowest velocity component of \lya\ and \ovi\ consistent with CIE at $\sim$10$^{5.3}$~K or at $\sim$10$^{5.9-6}$~K\\
           &  The \ion{N}{5}/\ion{O}{6} ratio for the higher velocity components ($v \sim -100$\kms) is consistent with photoionization\\
           & between $10^{-4} < n(h) < 10^{-3.5} cm^{-3}$ \\
J1339+5355 &  Insufficient data for modeling  \\
J1343+2538 &  Consistent with CIE at $\sim$10$^{5.3-5.4}$~K or $\sim$10$^{5.9}$~K \\
J1348+4303 &  \ion{C}{2}/\ion{Si}{3} ratio consistent with photoionization between $10^{-2.5} < n(h) < 10^{-2} cm^{-3}$  \\
J1424+4214 &  Low velocity component ($v \sim -100$\kms) inconsistent with photoionization, but is consistent with CIE at $\sim$10$^{5.2-5.3}$~K\\
           &  Higher velocity component ($v \sim -55$\kms) \ion{C}{2}/\ion{Si}{3} ratio consistent with photoionization\\
        & between $10^{-5} < n(h) < 10^{-4.5} cm^{-3}$ or $10^{-3} < n(h) < 10^{-2.5} cm^{-3}$ (2)\\
J1428+3225 &  Inconsistent with photoionization  \\
\enddata
\label{tab:ionization}
\tablecomments{ (1) Since individual \ovi\ components may be blended, we cannot definitively rule out photoionization based upon \siiii\ components. (2) There may be an unresolved, blended component of \cii, which increases the uncertainty in the density required for photoionization to be confirmed.}
\end{deluxetable*}
\normalsize

Some sightlines show absorbers that match the ratios and strengths predicted by CIE \citep{Gnat2007} for a hot $\approx \rm 10^{5.5}~K$ medium. For example, the sightline J1343+2538 (Figure 23) passing through a group at an impact parameter of 346~kpc ($\equiv \rm 0.9~R_{vir}$), shows broad \lya\ ($b_{Ly\alpha}$=47~\kms) along with \ovi\ suggestive of hot media \citep{Richter06}. The ratio of \ovi\ to \lya\ column density of 0.65 dex is consistent with temperatures of $\rm 10^{5.3-5.4}$~K or 10$^{5.9}$~K for 0.5$-$[1~Z/H]$_{\odot}$ with temperature inverse proportional to the metallicity for the same column density ratio. The choice of this metallicity range is based on measurement from X-ray studies for groups of galaxies that typically find the average metallicities of the X-ray bright IGrM to be 0.4$-$0.6~[Z/H]$_{\odot}$ \citep{helsdon_ponman00}. We do not have strong metallicity constrains for non-X-ray bright groups, so we adopt the metallicities seen in X-ray studies.

Another example of collisional ionized gas are seen in the sightline towards group J1301+2819 (Figure 21). 
In addition to tracing hot gas, this sightline shows a mix of multiple ionization states at slightly different velocities possibly tracing a multi-phase medium. 
The \lya\ feature shows three components - two strong components with associated \lyb\ and one weak component with log~N(HI)=13.05. The strong components are seen in both \nv\ and \ovi\ (the components are blended in \ovi), whereas the weakest component is most prominent in \ovi. This indicates that the different components trace different ionization states. For weakest component the ratios of \ovi\ and \lya\ are in agreement with collision ionization equilibrium model. The ratio of column densities, $\rm  log N(OVI) - log N(HI) =0.60$, corresponds to gas at 10$^{5.3}$~K or 10$^{5.9-6}$~K at 50\% solar metallicity. At lower metallicities, the observed ratio of column densities between \ovi\ and \lya\ would indicate a slightly higher temperature. 

On the other hand, the stronger components are quite puzzling. If photoionization was responsible for the observed ionization states, there should be other low-ionization transitions detected besides \hi\ such as \siii, \cii\ and \siiii. Despite these transitions not being present at detectable levels in the spectra, the \nv\ to \ovi\ ratio is consistent with photoionization (Table \ref{tab:ionization}). Therefore, we are unable to conclusively state the process behind the observed column densities, it is most likely a mixture of multiple ionization processes.

A similar case of multiphase media is seen in J1424+4214 (Figure 25), which shows two distinct ionization states with a velocity separation of about 45~\kms: a less ionized system at $\sim$~-55~\kms\ and a highly ionized state at $\sim$~-100~\kms. One component is seen in lower ionization transitions like  \lya, \cii, and \siiii, while the  second components is seen in higher ionization transitions like \nv, \ovi, as well as \siiii\ that show a weak feature suggesting \siiii\ is not the dominant ionization state of silicon.
The ratio of these lines indicate that the two components are at very different ionization states, thus suggesting that the \igrm\ is multiphase and cannot be described by a single ionization state. While the ionization processes for each of the two components cannot definitively be determined based upon the data at hand, the component centered at $\sim$~-100~\kms\ is consistent with CIE at $\sim$10$^{5.2-5.3}$~K, while the component at $\sim$~-55~\kms\ is consistent with photoionization based upon the \cii\ to \siiii\ line ratio.

Another sightline of interest is towards the group J1127+2654 (Figure 19). This sightline exhibits a saturated \lya\ profile with column density, logN(HI)$>$18.3, which makes it a Lyman-limit system \citep[LLS,][and references therein]{Lanzetta95}. The absorber complex shows multiple components commonly associated with extended disk \citep{lehner09}, inner CGM \citep{Werk2014,Armillotta2017,Fielding2020}, and/or tidal structures \citep{frye19}. This QSO passes within $\sim$119~kpc from a known group member; however, higher resolution spectroscopy and a rotation curve is necessary to confirm the connection between this LLS and the member galaxy. Similarly, due to the blending of \ovi\ components in this QSO spectra, we cannot rule out photoionization as the primary ionization mechanism for these absorption lines.

Lastly, the sightline towards J1017+4702 (Figure 16) shows that \lya\ is saturated at the same position as \cii, \siiii, and broad, shallow \ovi. Photoionization alone cannot produce broad, shallow \ovi\ and CIE does not predict the existence of saturated \lya\ and broad \ovi\ at a single temperature. Since this is an \igrm\ sightline, the broad \ovi\ may be tracing a hotter component; however, the lower transitions show evidence of cooler gas at the same velocity. Photoionization modeling with CLOUDY \citep{Ferland2013} showed that the lower velocity components of \cii\ and \siiii\ are consistent with photoionization, while the higher velocity components are inconsistent with photoionization. Overall, it is clear that the ionization states of these groups are complex and the ionization processes behind the multiphase gas cannot always be explained by either photoionization or CIE. Future studies with better modeling, higher resolution observations, and broader wavelength coverage can help shed insight into these ionization processes.

\subsection{Origin of \ovi\ Absorbers} \label{sec:coshalos}

\begin{figure}
\centering
\plotone{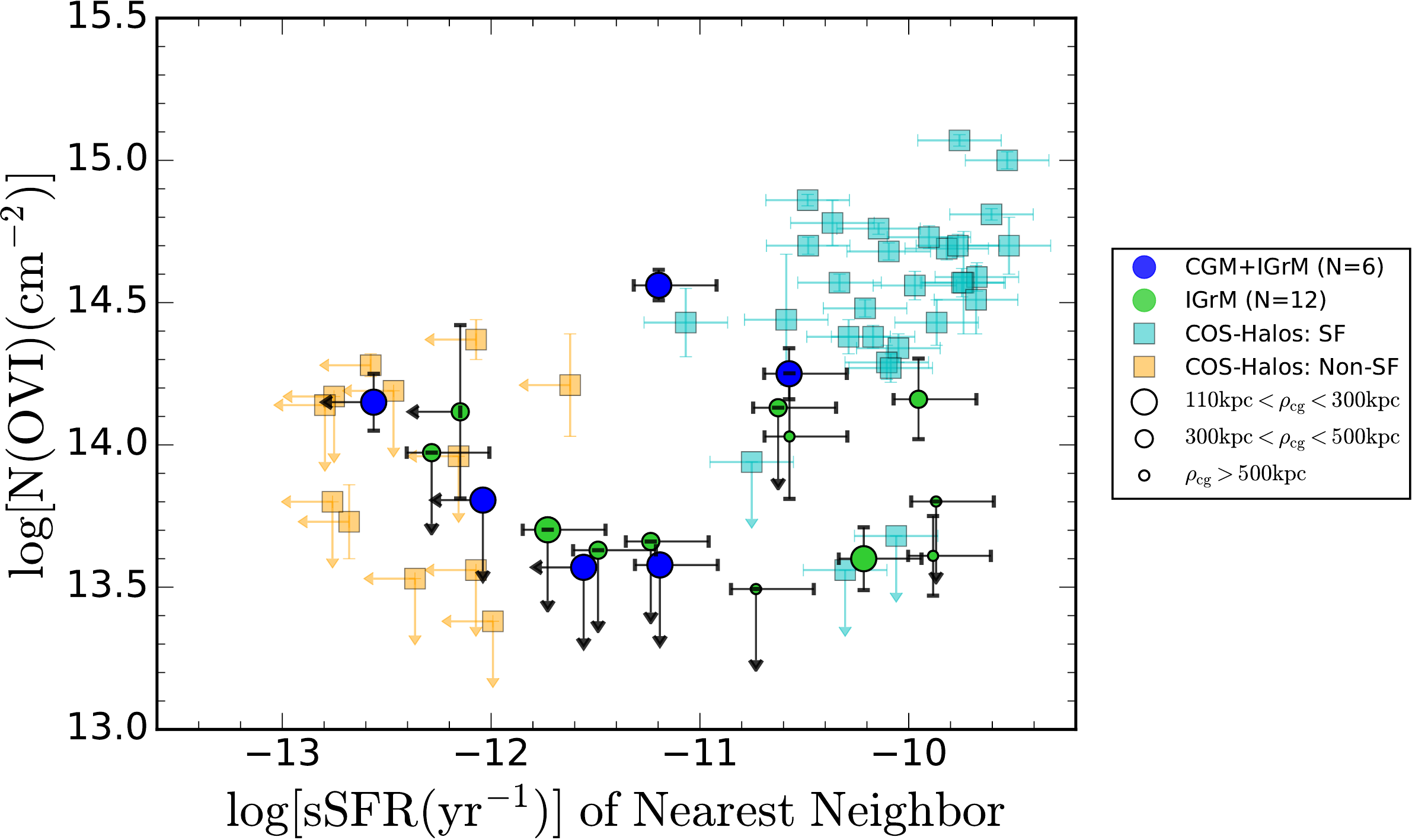}
\caption{\ovi\ column density as a function of the specific star formation rate of the nearest member galaxy to the QSO sightline. The blue (CGM~+~IGrM) and green (IGrM) points show the COS-IGrM sample and the size of the points correspond to the impact parameter from the closest, spectroscopically confirmed member galaxy. The cyan and orange data points show the results from the COS-Halos survey \citep{Tumlinson2011}, which robustly observed an increased amount of \ovi\ in the CGM of star forming galaxies.}
\label{fig:coshalos}
\end{figure}

We differentiate between CGM and \igrm\ absorption in galaxy groups by comparing our \ovi\ detections to those detected in the COS-Halos survey \citep{Tumlinson2011}. The COS-Halos survey discovered that a strong correlation between \ovi\ in the CGM and the star formation rate of galaxies existed. In order to compare the our data with the COS-Halos sample, we determined the galaxy closest to the QSO sightline and then matched the galaxy to the star formation rate from the MPA-JHU DR7\footnote{https://wwwmpa.mpa-garching.mpg.de/SDSS/DR7/} galaxy catalog \citep{Brinchmann2004}. 

Figure \ref{fig:coshalos} shows our data along with those from the COS-Halos survey. The CGM+IGrM sightlines, shown as deep blue circles, have a clear host galaxy as the sightline passes within the viral radius (assuming an isolated halo) of a member galaxy. The pure IGrM sightlines do not pass through the CGM of the nearest galaxy (shown in green circles). Therefore, they are not applicable for comparison with the COS-Halos sample; nevertheless, we show them on the plot for comparison with the CGM+IGrM sub-sample. It is worth noting that even the blue points that do probe the CGM pass through the outer CGM ($\rho >$110~kpc) and not the inner CGM like COS-Halos sample.

Overall, we do not see a trend of higher \ovi\ levels as a function of the specific star formation rate (sSFR) or the star formation rate of the nearest galaxy. This is not surprising considering the impact parameters. However, it does suggest that the origin of our \ovi\ absorbers are probably not related to the star formation activity of individual galaxies and therefore, we are most likely not tracing the CGM gas physics as seen in the COS-Halos survey, but instead, a more group-related phenomena.

\begin{figure}[t]
\centering
\plotone{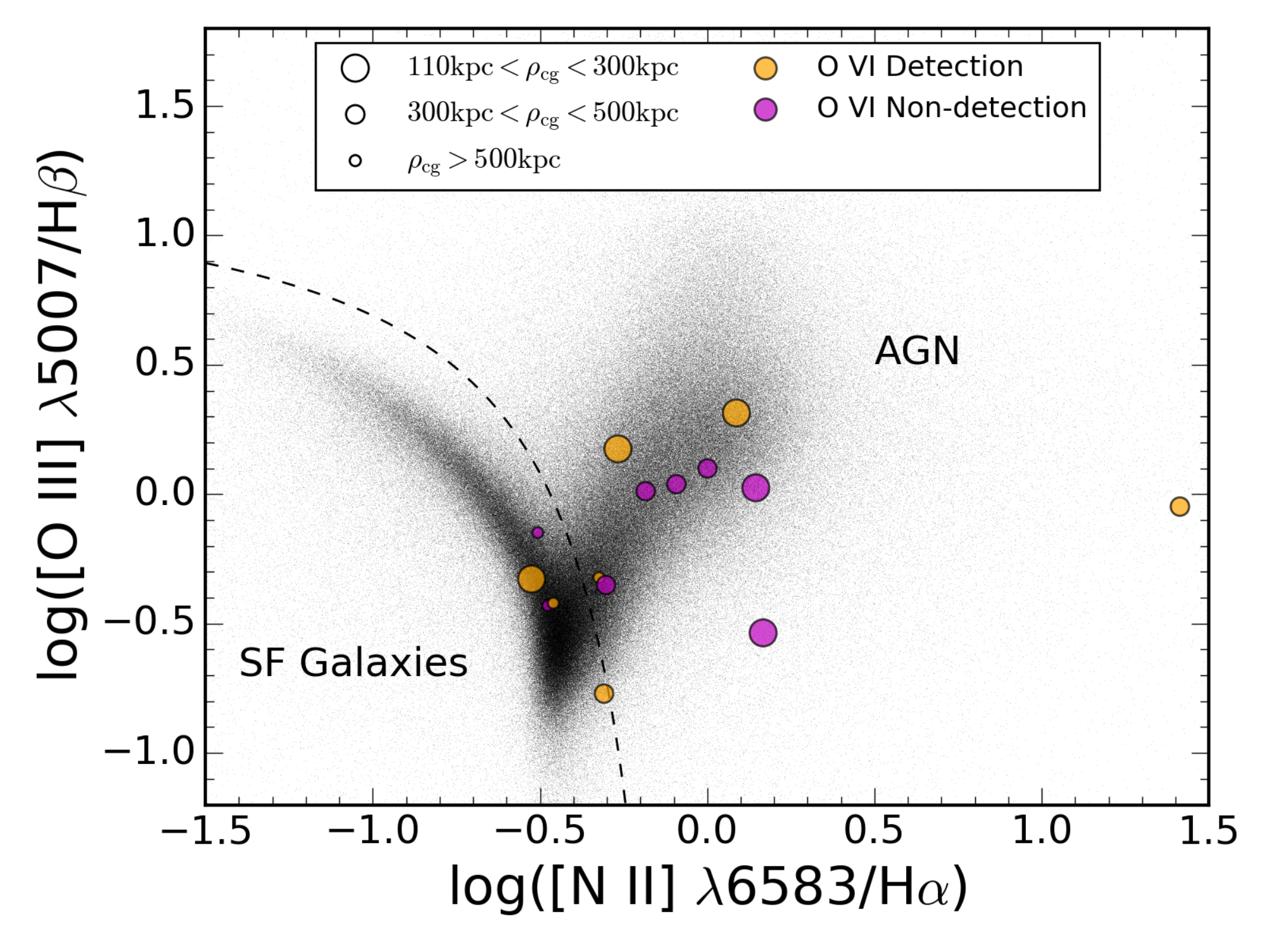}
\caption{BPT diagram showing the locations of the COS-IGrM sample where reliable flux measurements from the MPA-JHU DR7 catalog were present. The division between phase space pertaining to either star forming or AGN galaxies is marked by the dashed line \citep{Kauffmann2003}. The colored circles represent sightlines with \ovi\ detections vs. non-detections and the size of the circles represents the projected impact parameter to the closest galaxy to the sightline.  }
\label{fig:bpt}
\end{figure}

While we can confidently rule out the CGM of \Lstar\ galaxies as the source of \ovi\ absorbers, there could potentially be smaller galaxies that may be present closer to the sightline. A much deeper redshift survey of galaxies in the vicinity of the QSO sightlines would enable us to quantify the presence of low-mass galaxies. Nevertheless, sub-\Lstar\ galaxies are not expected to have significant metal reservoirs beyond their inner CGM ($\rho \rm > 0.5~R_{vir}$) \citep{bordoloi14}. Hence, it is not likely that the CGM of sub-\Lstar\ galaxies could dominate the \ovi\ detected in our sample. On the other hand, material spread out by tidal interactions can have a large cross-section on the sky and may survive as faint diffuse partially neutral gas in the IGrM for hundreds of millions of years \citep{Borthakur10,borthakur15}.

\begin{figure*}[t]
\centering
\includegraphics[width=2.25in]{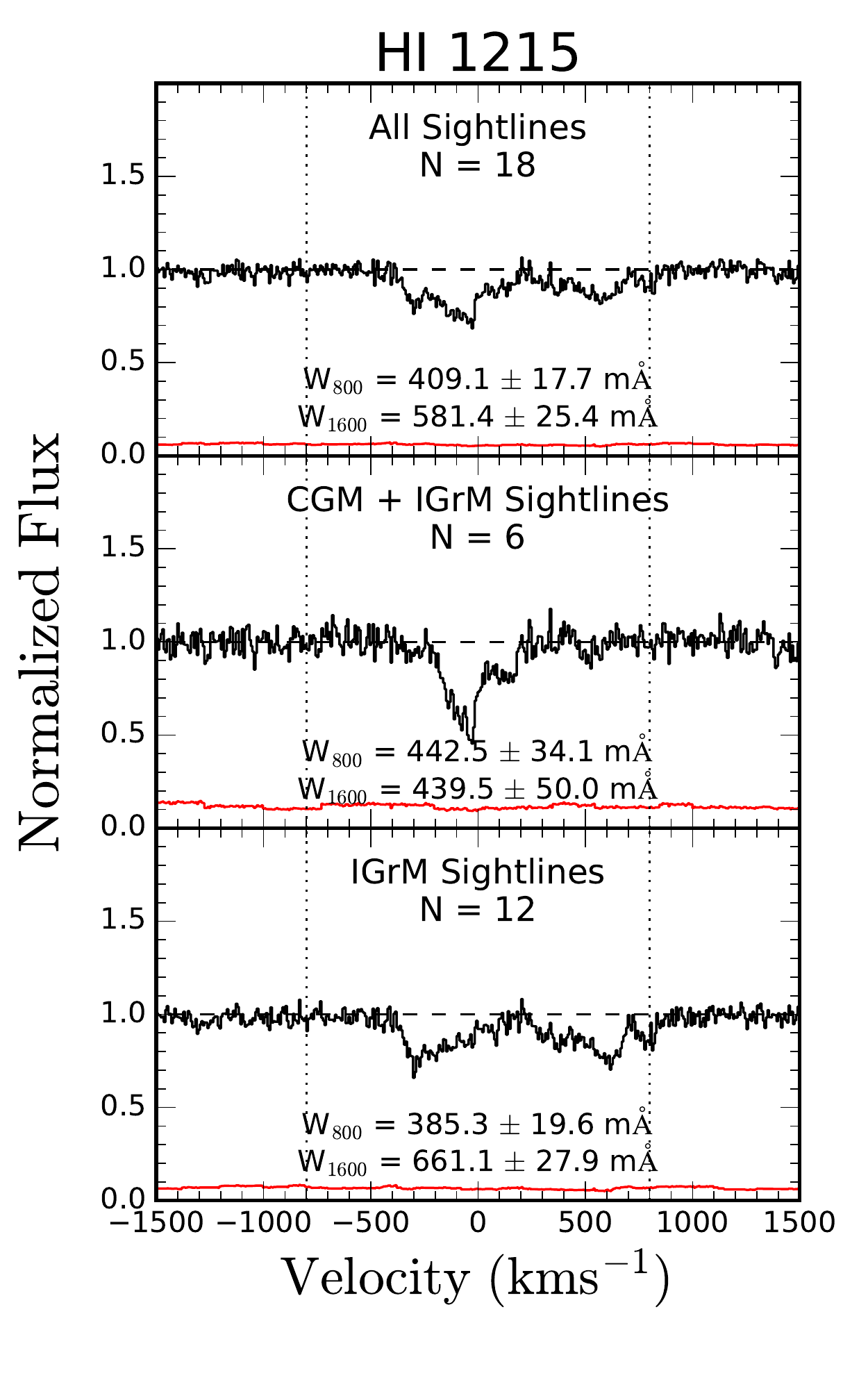}
\includegraphics[width=2.25in]{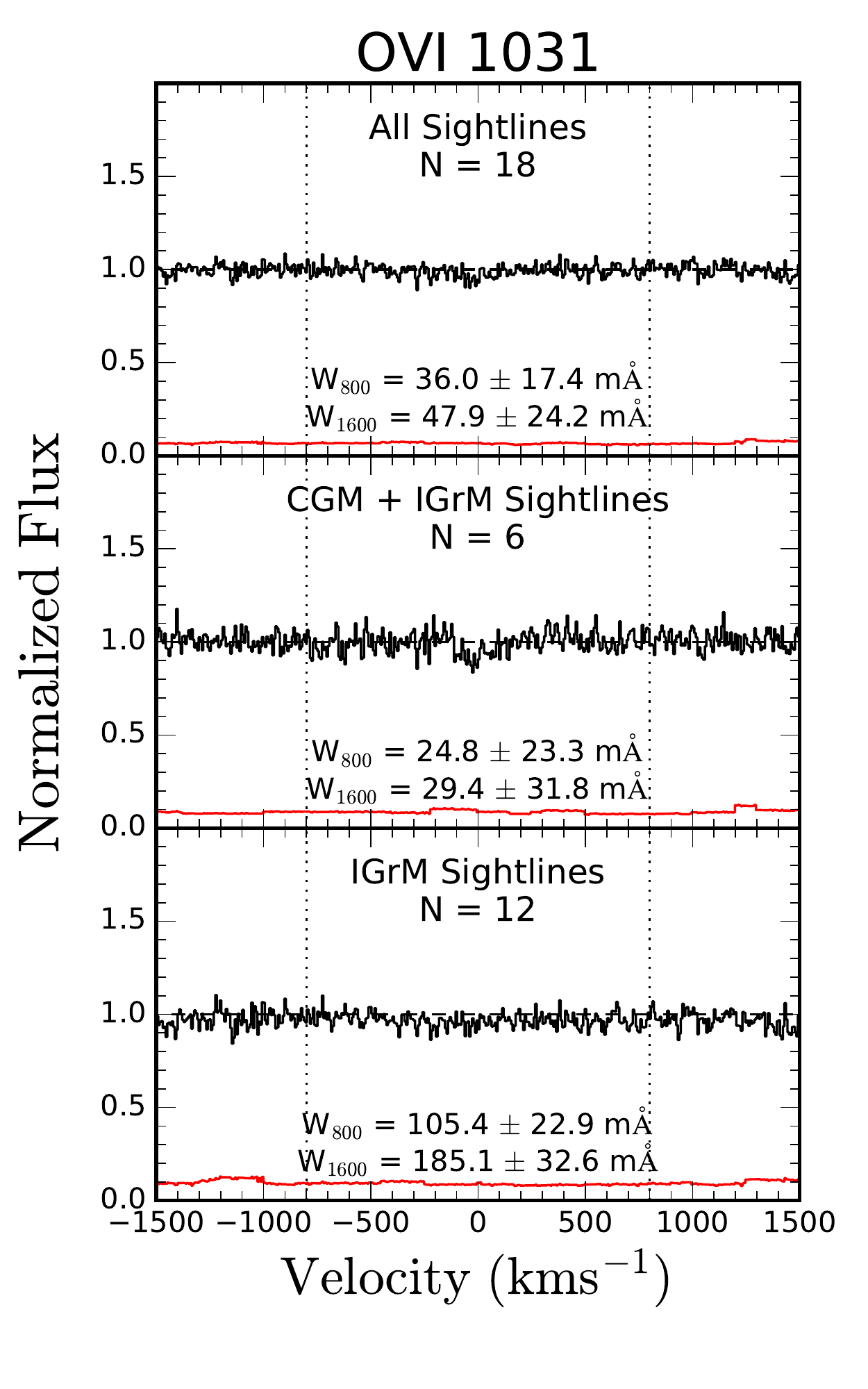}
\vspace{-.50cm}
\caption{Stacked spectra for \lya\ (left) and \ovi\ (right) for each of the 18 sightlines in the \cosigrm\ sample (top). The sightlines were also divided into the \igrm\ and CGM sightlines and stacks of each were created (middle and bottom respectively). The stacks are centered on the center of mass velocity of the group and all intervening absorption features were removed. }
\label{fig:stacks}
\end{figure*}

Our sample shows a larger fraction of green valley galaxies than typically observed in the Universe. \cite{Jian2020} finds that on average, 20\% of galaxies populate the green valley and the majority of those are field galaxies and not those found in more dense environments. Observing $\sim$33\% of the closest galaxies to the QSO sightline in our sample to be in the green valley reinforces the idea that galaxy group environments may act as important sites where the process of quenching is active \citep{Wetzel2012, Wetzel2013}. The role of the IGrM or the CGM in turning these galaxies green is still unclear.

Another possibility for the origin of \ovi\ in the IGrM could be due to AGN activity. In order to address this, used the emission line ratios from the MPA-JHU\footnote{\url{https://www.sdss.org/dr14/spectro/galaxy_mpajhu/}} DR7 catalog to construct a Baldwin, Phillips \& Terlevich (BPT) diagram \citep{Baldwin1981} so that star forming galaxies could be separated from AGN using the demarcation as defined by \citet{Kauffmann2003}. The locations of the \cosigrm\ sample compared to the SDSS DR7 sample from the MPA-JHU catalog are shown in Figure \ref{fig:bpt} as colored circles. 15 out of the 18 groups in our sample had emission line measurements for the closest galaxy to the QSO sightline and therefore, could be included in the BPT diagram. The color of the circle represents \ovi\ detections (orange) vs. non-detections (magenta). The size of the symbol represents the impact parameter of the sightline, where larger sizes indicate small impact parameter. We do not find any systematic over-density of \ovi\ detection or non-detection in sightlines with or without AGN. Therefore, we conclude that AGN activity is not the primarily contributor of \ovi\ in the IGrM.

\subsection{Stacked Spectra} \label{stacks}

In order to look for fainter gas associated with the \igrm, we stacked sightlines centered around the group systemic velocity for \lya\ and \ovi. For each species, stacks were created using all 18 sightlines as well as subsets of CGM $+$ \igrm\ or \igrm\ only sightlines. These stacks are shown in Figure \ref{fig:stacks} along with the number of sightlines going into each subset. The equivalent widths were measured for velocities within $\pm400$ \kms\ and $\pm800$ \kms\ from the group's systematic redshift. These values are listed as $\rm{W_{800}}$ and $\rm{W_{1600}}$ respectively. 

The \lya\ stacks show net absorption centered around zero velocity for the CGM $+$ \igrm\ subset, and absorption corresponding to higher velocity offsets in the \igrm\ stack. When we stack the full \cosigrm\ sample, we observe a combination of the two subsets meaning that the \igrm\ in our sample is traced by two distinct regions: gas at the systematic velocity of the group as well as gas that is at larger velocities than the group's systemic velocity. This could perhaps be a result of warmer gas condensing in the outskirts and falling back towards the center of the group. 

The \ovi\ stacks show weak net absorption throughout all the sightlines. However, there is  absorption in the \igrm\ sightlines. Both the CGM $+$ \igrm\ stack and the pure IGrM stack show that the majority of the absorption is within of the central 800 \kms\ of the group. The covering fraction of the IGrM stacks is more uniform than the CGM stacks as there is net absorption throughout the $\pm800$ \kms. Non-detection of \ovi\ in the full stack indicates that there is not a volume filling phase of the IGrM, but instead, an \ovi\ traced IGrM is a more transient phenomena.

\begin{figure*}
\centering
\plottwo{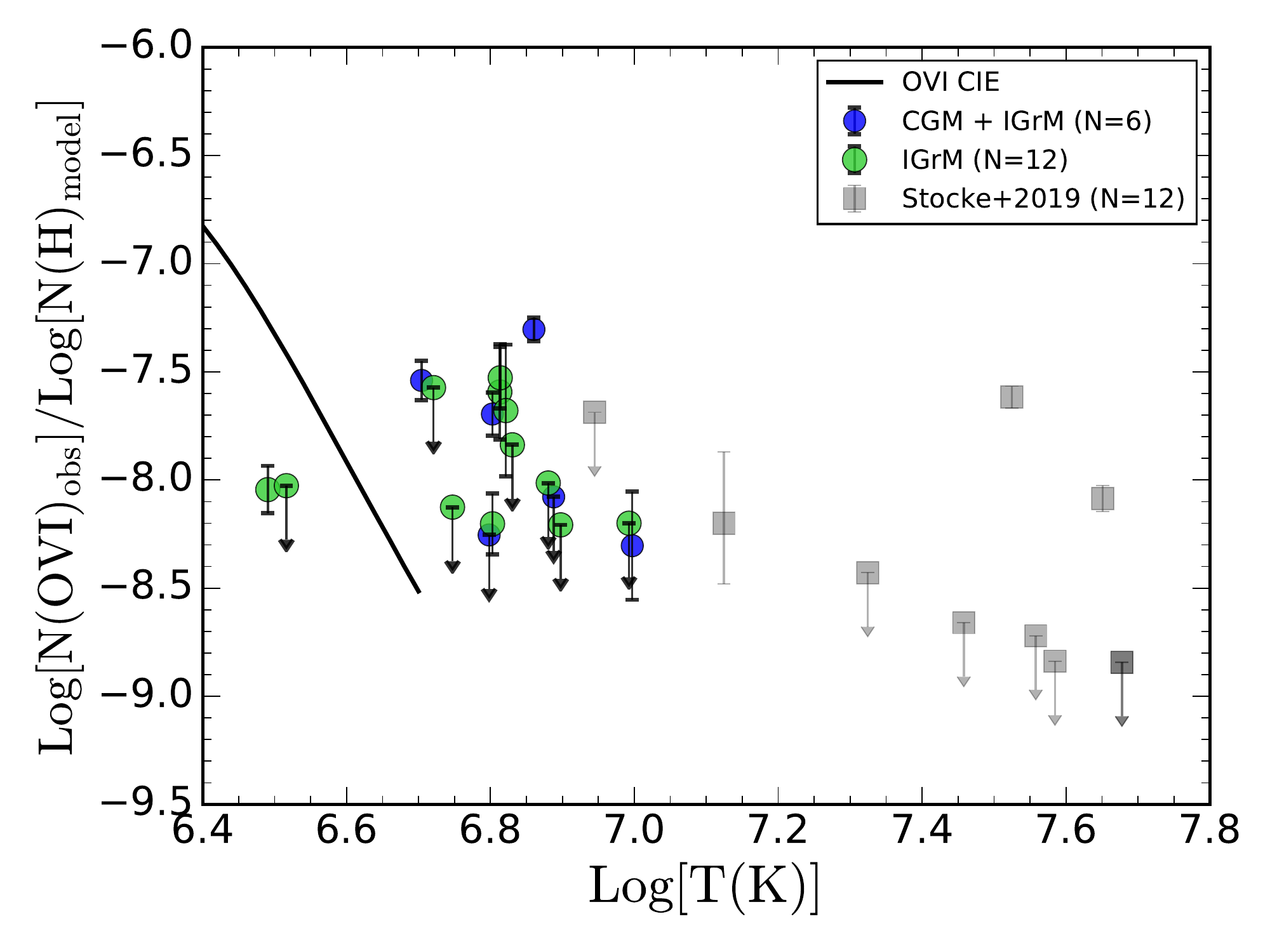}{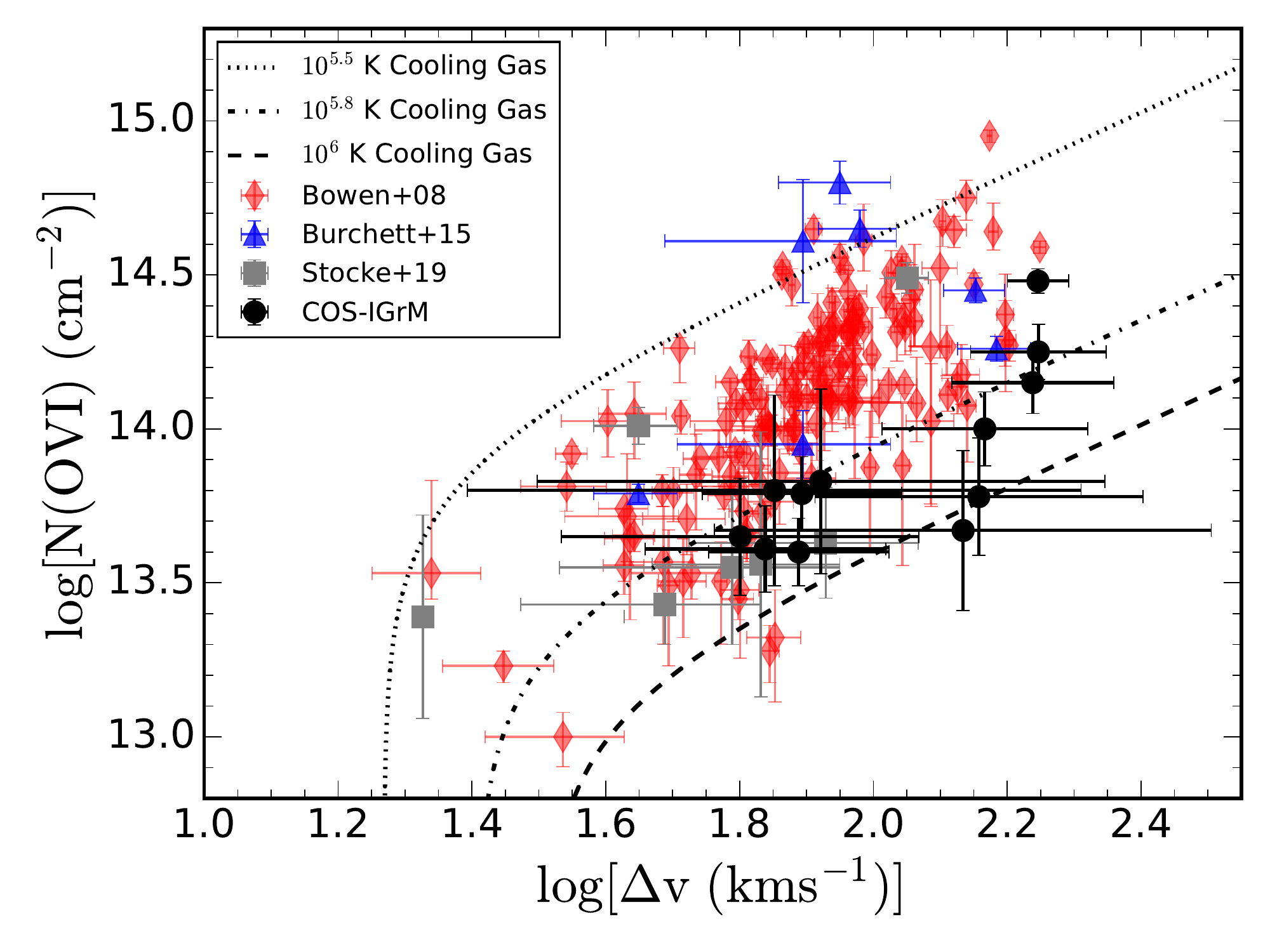}
\caption{(Left:) Observed \ovi\ column densities normalized by the QSO path length through the group as a function of group virial temperature (N(\ion{H}{1})$_{model}$). The blue and green data points are from the \cosigrm\ sample, while the gray squares show the results from \cite{Stocke2019}. The solid, black line represents theoretical predictions based on collisional ionization equilibrium models from \cite{Gnat2007} assuming 50\% solar metallicity, assuming a total hydrogen density of $10^{-3}~\rm{cm^{-3}}$ (Right:) \ovi\ column density as a function of the detected \ovi\ line width ($\Delta v$). The dotted, dot-dashed, and dashed lines show radiative cooling models from \cite{Bordoloi2017} at $10^{5.5}$~K, $10^{5.8}$~K, and $10^{6}$~K, respectively. }
\label{fig:temps}
\end{figure*}

\section{Discussion} \label{sec:discussion}

From the \cosigrm\ survey, we have observed no significant trends between the column density of \lya\ or \ovi\ and the physical parameters of the group such as virial radius, impact parameter, and halo mass. This may be an indication that we are not observing a hot, volume filling IGrM; instead we are detecting cooler pockets of gas that are perhaps in pressure confinement within the \igrm. This would line up more closely with what was concluded in \cite{Stocke2017, Pointon2017} and \cite{Stocke2019} for more massive groups. If this is indeed correct, X-ray spectroscopy of \ion{O}{7} and \ion{O}{8} would be required to observe the hotter component of the \igrm, even for lower mass groups ($10^{12.8}-10^{13.7}$~\msun). 

This idea is further reinforced by looking at the virial temperatures of the groups compared to the predicted \ovi\ column densities from collisional ionization equilibrium models \citep{Gnat2007}. Figure \ref{fig:temps} (left) shows the predicted and observed column densities of \ovi\ normalized by the total hydrogen column density through the group as a function of virial temperature (denoted by N(\ion{H}{1})$_{model}$). The column density of hydrogen was estimated by using the IGrM gas density of $n=10^{-3}$~cm$^{-3}$, and multiplying it by the total path length through each group in our sample, which is approximated by a sphere of radius, 2R$_{vir}$. The gas density was selected as a conservative estimate based upon electron density profiles from X-ray data of galaxy groups \citep{Sun2003, Khosroshahi2004} and from density measurements of the IGrM from double bent radio jets \citep{Freeland2011}. From this figure, it is evident that we are primarily observing cooler gas than what would be at the group's virial temperature based upon the amount of \ovi\ observed, which provides more support to our previous statements.

To investigate the theory that the observed \ovi\ is due to cooler gas than the hotter \igrm, we looked at the relationship between the \ovi\ column density and the \ovi\ line width for our sample and other samples from various environments (right panel of Figure \ref{fig:temps}). \cite{Heckman2002} demonstrated that \ovi\ absorption lines in various environments such as the Milky Way, high velocity clouds, Magellanic Clouds, starburst galaxies, and the intergalactic medium all can be described by radiatively cooling gas through the relationship between column density and the Doppler '$b$' parameter. \cite{Bordoloi2017} revisited these models to show that the line width, $\Delta v=3b_D/\sqrt{2}$, is a more appropriate tracer of the flow velocity than the Doppler '$b$' parameter in describing the radiatively cooling \ovi. We show data from \cite{Bowen2008}, \cite{Burchett2015}, and \cite{Stocke2019} along with the COS-IGrM survey \ovi\ detections\footnote{Since the line width is related to the Doppler `$b$' parameter and therefore a Voigt profile fit, there is no physical upper limit for non-detections.} to show that the trends observed from \ovi\ in the Milky Way and the IGM also largely agree with \ovi\ detected in the IGrM, respectively. On average, the COS-IGrM data can be described by radiatively cooling gas between $10^{5.8}$~K, and $10^{6}$~K. This may be indicating that the \ovi\ detected in our sample originates from gas falling towards member galaxies and cools radiatively as it passes through the CGM of a group member or passes through cooler pockets within the hotter IGrM. 

The cooling models described in \cite{Bordoloi2017} predict that \nv\ column densities should be about an order of magnitude lower than those predicted for \ovi. This prediction is consistent with our three \nv\ detections as well as our upper limits in this sample based upon the $10^{5.27}$~K cooling curves in \cite{Bordoloi2017}. Since many of the \ovi\ detections are relatively close to the detection limit, the lack of \nv\ detections is not unexpected due to this prediction. 

\begin{figure}[t]
\centering
\plotone{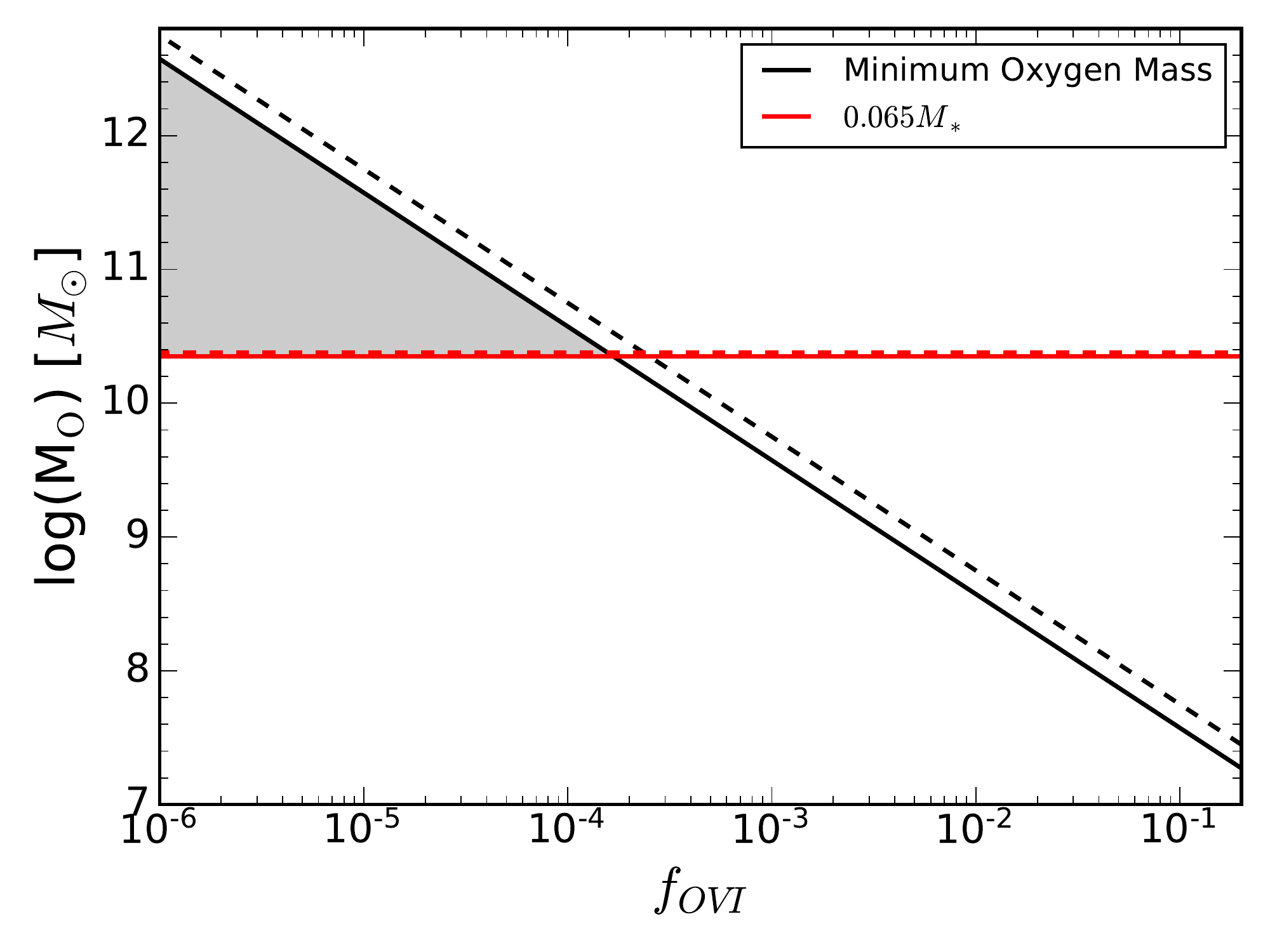}
\caption{The minimum mass of oxygen as calculated by Equation \ref{eq:mass} compared to the mass of oxygen in the member galaxies. The solid lines show the masses determined by median values while the dashed lines show the masses determined through the mean of the stellar mass, virial radii, and \ovi\ column densities. The gray shaded region shows the mass of oxygen that can be attributed to the IGrM at various values of $f_{OVI}$, the fraction of oxygen that is in \ovi.}
\label{fig:oxygenmass}
\end{figure}

Lastly, we can make an estimate as to the total amount of oxygen in these galaxy groups. Following equation 1 in \cite{Tumlinson2011}, we can calculate the minimum mass of oxygen in galaxy group halos by:

\begin{equation}
M_O = 5\pi \langle R_{vir} \rangle^2 \langle N_{OVI} \rangle m_O f_{hit} (\frac{0.2}{f_{OVI}})
\label{eq:mass}
\end{equation}

where $f_{OVI}$ is the fraction of oxygen that is in \ovi\ based upon CIE models \cite{Gnat2007}. Using both the mean and median values of the \ovi\ column densities and group virial radii, we can determine the minimum amount of oxygen mass in our galaxy groups. This can be compared to the amount of oxygen in the member galaxies by assuming $M_O \sim 0.065M_*$ \citep{Peeples2014,Tumlinson2017}. This difference (gray shaded region) is shown in Figure \ref{fig:oxygenmass} for both the mean (dashed lines) and median (solid lines) values of the stellar masses of group members. Based upon the virial temperature of these galaxy groups in Figure \ref{fig:temps}, the corresponding fraction of oxygen in \ovi\ is $<10^{-4}$. Therefore, we can estimate that over the narrow temperature range corresponding to $f_{OVI}$ of $10^{-4} - 10^{-5}$ ($10^{6.55}-10^{6.75}$~K), there is upwards of $10^{11.6}~M_{\odot}$ of oxygen in the IGrM.

\subsection{Future Outlook}

In order to accurately and completely characterize the \igrm, higher ionization species should be targeted in future studies. From the \cosigrm\ survey, it is clear that \ovi\ is not an ideal tracer of the \igrm. Since \ovi\ is only observed in 8 out of our 18 groups, the predominant, volume filling component of the \igrm\ should exist at a hotter temperature for galaxy groups at halo masses between 12.~\msun $<$ log[M$_{\rm halo}$] $<$ 14.7~\msun. We can rule out a pervasive media of the IGrM at cooler temperatures due to the weak low and medium ionization potential lines observed in our data. Therefore, to observe the dominant phase of the \igrm, future studies should look to \ion{O}{7}, \ion{O}{8}, \ion{Ne}{8} and \ion{Mg}{10}, which are stronger transitions at temperatures of $10^{6.5}-10^{7.5}$~K. 

Once the pervasive phase of the \igrm\ is observed, it can be combined with other studies to fully characterize the \igrm\ of galaxy groups. Simulations by \cite{Dave2002, LeBrun2017,Farahi2018} have made substantial progress in determining consistent scaling relations for lower mass halos that are consistent with observational programs such as those by \cite{Sun2003,Eckmiller2011,Babyk2018,Lovisari2020}. Additionally, the thermal SZ effect is being utilized in order to determine the baryonic content of lower mass galaxy clusters and groups \citep[and references therein]{Vikram2017, Henden2019,Pratt2020}. By the combination of these results, these hot halos can be fully characterized.

\section{Conclusions} \label{sec:conclusion}

We present the results of the \cosigrm\ survey, where 18 QSO sightlines passing through galaxy groups (0.2R$_{vir}\leq \rho \leq$1.6R$_{vir}$) were studied in an effort to characterize the \igrm. Our conclusions are as follows:
\begin{enumerate}
\item We detect \lya\ absorption in 12 of the 18 galaxy groups, with 4 of those groups also having corresponding \lyb\ absorption. However, we detect no statistically significant trend between \lya\ column density and halo mass or QSO impact parameter. 

\item 8 of the 18 groups show the presence of \ovi\, thus the covering fraction of \ovi\ is 44$\pm$5\%. The lack of \ovi\ absorption in over 50\% of our sample indicates that the volume filling \igrm\ at (or near) the virial temperature of galaxy groups is not primarily traced by \ovi. We also find no correlation between column density of \ovi\ and halo mass or QSO impact parameter. 

\item \cii, \siii, \siiii, and \nv\ absorption was detected in 5, 1, 5, and 2 groups, respectively. These lead to covering fractions of 28$\pm$5\%, 6$\pm$5\%, 28$\pm$5\%, and 11$\pm$5\% for \cii, \siii, \siiii, and \nv, respectively. These data suggest that the low-ionization transitions are primarily due to photoionization or other non-equilibrium processes. 

\item We find evidence that the \igrm\ is multiphase and has a complex structure. While higher resolution spectra and coverage of more intermediate ionization transitions are necessary for complete ionization modeling, we find five instances where CIE explains the observed spectra and four instances where photoionization is consistent with the transitions present. 

\item We find that 9 out of 70 absorbers (13$\pm$1\%) have sufficient velocities, relative to the group, to escape the group's gravitational potential. Therefore, we conclude that galaxy groups are primarily ``closed boxes'' for galaxy evolution at low redshifts (0.1 $\le z \le$0.2).

\item We show that the \ovi\ absorbers can be described by radiatively cooling gas between $10^{5.8}$~K, and $10^{6}$~K. This might indicate that the \ovi\ detected in our sample originates from pockets of gas cooling within the hotter component of the IGrM. 

\item We do not find evidence of AGN activity having an impact on whether or not \ovi\ is detected within a group or not. Similarly, we do not observe the star formation of the nearest spectroscopically confirmed neighbor to be a driver for \ovi. 

\item We observe some \ovi\ absorption in our stacked data. This shows evidence of \ovi\ traced \igrm\ throughout our sample. Despite \ovi\ not being the dominate form of oxygen at the virial temperature of these galaxy groups, we see evidence that we are observing gas cooler than the hot, volume filling component of the IGrM that could be observed in X-rays via \ion{O}{7}, \ion{O}{8}, \ion{Na}{12}, or extreme-UV lines such as \ion{Ne}{8}.

\end{enumerate}

Since the \ovi\ detections are determined to be primarily tracers of cooler pockets of gas and not the \igrm\ at the virial temperature of the group,
full accounting for the amount of baryonic matter in these groups cannot be accurately measured with the data in hand. In order to complete the baryon census for galaxy groups, future studies should try and observe higher ionization states such as \ion{O}{7} and \ion{O}{8}, which will trace gas closer to the virial temperature of galaxy groups. 

\acknowledgements
The authors would like to thank the anonymous referee for their feedback and suggestions to improve this manuscript. TM, SB, TH are supported by HST-GO-13314 from STScI. The authors would also like to thank Mark Voit along with Chris Dupuis, Mansi Padave, Ed Buie II, Hansung Gim, Jacqueline Monkiewicz, and Martin Flores from the STARs Lab at ASU for thoughtful discussion and input throughout this project. The Arizona State University authors acknowledge the twenty two Native Nations that have inhabited this land for centuries. Arizona State University’s four campuses are located in the Salt River Valley on ancestral territories of Indigenous peoples, including the Akimel O’odham (Pima) and Pee Posh (Maricopa) Indian Communities, whose care and keeping of these lands allows us to be here today. We acknowledge the sovereignty of these nations and seek to foster an environment of success and possibility for Native American students and patrons.

\bibliography{References}
\bibliographystyle{aasjournal}

\startlongtable
\begin{deluxetable*}{llCCCC}
\tabletypesize{\scriptsize}
\tablecolumns{6}
\tablecaption{Absorption Line Measurements within $\pm 800\ \rm{kms^{-1}}$ of Group Center}

\tablehead{
\colhead{Species}  & \dcolhead{\lambda_{rest}} 	& \dcolhead{\rm{W_{rest}}} 		& \dcolhead{v_{obs}} 	 		& \colhead{logN} 					& \dcolhead{b}					\\[-0.3cm]
\colhead{ }		   &\dcolhead{(\text{\AA})} 	& \dcolhead{(\text{m\AA})}   	& \dcolhead{(\rm{kms^{-1}})}	& \dcolhead{(\rm{log[cm^{-2}}])} 	& \dcolhead{(\rm{kms^{-1}})} 	\\[-0.5cm]
}
\startdata
\cutinhead{J0841+1406 at $z_{gp}$ = 0.125}
\hi\		&	1215	&	113.5 \pm 20.1		&	148.8\pm7.3		&	13.33\pm0.11	&	36.6~^{+15.5}_{-10.9}	\\
		&		&					&	218.7\pm10.1	&	12.80\pm0.29	&	19.7~^{+47.8}_{-14.0}	\\
\cii\		&	1036	&	-8.4\pm17.1			&		$-$		&	\le 13.63		&			$-$		\\
\nv\ 		&	1238	&	0.7\pm18.0			&		$-$		&	\le 13.41		&			$-$		\\
\siii\ 	&	1260	&	48.7\pm18.6			&		$-$		& 	\le 12.53		&			$-$		\\
\siiii\	&	1206	&	38.3\pm15.3			&		$-$		&	\le 12.33		&			$-$		\\
\ovi\ 	&	1031	&	107.5\pm19.9		&	-752.9\pm14.7	&	14.15\pm0.10	&	81.6~^{+26.2}_{-19.8}	\\
\cutinhead{J1017+4702 at $z_{gp}$ = 0.164}
\hi\		&   1215   	&  	1602.1\pm29.9	      &   340.1\pm11.8        &   14.73\pm0.34    &   41.6~^{+20.2}_{-13.6}   \\
            &           &    			            &   499.0\pm24.3        &   14.89\pm0.11    &   113.4~^{+41.3}_{-30.3}  \\
            &           &                             &   645.6\pm6.8         &   14.39\pm0.29    &   34.6~^{+29.1}_{-15.8}   \\
\cii\	      &   1036	&   67.5\pm18.7               &   502.8\pm5.9         &   13.41\pm0.21    &   12.5~^{+29.0}_{-8.7}    \\
            &           &                             &   587.5\pm4.3         &   13.82\pm0.10    &   21.1~^{+7.9}_{-5.7}     \\
\nv\		&	1238	&	6.4\pm24.0			&		$-$		&	\le 13.53       &			$-$				\\
\siii\      &     1260$^a$      &           $-$       &           $-$         &           $-$         &                 $-$               \\
\siii\		&	1193    &	12.8\pm11.7		&		$-$		&	\le 12.68		&			$-$			\\
\siiii\     &   1206	&   250.5\pm28.0               &   496.9\pm5.4              &   12.30\pm0.20        &   11.6~^{+22.5}_{-7.7}    \\
            &           &  	                         &   581.2\pm3.7              &   13.20\pm0.06        &   36.1~^{+5.4}_{-4.7}     \\
\ovi\	      &   1031    &   103.1\pm24.5               &   463.2\pm30.4             &   13.67\pm0.26        &   64.2~^{+86.6}_{-36.9}   \\
            &           &                              &   615.7\pm23.8             &   13.78\pm0.19        &   67.8~^{+51.4}_{-29.2}   \\
\cutinhead{J1020+1003 at $z_{gp}$ = 0.123}
\hi\        &   1215    &   75.8\pm21.2               &   508.5\pm3.1               &   13.23\pm0.12        &   11.7~^{+6.6}_{-4.2}     \\
\cii\		&	1036	&	27.0\pm56.9			&		$-$			&	\le 14.15		&			$-$				\\
\nv\		&	1238	&	32.8\pm24.6			&		$-$			&	\le 13.54		&			$-$				\\
\siii\      &     1260$^a$      &           $-$       &           $-$               &           $-$         &                 $-$                     \\
\siii\	&	1193    &	-12.6\pm23.2		&		$-$			&	\le 12.98		&			$-$				\\
\siiii\	&	1206	&	-18.3\pm21.7		&		$-$			& 	\le 12.48		&			$-$				\\
\ovi\		&	1031	&	14.9\pm56.4			&		$-$			& 	\le 14.13		&			$-$				\\
\cutinhead{J1025+4808 at $z_{gp}$ = 0.133}
\hi\		&	1215	&	6.4\pm11.6			&		$-$			&	\le 12.81		&			$-$				\\
\cii\		& 	1036	&	25.4\pm14.8			&		$-$			&	\le 13.57		&			$-$				\\
\nv\		&	1238	&	10.0\pm17.2			&		$-$			&	\le 13.39		&			$-$				\\
\siii\      &     1260$^a$      &           $-$       &           $-$               &           $-$         &                 $-$                     \\
\siii\	&	1193    &	6.5\pm14.0			&		$-$			& 	\le 12.76		&			$-$				\\
\siiii\	&	1206	&	20.9\pm11.7			&		$-$			&	\le 12.21		&			$-$				\\
\ovi\		& 	1031	&	-2.5\pm15.5			&		$-$			&	\le 13.57		&			$-$				\\
\cutinhead{J1102+0521 at $z_{gp}$ = 0.131}
\hi\		&	1215	&	12.2\pm16.8			&		$-$			&	\le 12.97		&			$-$				\\
\cii\		&	1036	&	8.0\pm13.7			&		$-$			&	\le 13.53		&			$-$				\\
\nv\		&	1238	&	18.2\pm16.9			&		$-$			&	\le 13.38		&			$-$				\\
\siii\      &     1260$^a$      &           $-$       &           $-$               &           $-$         &                 $-$                     \\
\siii\	&	1193    &	27.3\pm15.9			&		$-$			&	\le 12.81		&			$-$				\\
\siiii\	&	1206	&	32.5\pm16.5			&		$-$			&	\le 12.36		&			$-$				\\
\ovi\		&	1031	&	2.8\pm17.8			&		$-$			&	\le 13.63		&			$-$				\\
\cutinhead{J1126+1204 at $z_{gp}$ = 0.164}
\hi\        &   1215    &   269.8\pm21.6              &   -70.3\pm27.6              &   13.72\pm0.44        &   36.5~^{+28.5}_{-16.0}                  \\
            &           &                             &   -35.2\pm14.3              &   13.54\pm0.72        &   17.2~^{+19.5}_{-9.1}                   \\
\cii\		&	1036	&	-32.3\pm17.0		&		$-$			&	\le 13.63		&			$-$				\\
\nv\		&	1238	&	5.2\pm21.9			&		$-$			&	\le 13.49		&			$-$				\\
\siii\      &     1260$^a$      &           $-$       &           $-$               &           $-$         &                 $-$                     \\
\siii\	&	1193    &	-12.6\pm17.7		&		$-$			&	\le 12.86		&			$-$				\\
\siiii\	&	1206	&	16.0\pm18.7			&		$-$			&	\le 12.42		&			$-$				\\
\ovi\		&	1031	&	15.0\pm15.8			&		$-$			&	\le 13.58		&			$-$				\\
\cutinhead{J1127+2654 at $z_{gp}$ = 0.152}
\hi\        &   1215    &   1385.7\pm16.0       &   -58.9\pm8.8     &   15.51\pm0.10    &   57.2~^{+4.7}_{-4.3}     \\
            &           &                       &   32.1\pm8.3      &   18.34\pm0.13    &   13.4~^{+3.0}_{-2.4}     \\
            &           &                       &   115.6\pm4.1     &   15.26\pm0.07    &   36.7~^{+3.3}_{-3.0}     \\
\hi\        &   1025    &   1094.7\pm16.1       &   -58.9\pm8.8     &   15.51\pm0.10    &   57.2~^{+4.7}_{-4.3}     \\
            &           &                       &   32.1\pm8.3      &   18.34\pm0.13    &   13.4~^{+3.0}_{-2.4}     \\
            &           &                       &   115.6\pm4.1     &   15.26\pm0.07    &   36.7~^{+3.3}_{-3.0}     \\
\cii\       &   1036$^b$ &   \le 506.2\pm23.7   &   -78.4\pm10.0    &   14.21\pm0.29    &   31.1~^{+30.3}_{-15.3}   \\ 
            &           &                       &   -13.2\pm7.3     &   14.49\pm0.10    &   37.8~^{+8.9}_{-7.2}     \\
            &           &                       &   73.6\pm7.5      &   13.92\pm0.12    &   32.6~^{+16.3}_{-10.9}   \\
            &           &                       &   145.3\pm5.1     &   13.98\pm0.10    &   26.7~^{+9.1}_{-6.8}     \\
\nv\		&   1238	&	2.7\pm24.4	      &		$-$	  &	\le 13.54	    &			$-$  		  \\
\siii\      &   1260    &   460.0\pm46.7        &   -47.2\pm8.4     &   13.66\pm0.07    &   60.2~^{+10.7}_{-9.1}    \\
            &           &                       &   -1.8\pm1.1      &   13.29\pm0.10    &   7.1~^{+3.0}_{-2.1}      \\
            &           &                       &   53.0\pm7.5      &   12.91\pm0.21    &   25.5~^{+16.0}_{-9.8}    \\
            &           &                       &   122.4\pm6.6     &   12.90\pm0.12    &   28.5~^{+12.7}_{-8.8}    \\
\siii\      &   1193    &   324.0\pm26.9        &   -47.2\pm8.4     &   13.66\pm0.07    &   60.2~^{+10.7}_{-9.1}    \\
            &           &                       &   -1.8\pm1.1      &   13.29\pm0.10    &   7.1~^{+3.0}_{-2.1}      \\
            &           &                       &   53.0\pm7.5      &   12.91\pm0.21    &   25.5~^{+16.0}_{-9.8}    \\
            &           &                       &   122.4\pm6.6     &   12.90\pm0.12    &   28.5~^{+12.7}_{-8.8}    \\
\siii\      &   1190    &   267.8\pm27.4        &   -47.2\pm8.4     &   13.66\pm0.07    &   60.2~^{+10.7}_{-9.1}    \\
            &           &                       &   -1.8\pm1.1      &   13.29\pm0.10    &   7.1~^{+3.0}_{-2.1}      \\
            &           &                       &   53.0\pm7.5      &   12.91\pm0.21    &   25.5~^{+16.0}_{-9.8}    \\
            &           &                       &   122.4\pm6.6     &   12.90\pm0.12    &   28.5~^{+12.7}_{-8.8}    \\
\siiii\     &   1206    &   710.9\pm25.3        &   -67.8\pm8.1     &   13.29\pm0.09    &   43.3~^{+9.5}_{-7.8}     \\
            &           &                       &   -5.1\pm3.4      &   13.27\pm0.12    &   24.4~^{+7.3}_{-5.6}     \\
            &           &                       &   61.4\pm11.9     &   12.95\pm0.25    &   41.0~^{+26.9}_{-16.2}   \\
            &           &                       &   130.2\pm2.3     &   12.80\pm0.20    &   13.6~^{+6.6}_{-4.5}     \\
            &           &                       &   178.5\pm53.8    &   12.91\pm0.36    &   76.1~^{+98.7}_{-43.0}   \\
\ovi\       &   1031    &   322.9\pm23.7        &   -25.2\pm5.8     &   14.48\pm0.04    &   83.1~^{+9.3}_{-8.3}     \\
            &           &                       &   127.7\pm8.1     &   13.79\pm0.12    &   36.9~^{+15.1}_{-10.7}   \\
\ovi\       &   1037    &   171.3\pm25.4        &   -25.2\pm5.8     &   14.48\pm0.04    &   83.1~^{+9.3}_{-8.3}     \\
            &           &                       &   127.7\pm8.1     &   13.79\pm0.12    &   36.9~^{+15.1}_{-10.7}   \\  
\cutinhead{J1216+0712 at $z_{gp}$ = 0.136}
\hi\        &   1215    &   64.9\pm18.9         &   83.4\pm4.4      &   13.20\pm0.14    &   15.6~^{+9.3}_{-5.8}     \\
\hi\		&   1215    &   401.8\pm26.6        &   794.2\pm2.2     &   14.56\pm0.06    &   31.7~^{+2.5}_{-2.3}     \\
            &           &                       &   880.8\pm10.5    &   12.87\pm0.31    &   17.0~^{+129.3}_{-15.1}  \\
\hi\        &   1025    &   131.8\pm35.3        &   794.2\pm2.2     &   14.56\pm0.06    &   31.7~^{+2.5}_{-2.3}     \\
            &           &                       &   880.8\pm10.5    &   12.87\pm0.31    &   17.0~^{+129.3}_{-15.1}  \\
\cii\		&	1036	&	26.9\pm26.1			&		$-$			&	\le 13.81		&			$-$				\\
\nv\		&	1238	&	-9.0\pm27.2			&		$-$			&	\le 13.59		&			$-$				\\
\siii\      &     1260$^a$      &           $-$       &           $-$               &           $-$         &                 $-$                     \\
\siii\	&	1193    &	24.3\pm24.5			&		$-$			&	\le 13.00		&			$-$				\\
\siiii\	&	1206	&	6.0\pm36.8			&		$-$			&	\le 12.71		&			$-$				\\
\ovi\       &   1031    &   132.7\pm34.7        &   314.1\pm21.6    &   13.83\pm0.30    &   39.3~^{+65.3}_{-24.5}   \\
            &           &                       &   390.0\pm18.6    &   13.80\pm0.31    &   33.6~^{+62.8}_{-21.9}   \\
\cutinhead{J1301+2819 at $z_{gp}$ = 0.144}
\hi\        &   1215    &   563.7\pm20.2              &   -204.8\pm10.4             &   13.05\pm0.19        &   28.8~^{+23.4}_{-12.9}                 \\
            &           &                             &   -131.4\pm5.4              &   13.99\pm0.10        &   25.1~^{+7.1}_{-5.5}                   \\
            &           &                             &   -68.7\pm3.14              &   14.53\pm0.06        &   26.9~^{+2.8}_{-2.5}                    \\
\hi\        &   1025    &   251.2\pm24.8              &   -204.8\pm10.4             &   13.05\pm0.19        &   28.8~^{+23.4}_{-12.9}                 \\
            &           &                             &   -131.4\pm5.4              &   13.99\pm0.10        &   25.1~^{+7.1}_{-5.5}                    \\
            &           &                             &   -68.7\pm3.14              &   14.53\pm0.06        &   26.9~^{+2.8}_{-2.5}                    \\
\cii\		&	1036	&	23.8\pm17.2			&		$-$			&	\le 13.63		&			$-$				\\
\nv\        &   1238    &   90.2\pm18.6               &   -135.7\pm5.6              &   13.10\pm0.18        &   15.5~^{+14.4}_{-7.5}                  \\  
            &           &                             &   -85.8\pm7.0               &   13.16\pm0.17        &   21.0~^{+16.7}_{-9.3}                  \\
\siii\	&	1260	&	-8.1\pm38.8			&		$-$			&	\le 12.85		&			$-$				\\
\siiii\	&	1206	&	9.9\pm16.3			&		$-$			&	\le 12.36		&			$-$				\\
\ovi\       &   1031    &   88.0\pm20.5               &   -238.2\pm10.7             &   13.65\pm0.19        &   29.8~^{+25.3}_{-13.7}                 \\
            &           &                             &   -96.7\pm15.9              &   14.00\pm0.12        &   69.2~^{+29.4}_{-20.6}                 \\
\cutinhead{J1339+5355 at $z_{gp}$ = 0.159}
\hi\        &   1215    &   536.9\pm46.3              &   504.6\pm26.8              &   13.55\pm0.33        &   43.3~^{+54.4}_{-24.1}                 \\
            &           &                             &   599.3\pm7.9               &   14.34\pm0.12        &   43.7~^{+10.1}_{-8.2}                  \\
\cii\		&	1036	&	-33.0\pm29.4		&		$-$			&	\le 13.86		&			$-$				\\
\nv\		&	1238	&	-10.5\pm42.6		&		$-$			&	\le 13.78		&			$-$				\\
\siii\      &     1260$^a$      &           $-$       &           $-$               &           $-$         &                 $-$                     \\   
\siii\	&	1193    &	-6.3\pm28.6			&		$-$			&	\le 13.07		&			$-$				\\
\siiii\	&	1206	&	26.3\pm29.6			&		$-$			&	\le 12.62		&			$-$				\\
\ovi\		&	1031	&	53.2\pm26.4			&		$-$			&	\le 13.80		&			$-$				\\
\cutinhead{J1343+2538 at $z_{gp}$ = 0.075}
\hi\        &   1215    &   52.2\pm3.8                &   -333.5\pm1.1              &   13.06\pm0.03        &   18.5~^{+1.8}_{-1.6}                   \\
\hi\        &   1215    &   48.9\pm6.0                &   5.2\pm4.8                 &   12.95\pm0.05        &   47.1~^{+7.3}_{-6.4}                   \\
\cii\		&	1036	&	12.8\pm8.3			&		$-$			&	\le 13.31		&			$-$				\\
\nv\		&	1238	&	9.0\pm4.0			&		$-$			&	\le 12.75		&			$-$				\\
\siii\      &     1260$^a$      &           $-$       &           $-$               &           $-$         &                 $-$                     \\
\siii\	&	1193    &	-1.7\pm5.6			&		$-$			&	\le 12.36		&			$-$				\\
\siiii\	&	1206	&	-4.8\pm5.0			&		$-$			&	\le 11.84		&			$-$				\\
\ovi\       &   1031    &   47.0\pm12.0               &   33.0\pm7.5                &   13.60\pm0.11        &   36.5~^{+13.3}_{-9.7}                  \\
\cutinhead{J1344+5546 at $z_{gp}$ = 0.155}
\hi\		&	1215	&	74.8\pm53.6			&		$-$			&	\le 13.47		&			$-$				\\
\cii\		&	1036	&	-1.8\pm25.3			&		$-$			&	\le 13.80		&			$-$				\\
\nv\		&	1238	&	15.2\pm25.6			&		$-$			&	\le 13.56		&			$-$				\\
\siii\      &     1260$^a$      &           $-$       &           $-$               &           $-$         &                 $-$                     \\
\siii\	&	1193    &	-4.3\pm43.4			&		$-$			&	\le 13.25		&			$-$				\\
\siiii\	&	1206	&	73.1\pm44.9			&		$-$			&	\le 12.80		&			$-$				\\
\ovi\		&	1031	&	-5.3\pm26.7			&		$-$			&	\le 13.81		&			$-$				\\
\cutinhead{J1348+4303 at $z_{gp}$ = 0.095}
\hi\        &   1215    &   678.4\pm14.4              &   -335.0\pm93.3             &   13.61\pm0.34        &   145.8~^{+92.3}_{-56.5}                \\
            &           &                             &   -248.0\pm1.7              &   14.61\pm0.06        &   50.4~^{+4.7}_{-4.3}                   \\
            &           &                             &   -12.5\pm18.6              &   13.49\pm0.12        &   95.1~^{+28.8}_{-22.1}                 \\
\cii\		&	1036	&	126.0\pm26.2		&	-336.6\pm12.1	      &	13.39\pm0.31	&	19.0~^{+33.5}_{-12.1}	           \\
		&		&				      &	-259.7\pm8.6	      &	13.80\pm0.16	&	27.2~^{+19.1}_{-11.3}	             \\
		&		&				      &	-210.6\pm3.6	      &	13.67\pm0.91	&	4.3~^{+4.0}_{-60.5}		        \\
\nv\		&	1238	&	11.1\pm13.4			&		$-$		      &	\le 13.28		&			$-$				\\
\siii\      &     1260$^a$      &           $-$       &           $-$               &           $-$         &                 $-$                     \\
\siii\	&	1193    &	13.9\pm9.7			&		$-$			&	\le 12.24		&			$-$				\\
\siiii\     &     1206    &   85.7\pm7.3              &   -275.6\pm1.6              &   12.73\pm0.04        &   18.9~^{+2.5}_{-2.2}                    \\  
\ovi\		&	1031	&	17.7\pm21.0	 		&		$-$			&	\le 13.70		&			$-$				\\
\cutinhead{J1408+5657 at $z_{gp}$ = 0.130}
\hi\		&	1215	&	14.9\pm12.6			&		$-$			&	\le 12.84		&			$-$				\\
\cii\		&	1036	&	22.8\pm18.1			&		$-$			&	\le 13.65		&			$-$				\\
\nv\		&	1238	&	22.1\pm17.1			&		$-$			&	\le 13.38		&			$-$				\\
\siii\      &     1260$^a$      &           $-$       &           $-$               &           $-$         &                 $-$                     \\
\siii\	&	1193    &	-25.9\pm18.6		&		$-$			&	\le 12.88		&			$-$				\\
\siiii\	&	1206	&	-36.4\pm21.8		&		$-$			&	\le 12.48		&			$-$				\\
\ovi\		&	1031	&	-2.3\pm19.1			&		$-$			&	\le 13.66		&			$-$				\\
\cutinhead{J1424+4214 at $z_{gp}$ = 0.100}  
\hi\        &   1215    &   521.0\pm9.8               &   -65.9\pm0.9               &   14.39\pm0.02        &   47.1~^{+1.4}_{-1.4}       \\
\cii\       &   1036    &   66.2\pm12.2               &   -26.6\pm7.3               &   13.88\pm0.09        &   44.1~^{+12.3}_{-9.6}      \\
\nv\        &   1238    &   40.7\pm6.9                &   -102.0\pm4.7              &   13.01\pm0.08        &   29.4~^{+7.9}_{-6.2}       \\
\siii\      &   1260$^a$     &           $-$          &           $-$               &           $-$         &                 $-$         \\
\siii\	&   1193    &	19.0\pm8.4			&		$-$			&	\le 12.18		&			$-$		\\
\siiii\     &   1206    &   109.6\pm8.3               &   -106.9\pm4.4              &   12.03\pm0.18        &   12.0~^{+13.9}_{-6.4}      \\
            &           &                             &   -50.3\pm2.9               &   12.76\pm0.05        &   30.7~^{+4.8}_{-4.1}       \\
\ovi\       &   1037    &   64.5\pm12.5               &   -91.6\pm13.7              &   14.25\pm0.09        &   83.2~^{+21.8}_{-17.3}     \\
\cutinhead{J1426+1955 at $z_{gp}$ = 0.109}
\hi\		&	1215	&	-4.8\pm13.6			&		$-$			&	\le 12.87		&			$-$		\\
\cii\		&	1036	&	0.8\pm12.6			&		$-$			&	\le 13.48		&			$-$		\\
\nv\		&	1238$^c$ 	&		$-$		&		$-$			&		$-$		&			$-$	      \\
\siii\      &     1260$^a$      &           $-$       &           $-$               &           $-$         &                 $-$         \\
\siii\	&	1193    &	31.1\pm11.7			&		$-$			&	\le 12.70		&			$-$		\\
\siiii\	&	1206	&	20.9\pm12.1			&		$-$			&	\le 12.25		&			$-$		\\
\ovi\		&	1031	&	-11.1\pm13.0		&		$-$			&	\le 13.50		&			$-$		\\
\cutinhead{J1428+3225 at $z_{gp}$ = 0.131}
\hi\        &   1215    &   606.8\pm24.4              &   -370.7\pm18.3             &   13.05\pm0.27        &   34.0~^{+40.8}_{-18.5}   \\
            &           &                             &   -296.5\pm1.40             &   14.71\pm0.05        &   27.0~^{+1.6}_{-1.5}     \\
            &           &                             &   -161.8\pm10.0             &   13.70\pm0.06        &   85.8~^{+16.6}_{-13.9}   \\
\hi\        &   1215    &   111.8\pm16.1              &   392.4\pm3.9               &   13.38\pm0.06        &   31.1~^{+6.0}_{-5.1}     \\
\hi\        &   1025    &   222.1\pm21.0              &   -370.7\pm18.3             &   13.05\pm0.27        &   34.0~^{+40.8}_{-18.5}   \\
            &           &                             &   -296.5\pm1.40             &   14.71\pm0.05        &   27.0~^{+1.6}_{-1.5}     \\
            &           &                             &   -161.8\pm10.0             &   13.70\pm0.06        &   85.8~^{+16.6}_{-13.9}   \\
\cii\       &   1036    &   48.8\pm12.1               &   -311.8\pm16.1             &   13.89\pm0.11        &   74.4~^{+27.4}_{-20.0}   \\
\nv\		&   1238	&	-4.3\pm15.3	            &		$-$	            &	\le 13.34		&			$-$	    \\
\siii\      &   1260$^a$&           $-$               &           $-$               &           $-$         &                 $-$       \\  
\siii\	&   1193    &	20.7\pm12.7		      &		$-$	            &	\le 12.36		&			$-$	    \\
\siiii\     &   1206    &   199.6\pm17.5              &   -290.6\pm14.0             &   13.03\pm0.08        &   103.1~^{+18.9}_{-16.0}  \\  
\ovi\       &   1031    &   45.4\pm13.4               &   -291.9\pm8.1              &   13.61\pm0.14        &   32.5~^{+16.7}_{-11.0}   \\  
\cutinhead{J1617+0854 at $z_{gp}$ = 0.099}
\hi\		&	1215	&	41.7\pm24.4			&		$-$			&	\le 13.13		&			$-$				\\
\cii\		&	1036	&	-5.2\pm31.4			&		$-$			&	\le 13.89		&			$-$				\\
\nv\		& 	1238	&	-9.8\pm25.0			&		$-$			&	\le 13.55		&			$-$				\\
\siii\      &     1260$^a$      &           $-$       &           $-$               &           $-$         &                 $-$                     \\
\siii\	&	1193  &	-31.8\pm23.7		&		$-$			&	\le 12.98		&			$-$				\\
\siiii\	&	1206	&	-11.8\pm25.8		&		$-$			&	\le 12.56		&			$-$				\\
\ovi\		&	1031	&	18.6\pm39.2			&		$-$			&	\le 13.97		&			$-$				\\
\enddata
\tablecomments{(a) There is no COS coverage of \siii\ 1260 at the observed wavelength. (b) \cii\ is blended with Milky Way \siii\ 1193, which was modeled and removed. Since the Milky Way \siii\ line was saturated, the current data do not allow for all \cii\ absorption to be measured.  (c) There is no COS coverage of \nv\ 1238 at the observed wavelength.}
\label{tab:lines}
\end{deluxetable*}
\normalsize

\appendix
\twocolumngrid

\section{Group Environments}\label{ap:env}
The remaining group environments are shown below in Figure \ref{fig:allenv}. These environments were constructed from the New York University Value-Added Galaxy Catalog (NYU-VAGC; \cite{Blanton2005}) based on SDSS Data Release 2. These environments were used to distinguish between QSO sightlines that pass through only the \igrm\ from those passing through the  CGM and the \igrm. 

\begin{figure*}
\centering
\plotone{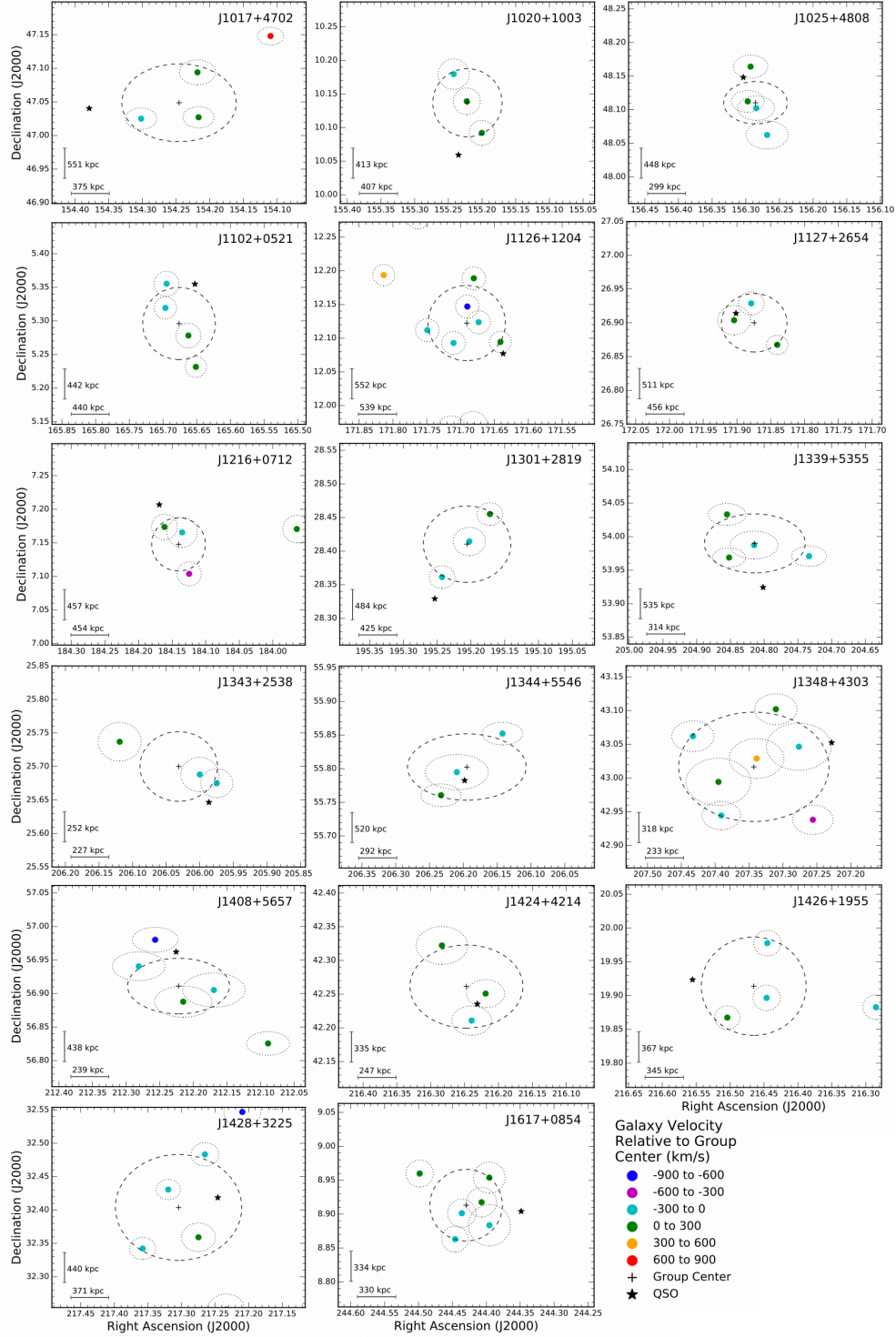}
\caption{Environment plots for the remaining groups in the \cosigrm\ sample that showed absorption lines within $\pm800$ \kms\ of the group center. The color of the points represent the velocity of the member galaxies relative to the center of the group. The thick, dashed line represents the virial radius of the group, the thin, dotted lines represent the virial radii of the group members, the QSO sightline is represented by the star, and the group center is marked by the plus sign.}
\label{fig:allenv}
\end{figure*}

\section{Absorption Line Detections}
The following plots show all of the sightlines for our sample. Intervening Milky Way absorption lines are marked in blue, while intervening absorption lines are labeled in green. However, without full wavelength coverage, the identity of all intervening lines cannot be determined. We have identified all lines that are robustly confirmed.

\begin{figure}
\centering
\plotone{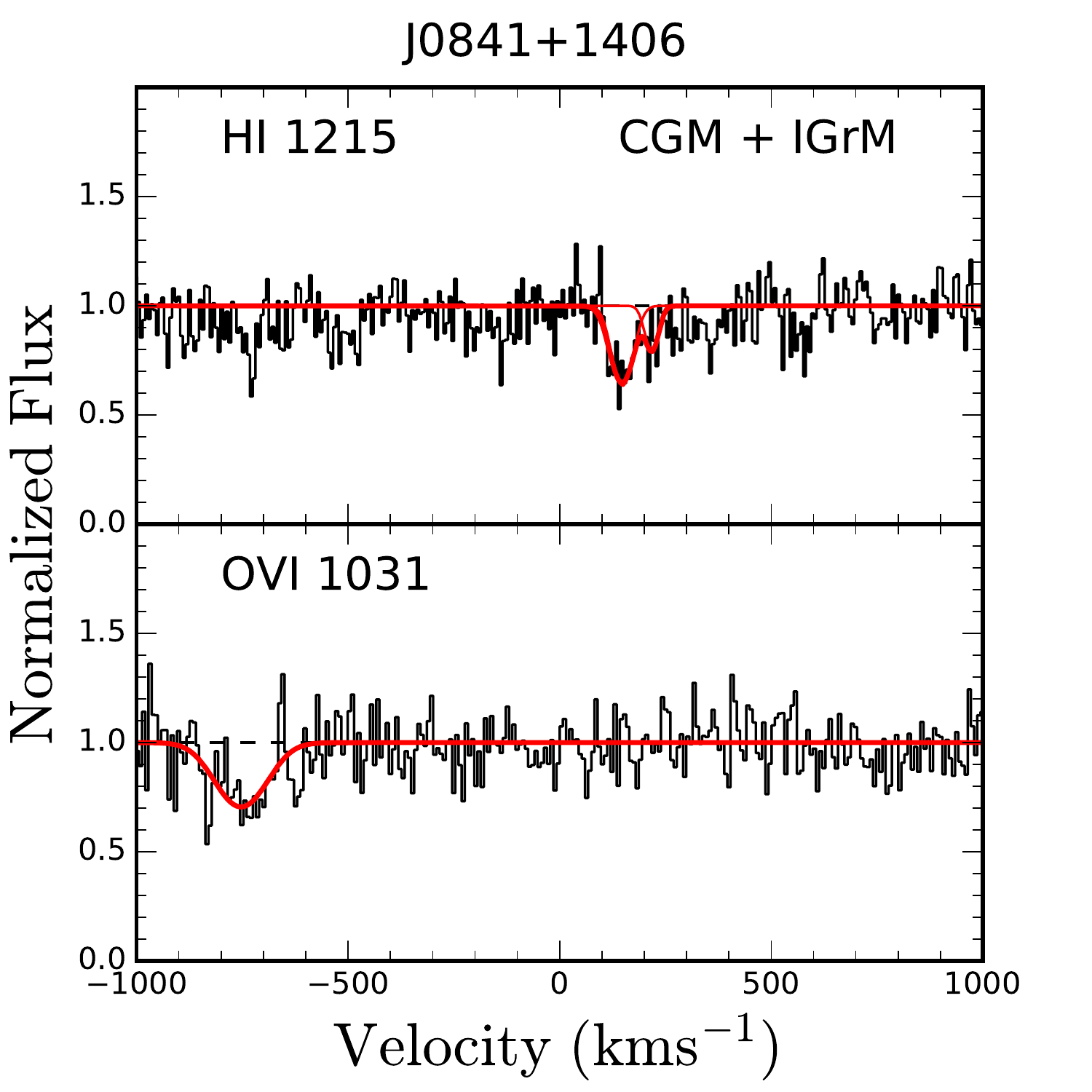}
\caption{The observed spectra of sightline J0841+1406 showing \hi\ and \ovi\ detections. The Voigt profile fits are shown in red.}
\label{J0841_spec}
\end{figure}

\begin{figure}
\centering
\plotone{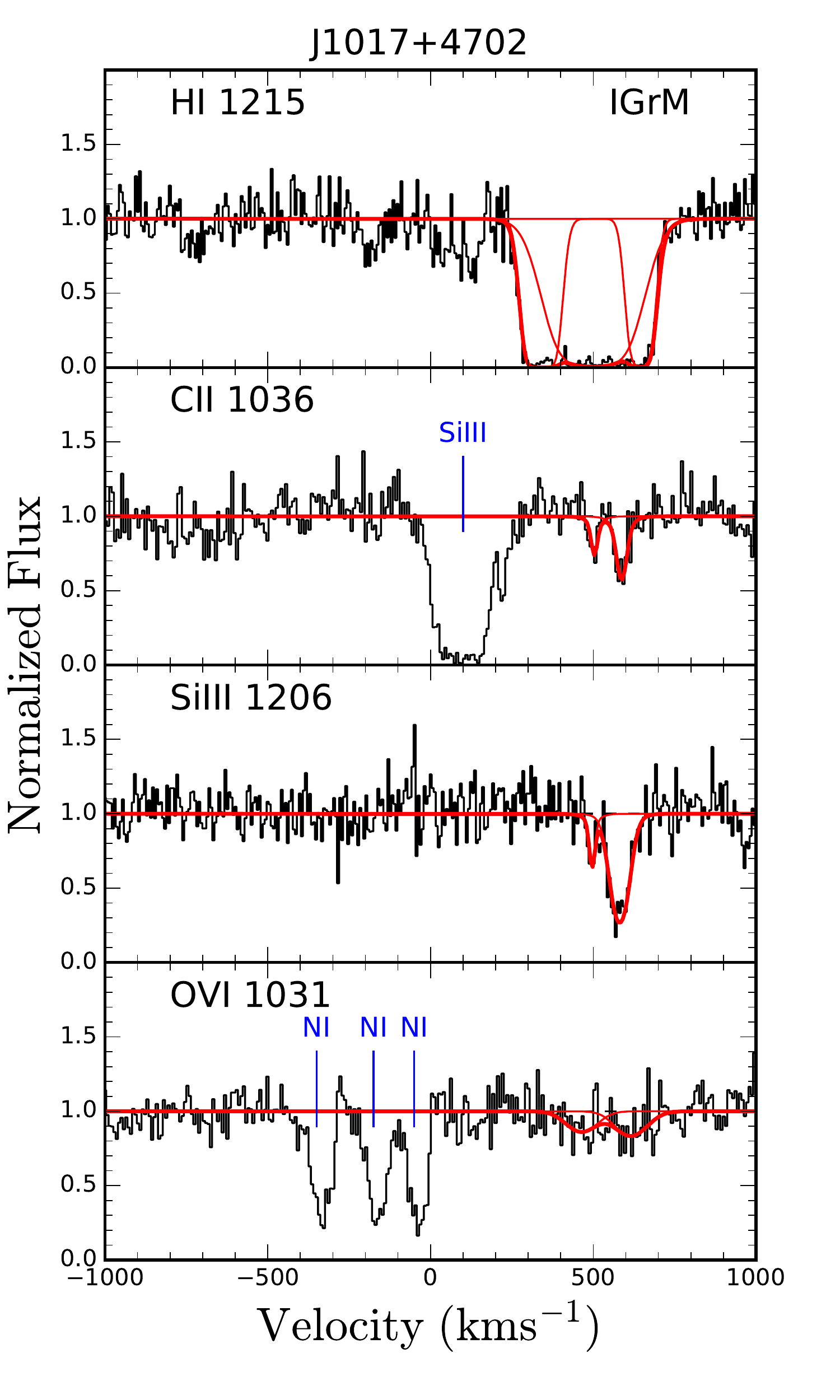}
\caption{The observed spectra of sightline J1017+4702 showing \hi, \cii, \siiii, and \ovi\ detections. The Voigt profile fits are shown in red and intervening Milky Way absorption lines are marked in blue.}
\label{J1017_spec}
\end{figure}

\begin{figure}
\centering
\plotone{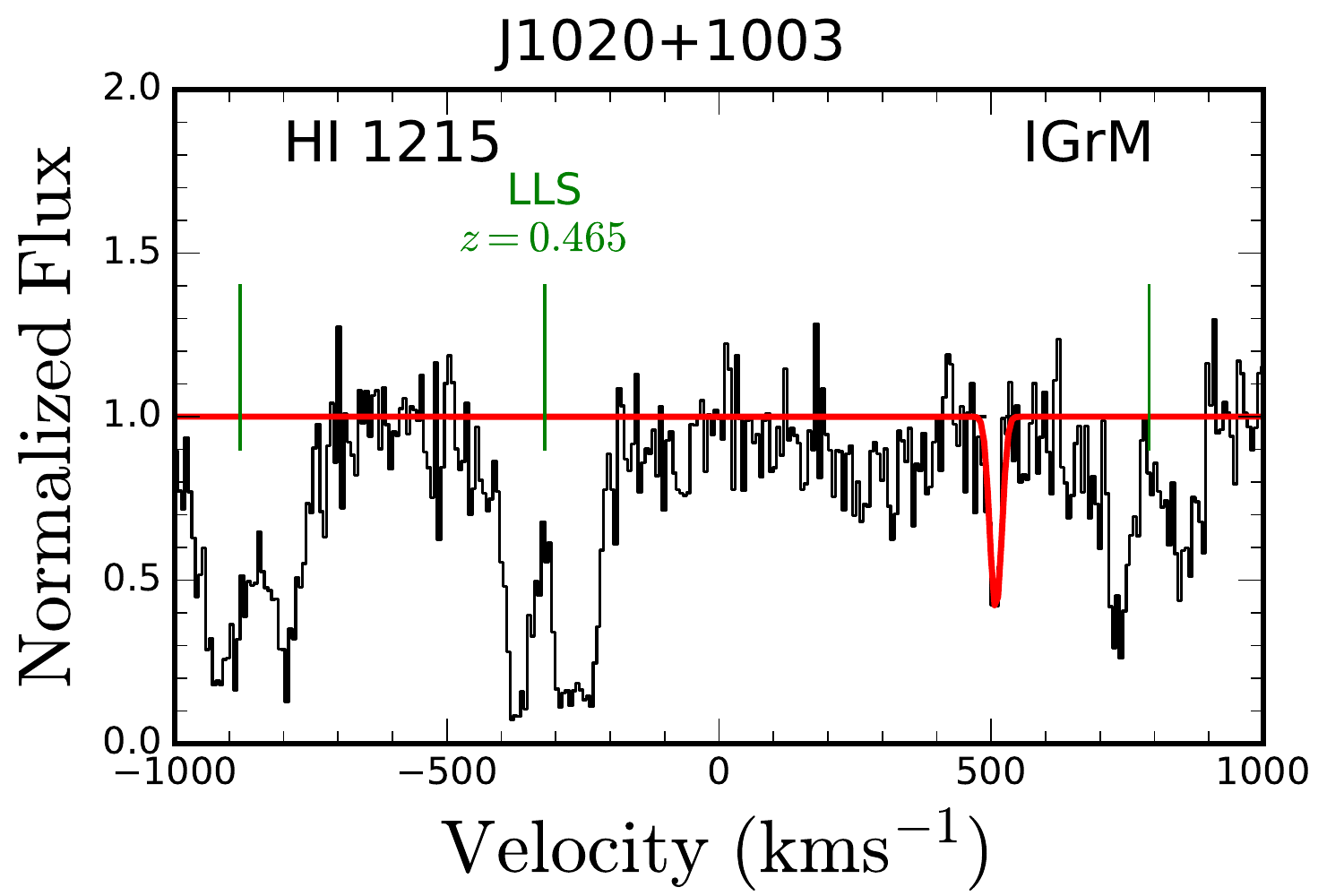}
\caption{The observed spectra of sightline J1020+1003 showing \hi\ absorption. The Voigt profile fit is shown in red and an intervening partial Lyman limit system line is labeled in green. }
\label{J1020_spec}
\end{figure}

\begin{figure}
\centering
\plotone{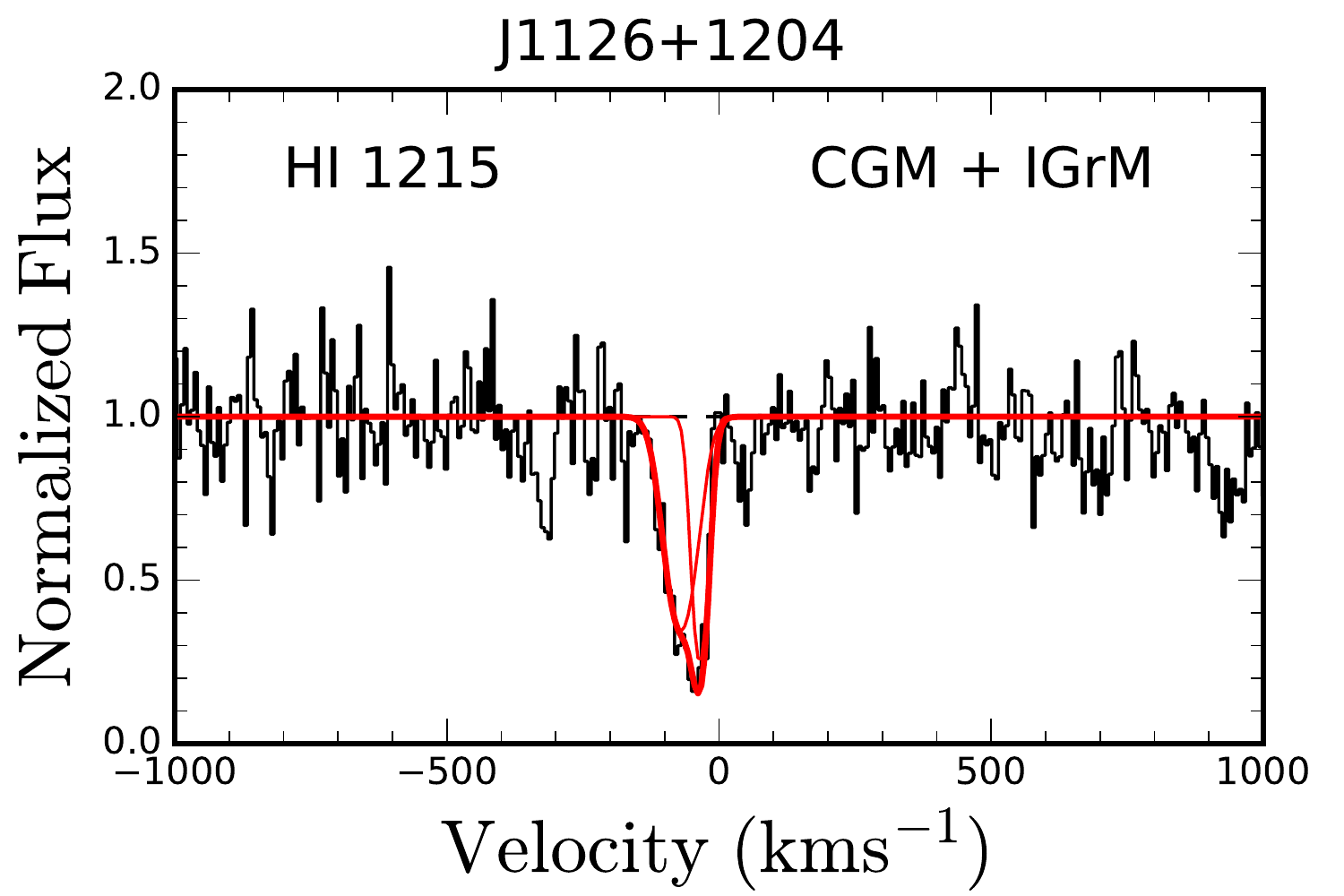}
\caption{The observed spectra of sightline J1126+1204 showing \hi\ absorption. The Voigt profile fits for each component are shown in red.}
\label{J1126_spec}
\end{figure}

\begin{figure}
\centering
\includegraphics[width=2.0in]{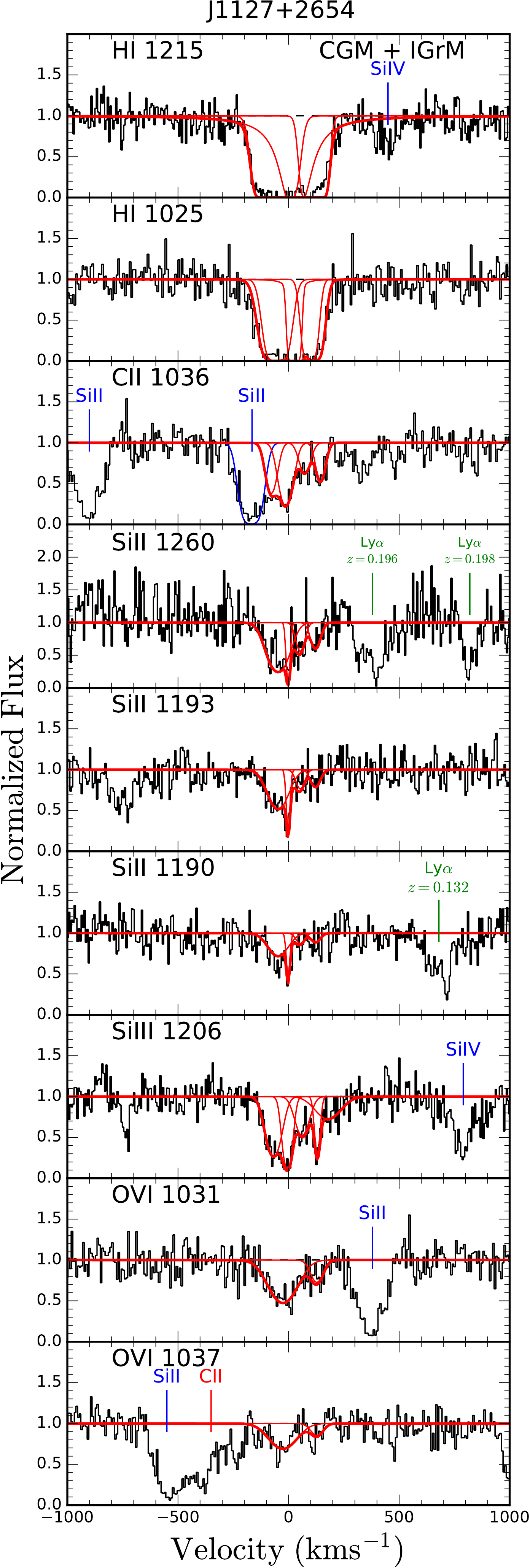}
\caption{The observed spectra of sightline J1127+2654 showing \hi, \cii, \siii, \siiii, and \ovi\ absorption. The Voigt profile fits for each component are shown in red, Milky Way lines are labeled in blue, and intervening \lya\ lines are labeled in green. The \cii\ absorption feature was blended with Milky Way \siii\ $\lambda$1193. The Milky Way \siii\ lines were modeled and the $\lambda$1193 was removed from the spectra, which allows us to model the \cii\ absorption. We place a lower limit on the \cii\ absorption since some flux might be lost during the removal of the Milky Way \siii\ $\lambda$1193 absorption feature. }
\label{J1127_spec}
\end{figure}

\begin{figure}
\centering
\plotone{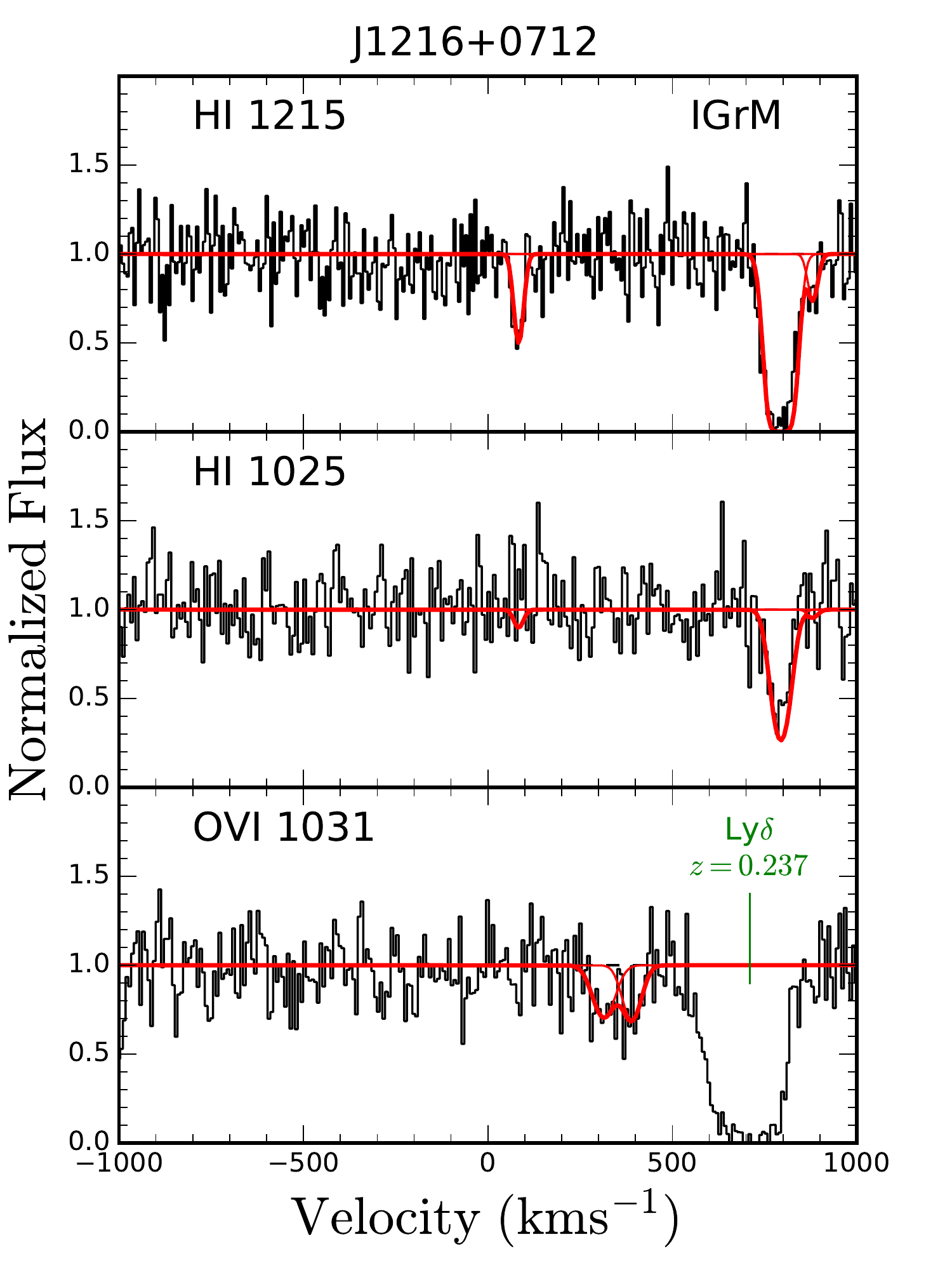}
\caption{The observed spectra of sightline J1216+0712 showing \hi\ and \ovi\ absorption. The Voigt profile fits for each component are shown in red and an intervening Lyman series line is labeled in green. }
\label{J1216_spec}
\end{figure}

\begin{figure}
\centering
\plotone{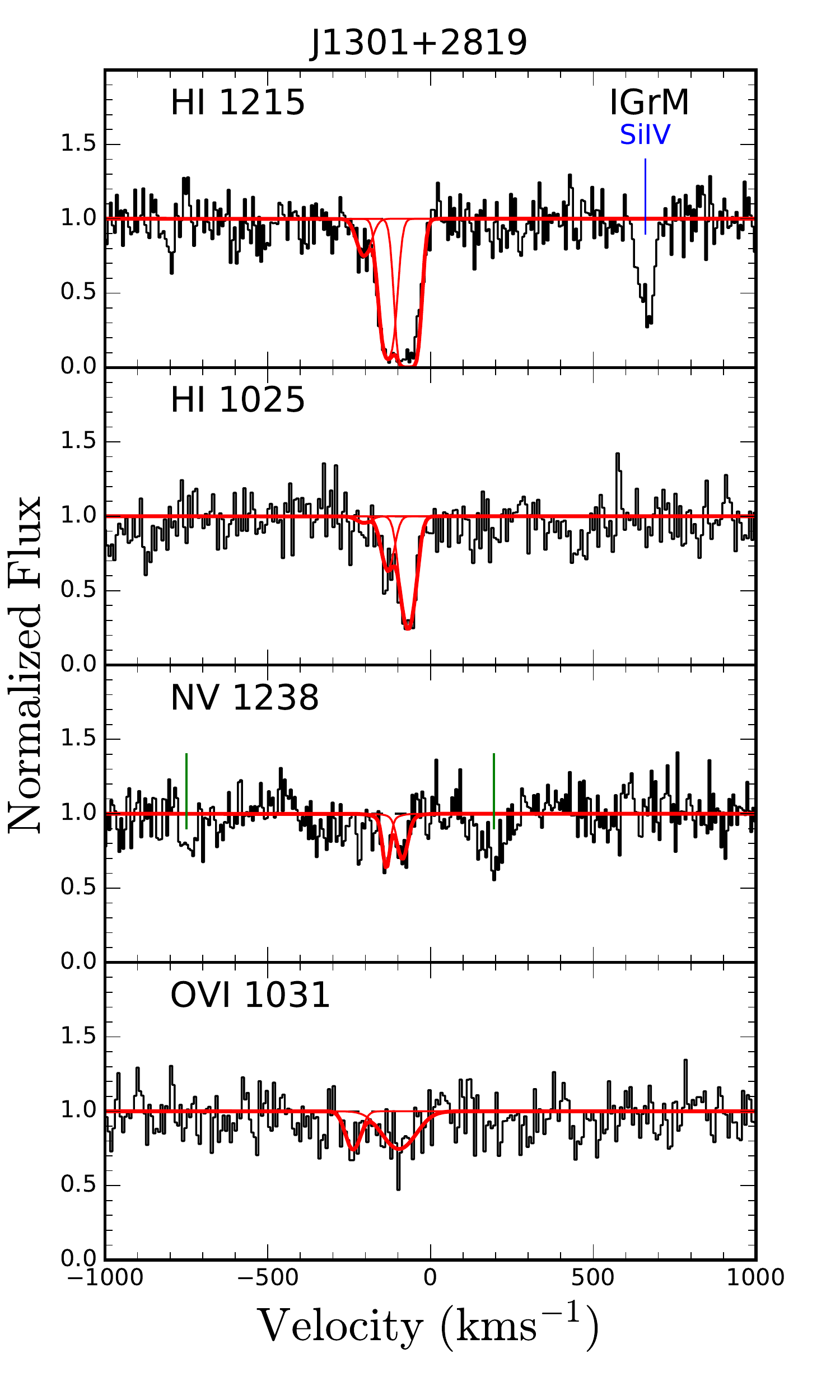}
\caption{The observed spectra of sightline J1301+2819 showing \hi, \nv, and \ovi\ absorption. The Voigt profile fits for each component are shown in red and Milky Way absorption is marked in blue. Two intervening lines could not be identified due to a lack of wavelength coverage and are shown in green.}
\label{J1301_spec}
\end{figure}

\begin{figure}
\centering
\plotone{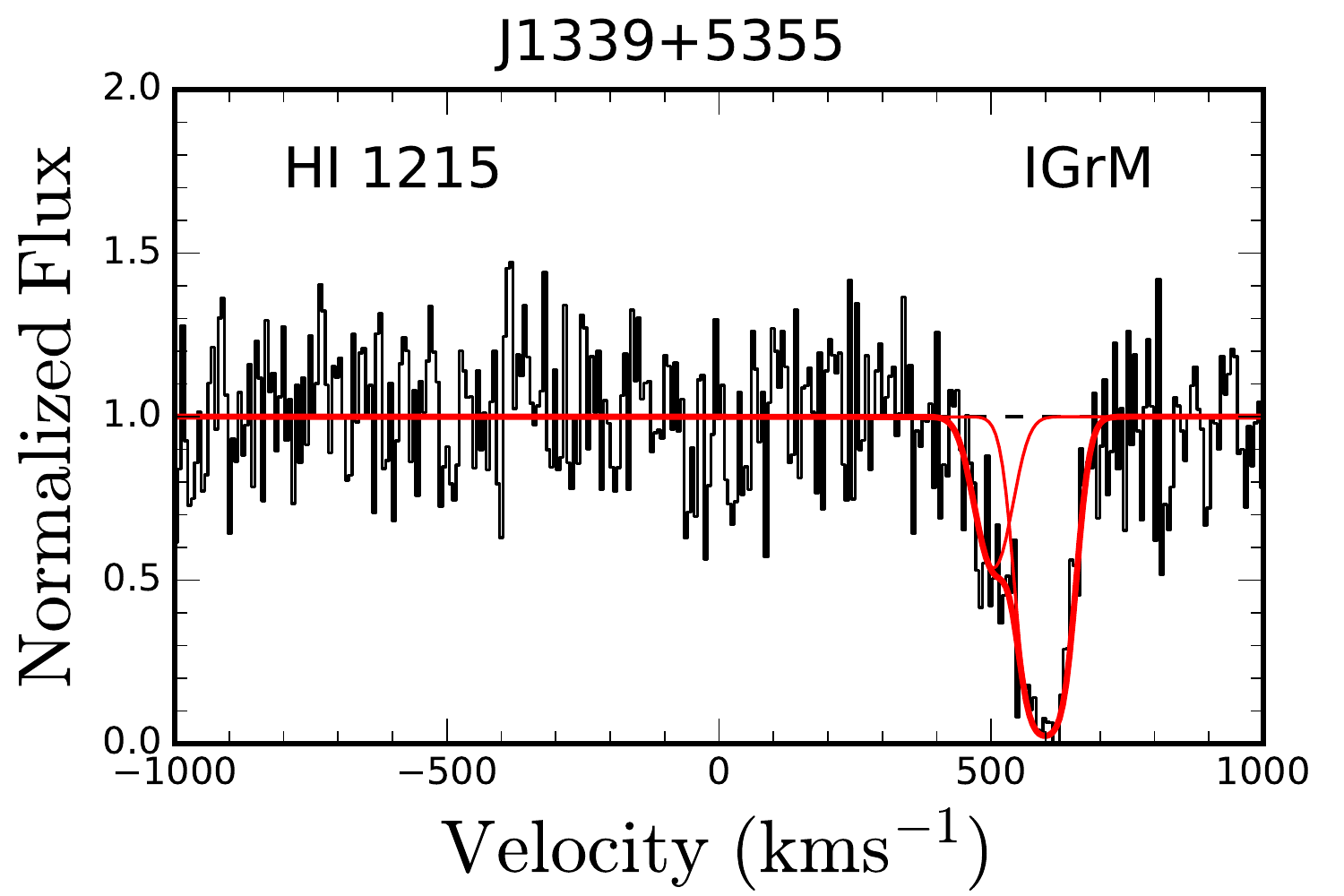}
\caption{The observed spectra of sightline J1339+5355 showing \hi\ absorption. The Voigt profile fits for each component are shown in red.}
\label{J1339_spec}
\end{figure}

\begin{figure}
\centering
\plotone{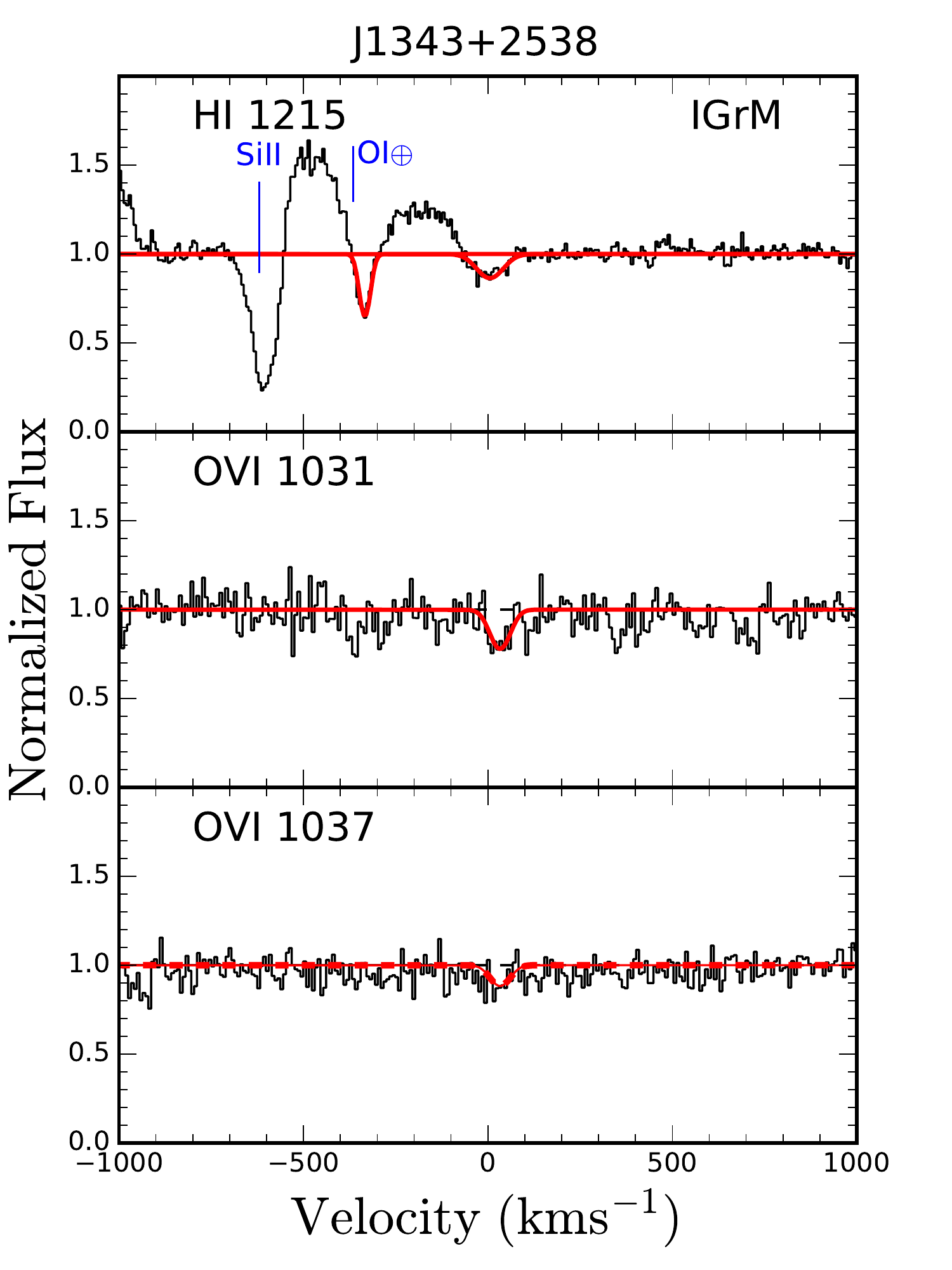}
\caption{The observed spectra of sightline J1343+2538 showing \hi\ and \ovi\ absorption. The Voigt profile fits for each component are shown in red and Milky Way lines are labeled in blue. The emission lines in the \lya\ spectrum are due to geocoronal \ion{O}{1} $\lambda$1304.}
\label{J1343_spec}
\end{figure}

\begin{figure}
\centering
\plotone{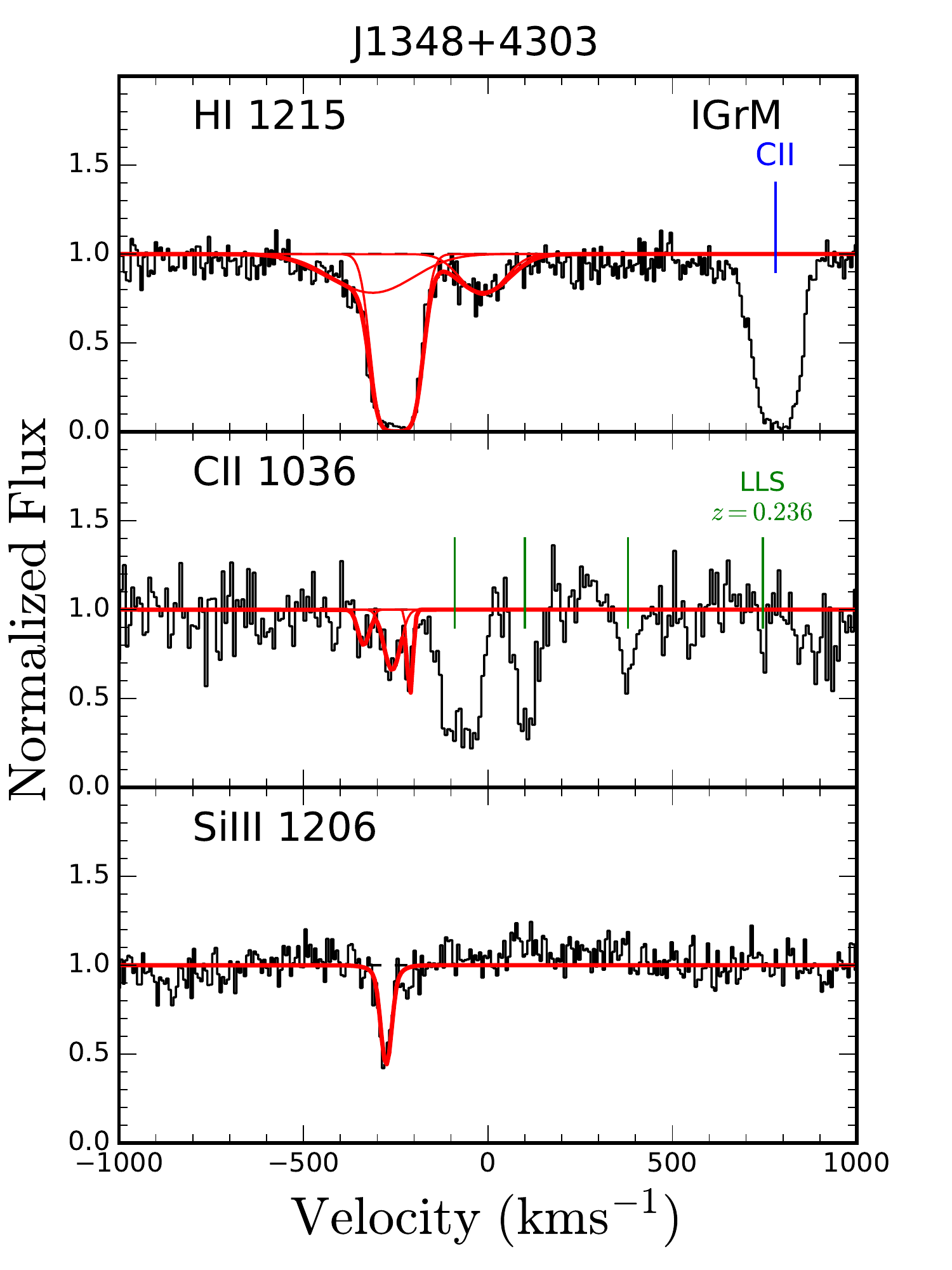}
\caption{The observed spectra of sightline J1348+4303 showing \hi, \cii, and \siiii\ absorption. The Voigt profile fits for each component are shown in red, Milky Way lines are labeled in blue, and intervening absorption lines are marked in green (and labeled if their identity is known).}
\label{J1348_spec}
\end{figure}

\begin{figure}
\centering
\plotone{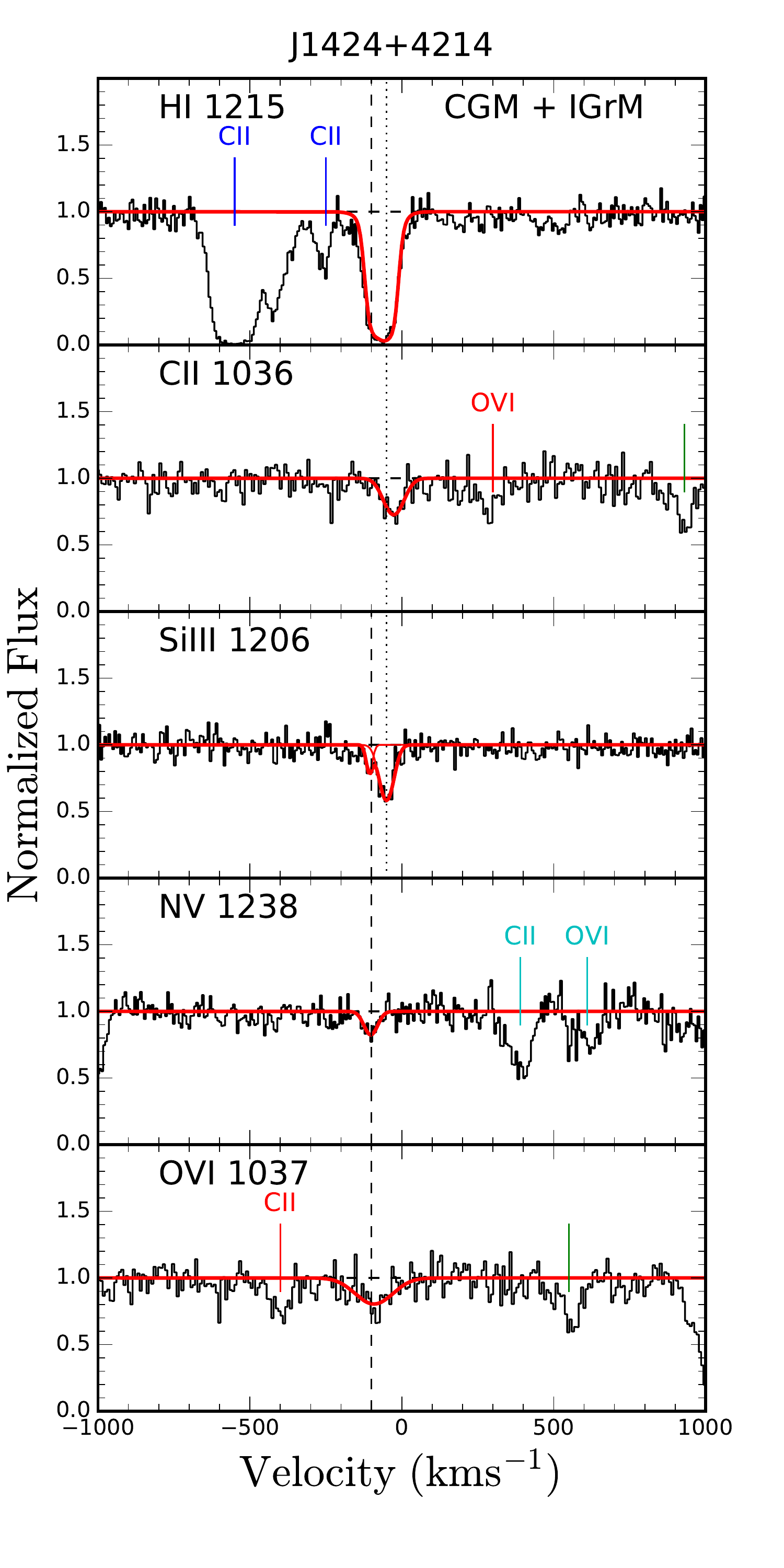}
\caption{The observed spectra of sightline J1424+4214 showing \hi, \cii, \nv, \siiii, and \ovi\ absorption. The Voigt profile fits for each component are shown in red, Milky Way lines are labeled in blue, intervening absorption lines are marked in green, and QSO absorption lines are labeled in cyan. There is no \ovi\ 1031 detection as it is contaminated by an intervening absorber. Two components were fit to the \lya\ profile; however, this dramatically increased the uncertainty in each of the two components. Therefore, to avoid over-fitting, we used a single Voigt profile.}
\label{J1424_spec}
\end{figure}

\begin{figure}
\centering
\plotone{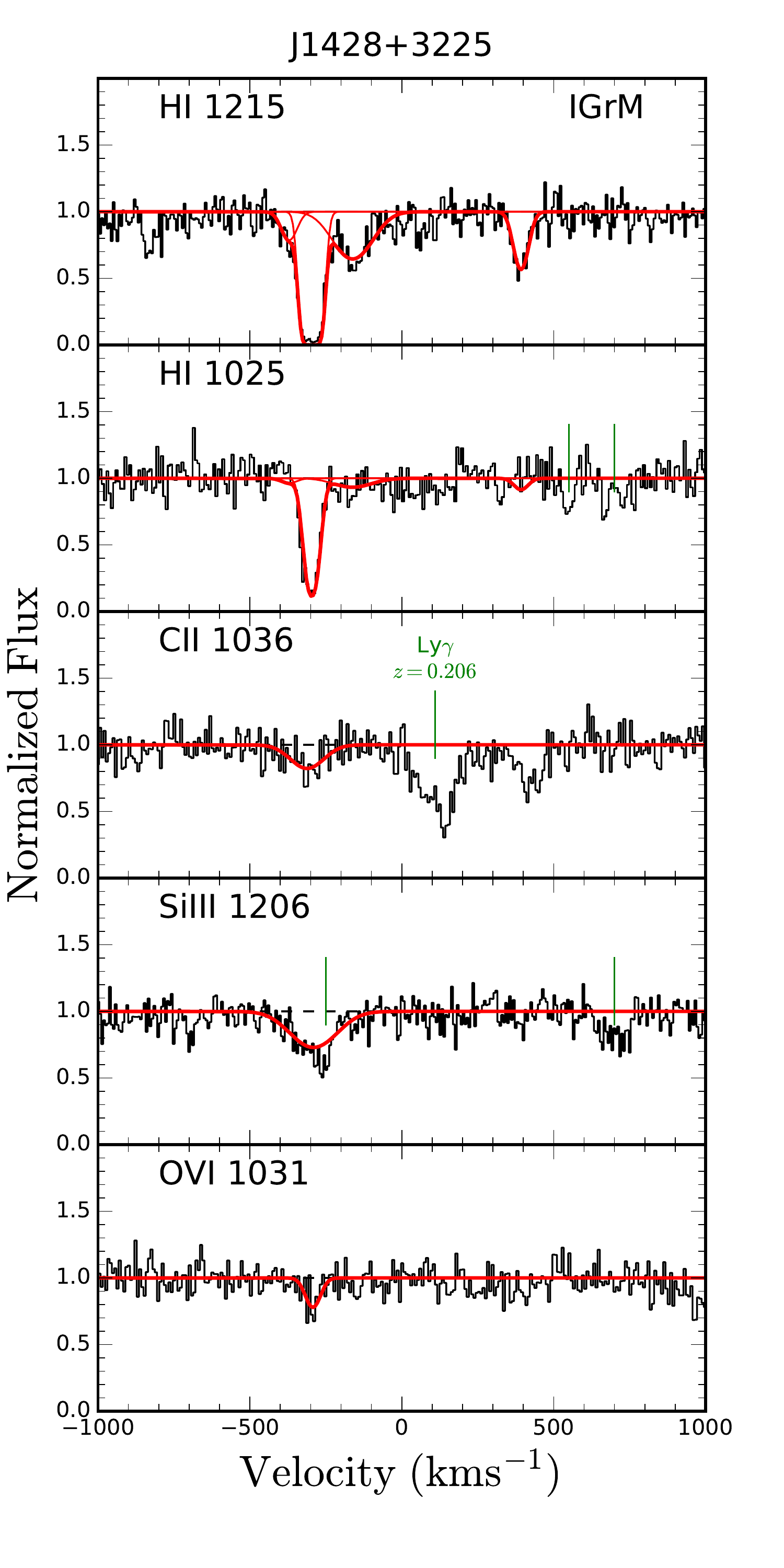}
\caption{The observed spectra of sightline J1428+3225 showing \hi, \cii, \siiii, and \ovi\ absorption. The Voigt profile fits for each component are shown in red and intervening absorption lines are marked in green. The intervening component of the \siiii\ profile is marked in green. Since there are not any other lines associated with this absorption feature, we marked it as intervening. }
\label{J1428_spec}
\end{figure}

\end{document}